\documentstyle[12pt]{article}

\begin{document}

\title{{\Large\bf Supersymmetric \\ 
Quantum Cosmology}\thanks{Topical
review for International Journal of Modern Physics -- A} \\ --- \\ 
{\Large\it Shaken not stirred}}
\author{{\Large\sf P.V. Moniz\thanks{\sf {\rm e-mail:}
 prlvm10@amtp.cam.ac.uk; PMONIZ@Delphi.com}~\thanks{
{\rm URL:} 
{\sf http://www.damtp.cam.ac.uk/user/prlvm10/}}}\\
DAMTP, University of Cambridge \\ Silver Street, Cambridge, CB3 9EW, UK}
\date{{\rm DAMTP -- R95/53}}

\maketitle

\vspace{-1cm}

\begin{abstract}
The canonical 
quantization of  $N=1$ and $N=2$ supergravity theories is reviewed 
  in  this
report.  A special emphasis is given  to  the topic 
of supersymmetric  Bianchi class-A and FRW 
minisuperspaces, namely  in the presence of supermatter fields. 
The quantization of the general theory (including supermatter) is also contemplated.
The issue of quantum physical states is subsequently analysed. 
A discussion on further  research problems still 
waiting to be addressed  is included. An extensive and updated 
bibliography  concludes this review.
\end{abstract}

\section{Introduction}

\indent

Research in supersymmetric  quantum gravity and cosmology using 
canonical methods started about 20 years ago  \cite{1}-\cite{4}. 
Since then, many   papers relating to the subject have
 appeared in the literature \cite{A1}-\cite{OO16a}. In  this review we will  
describe how some difficulties 
(which seemed to disclose a disconcerting future) were sensibly 
subdued. However, there are still other problems 
which have not yet been satisfactorily addressed or even considered.
In fact, 
these may constitute additional  tests that supersymmetric 
quantum gravity and cosmology has to confront. By presenting this review 
  in this way, we aim to further motivate the interested reader 
in  following our steps. 

The  canonical quantization of supergravity theories 
constitutes a fascinating topic. 
A theory of quantum gravity  
is one of the foremost aspirations  in 
theoretical physics 
\cite{5} and a 
promising line of approach is the use of  nonperturbative 
methods \cite{A}-\cite{9}.
The  inclusion of    supersymmetry  seems to yield significative  benefits  as well.

Supersymmetry is an attractive concept whose 
 basic feature is a transformation which relates bosons to fermions and vice-versa 
\cite{6}-\cite{8}. Its  promotion to a gauge symmetry 
has resulted in   an elegant  field theory: {\bf supergravity} \cite{6}-\cite{21}.
One of its more significant features is that the presence of local supersymmetry
  naturally 
implies    space-time to be  curved. Hence, gravity must {\em necessarally} be present.
Supergravity constitutes an extension the general relativity theory 
of gravity \cite{DFSA}. At large scales, supergravity allows to make the same 
predictions for classical tests as general relativity. But at small (microscopic) scales, 
supergravity quantum effects may instead bring about the cancelation of 
infinities otherwise present in several appraches to a quantum gravity theory. 
In particular, supersymmetry may   play an important role when dealing with (ultra violet) 
divergences in quantum cosmology and gravity \cite{9} 
 and removing Planckian masses 
induced by wormholes \cite{A10a,A10,HWH}.
For these and other reasons \cite{DFSA}, some researchers hope that 
nature has reserved a rightful place for supersymmetry and 
supergravity 
  \cite{6}.  
Furthermore,  
it 
would probably be adequate 
for the purpose of studying the  very early universe 
to consider scenarios 
 where both bosonic and  fermionic matter fields
would be present on an equal footing.

The  canonical formulation of N=1 (pure) supergravity was 
presented in  ref. \cite{4}, 
following ref. \cite{1}-\cite{3}.
N=1 supergravity is a theory with a gauge invariances. Namely, 
it is invariant under local coordinate, Lorentz and supersymmetry 
transformations. As standard for theories of this type, 
these gauge invariances are translated into constraints 
for the physical variables. Adopting the Dirac quantization procedure 
(see e.g., ref. \cite{N=2a}-\cite{9}), these constraints become 
operators applied to physical wave functionals, $\Psi$, and 
are subsequently   equated to zero. 
A fundamental feature of the canonical formulation is that 
in finding a physical state, it would be sufficient to just solve
the Lorentz and supersymmetry constraints: the 
algebra of constraints  
implies    that    $ \Psi $  will consequently  obey 
 the Hamiltonian and momentum constraints\footnote{The factor 
ordering in the  Hamiltonian constraint 
is determined by how 
 fermionic derivatives are ordered in the supersymmetry constraints, 
through  their 
anti-commutation relations. When establishing  the ordering of 
the fermionic derivatives   we usually assume that the 
supersymmetry constraints should describe 
 the left and right handed supersymmetry transformations \cite{4}. 
When considering reduced minisuperspace models, different factor ordering 
have been chosen 
\cite{A12}-\cite{A23b}, \cite{A16}-\cite{A18a}.} as well \cite{2,4}. 
For this reason,  
N=1 supergravity is said to constitute  a (Dirac-like) square root of 
  gravity \cite{2}. 
Notice that the Lorentz and supersymmetry 
constraints induce  a set of coupled  
{\em first-order}   differential equations
for $\Psi$ to satisfy. 
However, the analysis (in particular,  obtaining quantum physical solutions) 
of 
the general theory of supergravity is a laborious assignement: one has {\em infinite} 
degrees of freedom. Hence, a   sensible option is to consider    instead 
 simple truncated 
models.

Spatially homogeneous minisuperspace 
models have indeed proved to be a very valuable tool in supergravity 
theories. The study of minisuperspaces 
have  led to important 
and interesting results, pointing  out to  useful lines of research. 
Most of these features  have not been contemplated before in 
 far more complicated situations. Moreover,  we 
hope that some of the results  present 
in the minisuperspace sector will  hold in the full theory.

The reduction of N=1 supergravity in 4 dimensions 
to 1-dimensional models through 
suitable homogeneous ans\"atze  
(see ref. \cite{A5}-\cite{A11}, \cite{A22}-\cite{A18a}) 
 leads to minisuperspaces with N=4 
local supersymmetry. 
FRW models are the simplest ones.   Bianchi models enable us to 
consider  anisotropic gravitational degrees of freedom and thus more gravitino 
modes. 
An important
 feature is   that the fermion
number defined by the Rarita-Schwinger (gravitino) 
field  is then a good quantum
number. Hence,   each sector with a fixed fermion number may be treated
separately.
Nevertheless, we must be aware of the severe
reduction of degrees of freedom that a homogeneity truncation implies. 
The validity of the minisuperspace approximation
in  supersymmetric models is yet a  problem open for 
discussion\footnote{A detailed 
analysis regarding the validity of minisuperspace truncation in 
the context of ordinary 
quantum  cosmology was presented in ref. \cite{15}.}.

Using the triad ADM  
canonical formulation, Bianchi class-A 
models\footnote{ Supersymmetry (as well as other considerations) forbids mini-superspace
models of class $ B $ to be considered in this 
context.} obtained from 
 pure $N=1$ supergravity have been
studied in ref. 
\cite{A1}-\cite{A19} and receive a 
significant contribution in ref. \cite{A21,A20}.
Quantum states are described by a
 wave function of the form  $\Psi( e_{AA'i}
, \psi_{Ai})$ where $e_{AA'i}$ and $\psi_{Ai}$ denote, respectively, the two-component spinor form of 
the tetrad and the spin-$3 \over 2$ gravitino field. 
 The wave function is then   expanded in even powers of $\psi_{Ai}$,
symbolically represented by $\psi^{0}, \psi^{2}, \psi^{4}$ up to $\psi^{6}$, 
 because of the anti-commutations relations of the six spatial components of
 the gravitino fields (see  ref. {\cite{A5,A6,A14,A15,A22} and section 3).

Prior to ref. \cite{A21,A20},  solutions 
were {\it only} present 
in the empty $\psi^0$ (bosonic) and fermionic filled $\psi^6$
sectors. 
But this curious result was joined by yet 
another disturbing one.
When a cosmological constant ($\Lambda > 0$) was added,   it led to the undesirable 
situation that {\it no} 
physical states but the trivial one, $\Psi =0$, were found \cite{A13,A14,A15,D5}.
Regarding the $ k = +1$ FRW model, a bosonic 
state was found, namely  the Hartle-Hawking solution \cite{12} for a 
De Sitter case.

It seemed that the
gravitational and gravitino modes that were allowed to be excited
 contributed in such a way as to
 give only very simple states or even 
forbid any physical solutions. However,  we could {\it not}  identify both  a 
wormhole  \cite{13} and  a Hartle-Hawking \cite{12} state
in the same spectrum of solutions. 
Finding one {\it or}  the other 
depended on   homogeneity conditions imposed on 
the gravitino  (cf. ref. \cite{A7}). Furthermore, these solutions 
were shown (see ref. \cite{C1,C2})  to have no counterpart in the 
full theory:   states with {\em finite} number of 
fermions are {\em impossible} there. 
These results seemed then to suggest 
that minisuperspaces 
could be useless as models of full supergravity.

These problems were then properly subdued in ref. \cite{A21,A20}. 
The cause for the disconcerting results 
mentioned above
was the use (see ref. \cite{A5}-\cite{A12}, \cite{A13}-\cite{A15})
of an ansatz    too restrictive 
 for the $\psi^2$ and $\psi^4$ fermionic 
(middle) sectors. 
More precisely,   gravitational degrees of freedom 
were not   properly taken into account. Hence,   only two Lorentz invariants terms 
in each of the $\psi^2$ and $\psi^4$ sectors were allowed. 
These sectors included only the
modes of the gravitino field. However, 
there may be actually up to  15 such invariants in these sectors,  when
 the gravitational 
modes  are rightfully considered\footnote{{\rm For 
the case of a FRW model without supermatter and due 
to the restriction of the gravitino field to its  
spin-$\frac{1}{2}$ mode component \cite{A9}, the ``old'' ansatz 
 for 
the wave function 
used in ref. \cite{A5}-\cite{A12}, \cite{A13}-\cite{A15}remains valid.}}.
All the 15
amplitudes are assembled in terms of
a single one which must satisfy a Wheeler-DeWitt equation. 
As a consequence, nontrivial solutions were  then  found 
in ref. \cite{A21,A20} for  {\it all} 
the fermionic sectors, of which
infinitely many have fermion number 2 and 4.
Thus, these 
physical states may have direct
analogues  in {\it full} supergravity. 
Furthermore,   supersymmetric minisuperspaces 
recover their
significance as models of the general theory of  supergravity.
For a particular 
factor ordering, the (Hawking-Page) wormhole solution \cite{13}
 is obtained in the 
$\psi^0$ sector {\it and} the (Hartle-Hawking) no-boundary 
solution \cite{12} in the 
$\psi^4$ fermionic sector. I.e., 
in the same spectrum of solutions.

Nevertheless, the improved approach of ref. \cite{A21,A20} 
could not be straightforwardly employed 
to solve the cosmological constant   conundrum: 
terms with $\Lambda$  with violate fermionic number conservation in each 
fermionic sector of $\Psi$. 
Only  an {\it extension} of the ideas present in \cite{A21,A20} using 
Ashtekar variables \cite{14}  allowed   this problem to be solved 
(cf. ref.  \cite{A23b}). 
Solutions have the form of exponentials of 
the $N=1$ supersymmetric Chern-Simons functional, in consistency with ref.
\cite{D8,16a}.

With respect to $k=+1$ FRW models obtained from 
pure N=1 supergravity, 
specific ans\"atze  for the gravitational and
 gravitino fields  
 were employed in ref.
\cite{A8}-\cite{A11} (see ref. \cite{A22}-\cite{A18}, \cite{A19a}-\cite{A18a} as 
well).
The quantum constraint equations are  very simple  and the Hartle--Hawking 
wave-function was found.

The introduction of matter 
(usually denoted as {\it supermatter}) 
 in N=1 supergravity led to new and 
challenging results. A scalar supermultiplet, 
constituted by  complex scalar fields, $\phi,  \bar \phi$ and their 
spin$-{1 \over 2}$ partners, $\chi_A, \bar \chi_{A'}$ was considered in ref. 
\cite{A9}-\cite{A11}, \cite{A16}-\cite{A23}
for FRW models.
A vector supermultiplet, formed by  gauge vector fields $A^{(a)}_\mu$ and  
fermionic partners,  was added in ref. \cite{A17,A18}. 

A wormhole 
 state was found in ref. \cite{A10} but not in ref. 
\cite{A22}. The more general theory 
of N=1 supergravity with 
gauged supermatter (see ref. \cite{8})
 was  employed in ref. \cite{A22}. The reason for the 
discrepancy in ref. \cite{A10,A22} 
 was addressed in ref. \cite{A19a,A23} and identified 
with the type of Lagrange multipliers and fermionic derivative ordering 
that we could use.

As far as a Hartle-Hawking solution is concerned, most of the solutions found in the literature 
bear {\it some} of its properties. Unfortunately, 
the form of the supersymmetry 
constraints that were used could not  determine  the dependence of $\Psi$ on 
the scalar field (cf. ref. \cite{A19a,A23}). Some interesting improvements 
towards this direction were made in ref. \cite{cc,essay}. By employing 
$\phi = r e^{i\theta} = \phi + i\phi_2$ we were able to effectively 
decouple the two degrees of freedom associated with the complex 
scalar fields. The supersymmetry constraints became more manageable 
and a pleasant consequent of this approach was to 
provide a sensible framework where to discuss if conserved 
currents can be defined in supersymmetric quantum cosmology \cite{cc,essay}. 

Within the  more general matter content 
of ref. \cite{8}, the results found in ref. \cite{A17,A18} were quite 
unanticipated. The only allowed  physical state was $\Psi = 0$!
This 
motivated further research present in ref. \cite{A18a}, where some 
discussion concerning  the results 
in \cite{A17,A18}  can be found.
For a 
FRW model but solely in the presence of a vector supermultiplet
we  found  {\it non-trivial} solutions in different 
fermionic sectors \cite{A18a}. Among these we can identify 
(part of) the no-boundary
solution \cite{12, mb} 
and another state  which could be (up to some extent) interpreted as a quantum wormhole state 
\cite{13,4m}.
Overall, the results above mentioned strongly suggest  that the  
treatments of supermatter do 
need a revision. Moreover, it seems 
quite clear that this issue has not 
yet been sucessfully addressed. 

A Bianchi type-IX model 
coupled to a scalar supermultiplet was studied in ref. \cite{A20a,A21a}. 
This
model  bears important differences as far as  
 FRW models with 
supermatter are concerned.
Namely,  anisotropic gravitational degrees of 
freedom are now present. In addition,  
the gravitino spin-$3 \over 2$ modes  can now be  included   
\cite{A14,A15,A21a}. As a consequence, their presence    
 may  play an important role, 
revealing  some of the features of the full theory of N=1 
supergravity  with 
supermatter.

Models with a richer structure can be  found from
{\it extended} supergravities \cite{6,7}, \cite{16}-\cite{21}. 
These are supergravity theories 
with more gravitinos\footnote{$N=1$ supergravity is the simplest theory 
\cite{6,8} with {\it one}  real massless gravitino.  $N=0$ corresponds
 to ordinary general 
relativity. $N=2$ supergravity \cite{6,16,20} realises  Einstein's dream of 
unifying gravity with electromagnetism. This theory contains 
2 gravitinos besides the gravitational  and Maxwell fields. 
It was in this theory 
that finite probabilities for loop diagrams with 
gravitons were first obtained. In particular, the photon-photon 
scattering process which is divergent in 
a Einstein-Maxwell theory, was shown to be finite when $N=2$ supergravity was 
considered (cf. ref. \cite{6}, \cite{21} and references therein).}  
and 
have 
additional symmetries   which couple   several physical 
variables. 

The 
canonical quantization of 
Bianchi class-A models in  
$N=2$  supergravity 
was addressed in ref. \cite{A19c,A24}.
It was found that 
the presence of the Maxwell field in the 
supersymmetry constraints leads to a {\it non-}conservation 
of the fermion number. This then implies a 
 mixing between  Lorentz invariant fermionic 
sectors in the wave function. 
It should be stressed that the intertwining 
between different fermionic sectors in minisuperspaces obtained  
in  $N=1$ supergravity with supermatter is {\it different}   from 
the mixing now caused by the Maxwell field.

Another crucial step towards a better understanding of 
the 
canonical quantization 
of N=1 supergravity 
is to  relate the set of states found in supersymmetric minisuperspaces 
with any  physical states  obtained  in the full theory.
It was shown in ref. \cite{C1,C2} 
that the physical states of full N=1 pure  supergravity can only have
infinite fermionic number. 
Recently, a relation between the minisuperspace sector and the 
 general theory was 
proposed in ref. \cite{A21,A20}.
  A (formal)  quantum state  with infinite number of fermions 
was then found in ref. \cite{C3}. A wormhole state could be identified 
but   the same was not possible for a Hartle-Hawking  
solution. 
The generalization of the canonical formalism of N=1 supergravity \cite{4}  
to include supermatter fields was subsequently described in 
ref. \cite{A15a,A16,A21a}.

We hope   this rather detailed  introduction
has motivated the reader to bear with us for what 
follows. We will emphasize some of the technical aspects 
found in supersymmetric quantum cosmology. In this way, anyone 
enthusiastic to persue this line of research will get a fair view 
of {\it which} and {\it how} problems have been  dealt with. We  
we will  describe the main achievements and point out to 
further issues waiting for adequate explanations. 
In reviewing a subject of this type where there are a large number of 
related topics, solutions and approaches, a decision has to be taken 
with respect to how much information to include. We hope to have made this 
review self-contained and useful as a reference, while keeping it clear and 
readable.

The review henceforth 
presented is organized as follows. 
The canonical quantization of the general theory of N=1 supergravity 
is described  in section 2. In section 2.1 we summarize the
 basic results for pure N=1 supergravity (see subsection 2.1.1), 
and include the 
 generalization for  all matter fields (see subsection 2.1.2). The 
issue of finding physical states 
in the general theory is addressed in section 2.2 and a   solution 
is described in section 2.3.
In section 3 we will then analyse  the 
canonical quantization of supersymmetric minisuperspaces. 
We will employ here 
a ADM metric representation point of view, 
together with   a differential operator representation 
for the fermionic variables\footnote{It is our 
thinking that a differential operator representation for the 
fermionic variables constitute the rightfull approach. 
It is totally consistent with the existence of second-class constraints 
and subsequent Dirac brackets. These then imply that fermionic 
variables and their Hermitian conjugates are intertwined 
within a canonical  coordinate-momentum relation 
\cite{3,4,A5}-{A24}. There are, however, other approaches. 
Ashtekar and loop variables were used in \cite{D1}-\cite{OO16a}. 
The method employed in \cite{B1}-\cite{B5} is based 
on a $\sigma-$model approach to supersymmetric quantum 
mechanics. Finally, a matrix representation for the gravitinos was used 
in ref.  \cite{A1}-\cite{A4}. All these approaches share some 
similarities but also have specific differences in method and results. 
Moreover, a clear analysis establishing {\it if} 
(up to any extent) and {\it how} they are related is yet to 
be achieved. }
Subsection 3.1 includes  
  models obtained from pure N=1 supergravity, while 
a cosmological constant term is added in subsection 3.2. 
Bianchi class A models are discussed in subsections 
3.1.1 and 3.2.1, while 
the particular case of FRW models 
is  addressed in 3.1.2 and 3.2.2. 
FRW models are also  explored in subsection 3.3, 
 where supermatter fields 
are then brought into study. Their canonical 
quantization  is presented  in subsection 3.3.1. Models with scalar supermultiplets 
are used in 3.3.2, while in 3.3.3 
we add a vector supermultiplet. 
FRW models with only vector supermultiplets 
are described in subsection 3.3.4. A Bianchi-IX model 
in the presence of scalar supermultiplets is analysed in 3.3.5.
Bianchi class-A models derived from N=2 supergravity are explored in section 
4.   Finally, section 5 includes 
a discusion on the results obtained so far in 
supersymmetric quantum cosmology. In addition, 
an outlook on  further research projects in canonical quantum   
 supergravity  still waiting to be satisfactorily addressed  
is included. 


\section{Canonical quantization of N=1 supergravity}

\indent 

This subject is quite long and only a brief summary will be presented 
here. Some  problems and related issues will  be 
    discussed, together with some  
recent   results. Before we proceed, 
the following should be pointed out. The basic features and issues of 
supersymmetry, supergravity and related concepts 
will {\it not} be addressed here in detail. That would go far beyond 
the scope and aim of this review. The reader 
interested on such topics is therefore invited to consult, e.g.,   ref. 
\cite{6}-\cite{8}. 

\subsection{General formalism}

\indent 

In this subsection we will describe  some features  of 
 the Hamiltonian formulation of  N=1 supergravity  and its  
canonical quantization  \cite{4,A15a,A16,A21a,C1,C2,C3}. The original 
  treatments can be found in ref. \cite{1}-\cite{4}.

\subsubsection{Pure N=1 supergravity}

\indent

\indent

The action of pure N=1 supergravity is taken to be \cite{4,6}
\begin{equation}
S = \int d^4 x \left[ {1 \over 2 \kappa^2} (\det e) R 
+ {1 \over 2} \epsilon^{\mu \nu \rho \sigma} ( \bar  \psi^{A'}_{~~\mu} e_{A A' \nu}
D_\rho \psi^A_{~~\sigma} + {\rm H.c.} ) \right]~, 
\label{eq:2.7}
\end{equation}
where $\kappa^2$ denotes $8\pi$ times the gravitational constant.
Here $R$ represents the Ricci scalar curvature, calculated 
from $e^{AA'}_{~~\mu}$ and $\psi^A_{~\mu}, \bar\psi^{A'}_{~\mu}$. 
$e^{AA'}_{~~\mu}$ is the spinorial form of the tetrad 
$e^a_\mu$. 
$ e^{A A'}_{~~~~i} $ gives the three-metric according
to $ h_{i j} = - e^{A A'}_{~~~~i}~e_{A A' j} $. It is given by $ e^{A 
A'}_{~~~~i} = e^a_{~i} \sigma_a^{~A A'} $ , where $ e^a_{~i} $ are 
the spatial tetrad components and $ \sigma_a^{~A A'} $ are the Infeld--van der 
Waerden translation symbols \cite{3,4,6,7,8}. $\psi^A_{~\mu}, \bar\psi^{A'}_{~\mu}$ denote 
the gravitino (Rarita-Schwinger) fields and a ``overline''   represents 
the Hermitian conjugate (H.c.).

These variables transform under supersymmetry - $\delta_{(s)}$ -, Lorentz - $\delta_{(L)} $ -   and local coordinate transformations 
- $\delta_{(lc)} $ - 
as follows:
\begin{eqnarray}
\delta_{(s)} e^{AA'}_{~~~~\mu} &=& - i 
\kappa 
\left( \epsilon^A \bar \psi^{A'}_{~~\mu} + 
\bar \epsilon^{A'} \psi^A_{~~\mu} \right)~,  
\label{eq:2.1}     \\
\delta_{(s)} \psi^A_{~~\mu} &=  &{2 \over \kappa}   D_\mu \epsilon^A ~,
  \label{eq:2.2}      \\
\delta_{(L)} 
 e^{AA'}_{~~~~\mu} &= &N^A_{~~B} e^{BA'}_{~~~~\mu} + \bar N^{A'}_{~~B'}
e^{AB'}_{~~~~\mu}~, \label{eq:2.3}      \\ 
\delta_{(L)} \psi^A_{~~\mu} &=&N^A_{~~B} \psi^B_{~~\mu}~, 
\label{eq:2.4}     \\
\delta_{(lc)} e^{AA'}_{~~~~\mu} &= &\xi^\nu \partial_\nu e^{AA'}_{~~~~\mu} + 
e^{AA'}_{~~~~\nu} \partial_\mu \xi^\nu~, 
\label{eq:2.5}
\\
\delta_{(lc)} \psi^A_{~~\mu} &=& \xi^\nu \partial_\nu \psi^A_{~~\mu} + \psi^A_{~~\nu} \partial_\mu 
\xi^\nu~, \label{eq:2.6}
\end{eqnarray}
together with the H.c. of (\ref{eq:2.2}), (\ref{eq:2.4}), (\ref{eq:2.6}). 

The 
derivative operator $ D_\mu $
acts on spinor-valued forms and only notices their spinor 
indices, but not their space-time indices, e.g.,
\begin{equation}
D_\mu \psi^A_{~~\nu} = \partial_\mu \psi^A_{~~\nu} + \omega ^A_{~~B \mu} 
\psi^B_{~~\nu}~. 
\label{eq:2.8}\end{equation}
Notice the connection forms 
$ \omega ^{a b}_{~~\mu} = 
\omega ^{[a b]}_{~~~~\mu} $ in their 
 spinorial version $ \omega ^{A 
B}_{~~~\mu} $
(cf. \cite{3,4,6}) with 
$
 \omega ^{A B}_{~~~\mu} = {}^s\omega ^{A B}_{~~~\mu} + \kappa^{A B}_{~~~\mu} 
$. 
Here $ \kappa^{\nu \rho}_{~~\mu}$ is the 
contorsion tensor, related to the torsion  
 $ S^{A A'}_{~~~~\mu \nu} = - {i \kappa^2 \over 2} \bar
  \psi^{A'}_{~~[ \mu}
\psi^A_{~~\nu ] }$, 
and ${}^s\omega ^{A B}_{~~~\mu}$ is the usual 
torsion-free connection \cite{4}.


From the action (\ref{eq:2.7}) the canonical 
momenta to $\psi^A_{~i}, \bar\psi^{A'}_{~i}$ are 
\begin{equation}
\pi_A^{~~i} = - {1 \over 2} \varepsilon^{i j k} \bar  \psi^{A'}_{~~j} e_{A A' k}~,
~\bar \pi_{A'}^{~~i} = {1 \over 2} \varepsilon^{i j k} \psi^A_{~~j} e_{A A' k}~. 
\label{eq:2.9}\end{equation}
Expressions (\ref{eq:2.9}) constitute (second-class) constraints 
(see, e.g., ref. \cite{9} for more details on this 
and related topics). In fact, notice that 
(\ref{eq:2.7}) is linear in $D_\rho\psi^A_{~\mu}, D_\rho\bar\psi^{A'}_{~\mu}$.
We can then eliminate 
$
\pi_A^{~~i} $ and $ \bar \pi_{A'}^{~~i} $  as dynamical 
variables through the second-class constraints (\ref{eq:2.9}).
Consequently, the  basic dynamical variables in the 
theory can then be reduced to $ e^{A A'}_{~~~~i},~p_{A A'}^{~~~~i},~ 
\psi^A_{~~i} $, and $ \bar  \psi^{A'}_{~~i} $, where $ p_{A A'}^{~~~~i} $ is the momentum conjugate to $ e^{A A'}_{~~~~i}$.  The momentum $ p_{A A'}^{~~~~i} $ can be expressed in terms
of the extrinsic 
curvature  $ K_{i j} $. This one  includes, besides the usual symmetric part 
dependent on the tetrad variables, an anti-symmetric part due to 
torsion (i.e., gravitino fields)  \cite{4}.

Poisson brackets can be defined in a classical theory containing both bosons 
and fermions \cite{3,4,A9}. After  the 
elimination 
$ 
\pi_A^{~~i} $ and $ \bar \pi_{A'}^{~~i} $ we  find the 
following Dirac brackets $([~]_{D})$:
\begin{eqnarray}
\left[ e^{A A'}_{~~~~i} (x), e^{B B'}_{~~~~j} (x') \right]_{D} & = &  0~,
\left[ e^{A A'}_{~~~~i} (x), p_{B B'}^{~~~~j} (x') \right]_{D} =
\epsilon^A_{~~B} \epsilon^{A'}_{~~B'} \delta_i^{~j} \delta (x, x')~, 
\label{eq:2.11}
\\
\left[ p_{A A'}^{~~~~i} (x), p_{B B'}^{~~~~j} 
(x') \right]_{D} & = & {1 \over 4} 
\epsilon^{j l n} \psi_{B n} D_{A B' k l} 
\varepsilon^{i k m} \bar  \psi_{A' m} \delta (x, x') + {\rm H.c.}~, 
\label{eq:2.12} \\
\left[ \psi^A_{~~i} (x), \psi^B_{~~j} (x') \right]_{D} &= &  0~, 
\left[ \psi^A_{~~i} (x), \bar  \psi^{A'}_{~~j} (x') \right]_{D} = - D^{A A'}_{~~~~i 
j} \delta (x, x') 
\label{eq:2.13} \\
\left[ e^{A A'}_{~~~~i} (x), \psi^B_{~~j} (x') \right]_{D} &=& 0 ~,~
\left[ p_{A A'}^{~~~~i} (x), \psi^B_{~~j} (x') \right]_{D} =  {1 \over 2} 
\varepsilon^{i k l} \psi_{A l} D^B_{~~A' j k} \delta (x, x')~, \label{eq:2.14}
\end{eqnarray}
and Hermitian conjugate relations, where $\epsilon^{AB}$ is the alternating spinor and 
\begin{equation}
 D^{A A'}_{~~~~j k} = - 2 i h^{- 1 / 2} e^{A B'}_{~~~~k} e_{B B' j} n^{B
A'}~. \label{eq:2.15}
\end{equation}

Because of the invariance of the Lagrangian density in eq. (\ref{eq:2.7}) under 
local Lorentz transformations, these variables obey the primary constraints
$ J_{A B} = 0, \bar{J}_{A' B'} = 0~,$
where
\begin{equation} 
 J_{A B} = e_{(A}^{~~A' i} p_{B) A' i} + \psi_{(A}^{~~i} \pi_{B) i}~. 
\label{eq:2.10}
\end{equation}

The Hamiltonian takes a form
of the standard type for theories with gauge invariances:
\begin{equation} 
 H = \int d^3 x \left( N {\cal H}_\perp + N^i {\cal H}_i + \psi^A_{~~0} S_A + 
\bar S_{A'}
\bar  \psi^{A'}_{~~0} - 
M _{A B } J^{A B} - \bar  M_{A' B' } \bar 
J^{A' B'}\right)
~.
\label{eq:2.16}\end{equation}
The procedure to find explicit  expressions 
for the constraints 
is  simple. It requires   the calculation 
of the conjugate momenta of the
 dynamical variables and then evaluate the   Hamitonian (\ref{eq:2.16}). 
In the case of the supersymmetry constraits, e.g., 
  we  read out the coefficients of $\psi_{0}^{~A}$    and 
 $\bar \psi_{0}^{~A'}$   from this expression in 
order to get the $S_{A}$ and $\bar S_{A'}$ 
constraints, respectively. 
In fact,  $ N, N^i,~\psi^A_{~~0}$, $~\tilde 
\psi^{A'}_{~~0},~M _{A B } $ and $ \bar M _{A' B' } $ 
constitute Lagrange
multipliers for the generators $ {\cal H}_\perp,~{\cal H}_i,$ $ 
~S_A,~\bar 
  S_{A'},~J^{A B} 
$ and $ \bar  J^{A' B'} $, which are formed from the basic
variables. Here $ N $ and $ N^i $ are the lapse and shift 
\cite{4}, $ 
\psi^A_{~~0} $ and $ \bar  \psi^{A'}_{~~0} $ are the zero components of
 $  \psi^A_{~~\mu},~\bar  \psi^{A'}_{~~\mu} 
$ \cite{3,4}. $ {\cal H}_\perp $ gives the generator of (modified) 
normal
displacements applied to the basic Hamiltonian variables, $ {\cal H}_i $ gives the 
generator of (modified) spatial coordinate transformations, $ S_A $ and $ \bar  
S_{A'} $ are the generators of supersymmetry transformations, and $ J^{A B} $ 
and $ \bar  J^{A' B'} $ are the generators of local Lorentz transformations. Here 
`modified' indicates that a certain amount of 
  supersymmetry and local Lorentz transformations has been added
to the coordinate transformation. 
Classically, the dynamical variables obey the (first-class) constraints
\begin{equation}
 {\cal H}_\perp = 0, ~ {\cal H}_i = 0, ~ S_A = 0,  ~\bar  S_{A'} = 0, ~ J^{A B} = 0, 
~\bar  J^{A' B'} = 0~. 
\label{eq:2.17}\end{equation}

The symmetries of the theory in Hamiltonian form are most easily understood 
by rewriting $ H $ so that the Lagrange multipliers $M_{AB}, 
\bar M_{A'B'}$ of $ J^{A B},~\bar  J^{A'
B'} $ are minus the zero components $ \omega _{A B 0},~\bar  \omega _{A' B' 0} $ of the
connection forms \cite{3,4}. 
We then get  the explicit form of
\begin{eqnarray}
S_A & =& \varepsilon^{i j k} e_{A A' i}~^3D_j \bar  \psi^{A'}_{~~k} - {1 \over 2} i
\kappa^2 p_{A A'}^{~~~~i} \bar  \psi^{A'}_{~~i} \nonumber \\
& + & {1 \over 2} \kappa^2 h^{1 / 2} n_{A A'} \bar  \psi^{A'}_{~~i} e^{B B'}_i 
\bar  
\psi_{B'}^{~~[ j} \psi_B^{~~i ]} 
- {1 \over 4} i \kappa^2 \epsilon^{i j k} n_{A A'} \bar  \psi^{A'}_{~~j} n_{B B'}
\bar  \psi^{B'}_{~~k} \psi^B_{~~i}~. \label{eq:2.19}
\end{eqnarray}
Here the covariant spatial 
derivative $ ^3 D_i $ acts on spinor-valued spatial forms according
 to 
\begin{equation}  ^3 D_i \psi^A_{~~j} = \partial_i \psi^A_{~~j} +~^3\!\omega ^A_{~~B i}
\psi^B_{~~j}~. 
\label{eq:2.20}\end{equation}
The spatial connection forms $ ^3 \omega^{A B}_{~~~i} $ can be expressed as
$ ^3 \omega^{A B}_{~~~i} =~^{3 s}\!\omega^{A B}_{~~~i} +~^3\!\kappa^{A B}_{~~~i}$

In a quantum representation
Dirac brackets must be 
replaced by commutators $ [~,~] $ or anticommutators $ \{~,~\} $. 
The quantum state is described by a wave functional, $ \Psi $, 
depending on  $ e^{A A'}_{~~~~i} (x) $ 
but the fermionic variables $ \psi^A_{~~i}, \bar 
\psi^{A'}_{~~i} $ {\it cannot} be treated on the same footing. Eq.  
 (\ref{eq:2.13}) shows that one cannot have a 
simultaneous eigenstate of $ \psi^A_{~~i} $ and $ \bar  \psi^{A'}_{~~i} $. 
We may  choose $ \psi^A_{~~i} $, so that we have a wave functional $ \Psi \left[ 
e^{A 
A'}_{~~~~i} (x), \psi^A_{~~i} (x) \right] $, and the operator $ \psi^A_{~~i} (x) $
is given by multiplication on the left by $ \psi^A_{~~i} (x) $. Then, in 
order to satisfy the anticommutation relation following from eq. 
(\ref{eq:2.13}), $ \bar 
\psi^{A'}_{~~i} $ will be given by the   operator
\begin{equation} 
\bar  \psi^{A'}_{~~i} (x) = - i \hbar D^{A A'}_{~~~~j i} (x) {\delta \over \delta 
\psi^A_{~~j} (x)}~, \label{eq:2.21}\end{equation}
where as before $ \psi^A_{~~i} $ must be brought to the left in any 
expression before the {\it 
functional} derivative $\frac{\delta}{\delta \psi^A_j}$ 
is applied. 
A
 representation for  the operator $ p_{A A'}^{~~~~i} (x) $  consistent with 
(\ref{eq:2.11}), (\ref{eq:2.12}) is 
\begin{equation}  
p_{A A'}^{~~~~i} (x) =  - i \hbar {\delta \over \delta e^{A A'}_{~~~~i} (x)} 
- {1 \over 2} i \hbar \varepsilon^{i j k} \psi_{A j} (x) D^B_{~~A' l k} (x) {\delta 
\over \delta \psi^B_{~~l} (x)}~. 
\label{eq:2.22}
\end{equation}
At quantum level
\begin{eqnarray}
J_{A B} &= & - {1 \over 2} i \hbar \left[ e_{B~~i}^{~~A'} {\delta \over \delta e^{A 
A'}_{~~~~i}} + e_{A~~i}^{~~A'} {\delta \over \delta e^{B A'}_{~~~~i}}
+ \psi_{B i} {\delta \over \delta \psi^A_{~~i}} + \psi_{A i} {\delta \over \delta
\psi^B_{~~i}} \right] ~,
\label{eq:2.23}
\\
\bar  J_{A' B'} &=&  - {1 \over 2} i \hbar \left[ e^A_{~~B' i} {\delta \over \delta e^{A
A'}_{~~~~i}} + e^A_{~~A' i} {\delta \over \delta e^{A B'}_{~~~~i}} \right]~.
\label{eq:2.24}
\end{eqnarray}

The form of $ \bar  S_{A'} $ given by conjugating eq. (\ref{eq:2.19}) 
has extra $ \psi, 
\bar  \psi $ dependence involved through torsion in the derivative $ ^3D_j 
\psi^A_{~~k} $. When this is expanded out in terms of the torsion-free 
derivative $ ^{3 s}D_j \psi^A_{~~k} $ based on the connection $ ^{3 s} \omega ^{A 
B}_{~~~i} $ we find that all the $ \bar  \psi \psi \psi $ 
terms cancel to give classically
$
 \bar  S_{A'} = \varepsilon^{i j k} e_{A A' i}~^{3 s} D_j \psi^A_{~~k} + {1 \over 2} i
\kappa^2 \psi^A_{~~i} p_{A A'}^{~~~~i}~. $
Quantum mechanically we get 
\begin{equation} 
 \bar  S_{A'} = \epsilon^{i j k} e_{A A' i}~^{3s}D_j \psi^A_{~~k} + {1 \over 2}
\hbar \kappa^2 \psi^A_{~~i} {\delta \over \delta e^{A A'}_{~~~~i}}~, 
\label{eq:2.25}
\end{equation}
where $\hbar$ is the Planck constant. 
Correspondingly, the operator
$
 S_A = \varepsilon^{i j k} e_{A A' i}~^{3s}D_j \bar  \psi^{A'}_{~~k} - {1 \over 2} i
\kappa^2 p_{A A'}^{~~~~i} \bar  \psi^{A'}_{~~i}$,
given by the Hermitian adjoint of eq. (\ref{eq:2.25}), is of second order:
\begin{equation}
 S_A = i \hbar~^{3s}D_i \left[ {\delta \over \delta \psi^A_{~~i}} \right] + {1 \over 2} i
\hbar^2 \kappa^2 {\delta \over \delta e^{A A'}_{~~~~i}} \left[ D^{B A'}_{~~~~j i} {\delta \over 
\delta
\psi^B_{~~j}} \right]~.
\label{eq:2.26}\end{equation}

The supersymmetry and Lorentz  transformation properties
are all that is required of a physical state. 
Notice that  $ {\cal H}_{A A'} $ follows 
from the anti-commutator $\{ S_A,~\bar  S_{A'}
\} \sim~{\cal H}_{A A'} $,  which  classically differs from 
this one   by terms linear
in $ J^{A B} $ and $ \bar  J^{A' B'} $ \cite{4}.

\subsubsection{N=1 supergravity with supermatter}

\indent

The generalization of the canonical formulation of pure N=1 supergravity 
to include matter fields was described in ref. \cite{A15a,A16,A21a}.

Besides its dependence on the tetrad
$ e^{A A'}_{~~~~\mu} $ and 
 the gravitino field $ 
 \psi^A_{~~\mu}, \bar \psi^{A'}_{~~\mu}$,
the  most  general N=1 supergravity  theory coupled to gauged supermatter \cite{8}
 also includes  a vector field $ A^{(a)}_\mu $
labelled by an index $ (a) $, its spin-$\frac{1}{2} $ partners $ \lambda^{(a)}_A, 
\bar \lambda^{(a)}_{A'} $, a family of scalars $  \Phi^I, \Phi^{J^*}
 $ and their spin-$\frac{1}{2}$ partners $  \chi^I_A, \bar \chi^{J^*}_{A'} $.  
Its Lagrangian is given in eq. (25.12) of ref. \cite{8}.
The indices $ I, J^*$ are K\"ahler indices, and
there is a K\"ahler metric
$ g_{I J^*} =   K_{I J^*} $
on the space of $  \Phi^I, \Phi^{J^*}$ (the 
K\"ahler manifold), where $ K_{I J^*} $ is a shorthand
for $ \partial^2 K / \partial \Phi^I \partial \Phi^{J^*} $ with $ K $ the K\"ahler potential. 
Each
index $ (a) $ corresponds to an independent (holomorphic) 
Killing vector field $X^{(a)}$ 
of the
K\"ahler geometry.
Killing's equation implies that there exist real scalar functions $ D^{(a)} 
( \Phi^I, \Phi^{I^*} ) $ known as Killing potentials.

Analytic isometries that preserve the analytic structure of the 
manifold are associated with Killing vectors
\begin{equation} 
X^{(b)} = X^{I (b)} ( \phi^J )~{\partial \over \partial \phi^I}~, ~
X^{^* (b)} = X^{I^* (b)} ( \phi^{J^*} )~{\partial \over \partial \phi^{I^* }}~.
\label{eq:2.27}   \end{equation}
The index $(b)$ labels the Killing vectors and runs over the 
dimension of the isometry (gauge) group $\hat G$. Killing's equation integrability 
condition is equivalent to the statement that there exist scalar functions 
 $D^{(a)}$ such that
\begin{equation} 
g_{I J^*} X^{J^* (a)} = i~{\partial \over \partial \phi^I}~D^{(a)}~, ~
g_{I J^*} X^{I (a)} = - i~{\partial \over \partial \Phi^{J^*}}~D^{(a)}~. 
  \label{eq:2.28}\end{equation}
The Killing potentials $D^{(a)}$ are defined up to constants $c^{(a)}$.
They further satisfy 
\begin{equation}\left[ X^{I (a)} {\partial \over \partial\phi^I} + X^{I* (a)} {\partial 
\over \partial \phi^{I*}}\right]D^{(b)} = 
- f^{abc} D^{(c)}.~ 
\label{eq:2.21a}\end{equation} This fixes the constants $c^{(a)}$ for non-Abelian 
gauge groups.

Within the full theory of N=1 supergravity, 
our field variables 
are transformed as follows  (it should be stressed that eq. 
(\ref{eq:2.1}), (\ref{eq:2.3})--(\ref{eq:2.6}) 
 remain valid). 
Under supersymmetry transformations - $\delta_{(s)}$ -
we have 
\begin{eqnarray} 
\delta_{(s)} \psi^A_{~~\mu} &=  &{2\over \kappa} 
  \hat{D}_\mu \epsilon^A
 -{i \over 2} C_{\mu\nu}^{~~AB} \epsilon_B g_{IJ^*}\chi^{IC} e^{\nu}_{CC'}
\bar\chi^{J^{*}C'} + {i\over 2} \left(g_{\mu\nu} \varepsilon^{AB} + 
C_{\mu\nu}^{~~AB}\right)\epsilon_B \lambda^{(a)C} e^\nu_{CC'} \bar\lambda^{(a)C'} 
\nonumber \\
&- &{1 \over 4} \left( {\partial K \over \partial \phi^J} \delta_{(s)}\phi^J 
- {\partial K \over \partial \bar\phi^{J^{*}}} \delta_{(s)}\bar\phi^{J^{*}}\right)\psi^A_\mu 
+ ie^{K/2} P e_{\mu}^{ AB'} \bar\epsilon_{B'}~, 
\label{eq:2.22a}\\
\delta_{(s)} A_\mu^{(a)} &= & i \left(\epsilon^A e_{\mu AA'} \bar\lambda^{(a)A'} - 
\lambda^{(a)A} e_{\mu AA'} \bar\epsilon^{A'}\right) , 
\label{eq:2.23a}\\
\delta_{(s)}  \lambda^{(a)}_A   &= &- {1 \over 4} \left( {\partial K \over \partial \phi^J} \delta_{(s)}\phi^J 
- {\partial K \over \partial \bar\phi^{J^{*}}} \delta_{(s)}\bar\phi^{J^{*}}\right)\lambda^{(a)A} - iD^{(a)} \epsilon_A + \hat{F}^{(a)}_{\mu\nu} C^{\mu\nu~~B}_{~~~A} \epsilon_B
 ~, \label{eq:2.24a}\\
\delta_{(s)}\phi^I &=& \sqrt{2} \epsilon_A \chi^{IA},~ \label{eq:2.25a}\\
\delta_{(s)} \chi^{I}_A &= &i\sqrt{2} e^\mu_{AA'} \bar\epsilon^{A'} 
\left(\tilde D_\mu \phi^I 
-\frac{1}{2} \sqrt{2} \psi_{\mu C}\chi^{IC}\right) - \Gamma^I_{JK} \delta_{(s)}\phi^J \chi^K_A \nonumber  \\
&+ &{1 \over 4} \left( {\partial K \over \partial \phi^J} \delta_{(s)}\phi^J 
- {\partial K \over \partial \bar\phi^{J^{*}}} \delta_{(s)}\bar\phi^{J^{*}}\right)\chi^I_A 
-\sqrt{2} e^{K/2} g^{IJ^{*}}D_{J^{*}}P^*\epsilon_A
~, \label{eq:2.26a} \end{eqnarray}
and their Hermitian conjugates, $\epsilon^{AB}$ is the alternating spinor 
\cite{6,7,8}, 
where $ \epsilon^A $ and $ \bar \epsilon^{A'} $ are odd (anticommuting) fields, 
$C^{\mu\nu~~B}_{~~A} = {1 \over 4} 
(e^\mu_{AA'}e^{\nu BA'} - e_{AA'}^\nu e^{\mu BA'})$, 
$\hat{F}_{\mu\nu}^{(a)} = F^{(a)}_{\mu\nu} $ $- i\left(\psi^A_{[\mu}e_{\nu]AA'} \bar\lambda^{(a)A'}
\right.$ $\left. - \bar\psi_{A'[\mu}e_{\nu]}^{AA'}\lambda^{(a)}_A\right)$,
$\Gamma^I_{JK}$ is a Christoffel symbol derived from 
K\"ahler metric,  $P$ is a complex scalar-field dependent analytic 
potential energy term,  
$D_I = {\partial \over \partial \phi^{I}} + {\partial K \over \partial \phi^{I}}$, 
$\hat{D}_\mu = \partial_\mu + \omega_\mu + 
{1 \over 4} \left( {\partial K \over \partial \phi^J} \tilde D_\mu\phi^J 
- {\partial K \over \partial \bar\phi^{J^{*}}} \tilde D_\mu\bar\phi^{J^{*}}\right) 
+ {i\over 2} A^{(a)}_\mu Im F^{(a)}$, 
$\tilde D_\mu = \partial_\mu - A^{(a)}_\mu X^{I(a)}$,  
$F^{(a)} = X^{I(a)}{\partial K \over \partial \phi^I} + i D^{(a)}$.

For Lorentz transformations - $\delta_{(L)}$ - it follows that 
\begin{eqnarray}
\delta_{(L)} \phi^I &=& 0~,~~~ \delta_{(L)} \chi^I_A = N_{AB}\chi^{IB},~
\label{eq:2.27a} \\
\delta_{(L)} A^{(a)}_\mu &= &N_\mu^{~\nu} A^{(a)}_\nu, ~
\delta_{(L)} \lambda^{(a)}_A = N_{AB} \lambda^{(a)B} 
 \label{eq:2.28a} \end{eqnarray} with their 
Hermitian 
conjugates,  where $ N^{AB} = N^{(AB)} $ and $  N^{\mu\nu} = N^{[\mu\nu]} $.

Under local coordinate transformations - $\delta_{(lc)}$ -
\begin{eqnarray} 
\delta_{(lc)} A^{(a)}_\mu &= &\xi^\nu \partial_\nu A^{(a)}_\mu 
+ A^{(a)}_\nu \partial_\mu 
\xi^\nu~, \label{eq:2.29a}\\
\delta_{(lc)} \lambda^{(a)}_A &= &\xi^\nu \partial_\nu \lambda^{(a)}_A~,
\label{eq:2.30a}\\
\delta_{(lc)} \phi^I&= &\xi^\nu \partial_\nu \phi^I, ~ 
\delta_{(lc)} \chi^I_A = \xi^\nu \partial_\nu \chi^I_A, \label{eq:2.31a}
  \end{eqnarray} considering the Hermitians conjugates as well.

Finally,  for gauge transformations - $ \delta_{({\rm g})} $ - we get 
\begin{eqnarray}
\delta_{({\rm g})} \psi^A_\mu &=& -{i\over 2} \zeta^{(a)} Im F^{(a)} 
\psi^A_\mu,~ \delta_{({\rm g})} \phi^I = \zeta^{(a)} X^{I(a)},~
\label{eq:2.32a}\\
\delta_{({\rm g})} \chi^I_A &= &\zeta^{(a)} {\partial X^{I(a)} \over \phi^{J}} 
\chi^J_A + {i\over 2} \zeta^{(a)} Im F^{(a)} \chi^I_A,~ \label{eq:2.33a}\\
\delta_{({\rm g})} A^{(a)}_\mu &=& \partial_\mu \zeta^{(a)} + k^{abc} 
\zeta^{(b)} A_\mu^{(c)},~ \label{eq:2.34a} \\
\delta_{({\rm g})} \lambda^{(a)}_A &= &
k^{abc} \zeta^{(b)} \lambda ^{(c)}_A - {i\over 2}\zeta^{(b)} Im F^{(b)} 
\lambda^{(b)}_A, \label{eq:2.35a}
  \end{eqnarray}
and Hermitian conjugates.

The total  Hamiltonian includes now $A^{(a)}_0 Q_{(a)}$, where 
$ Q_{(a)} $ is the generator of gauge invariance.       
For the      
gravitino and spin-$\frac{1}{2}$ fields, the 
corresponding 
canonical momenta give again (second-class)      
constraints \cite{A21a,4a}. These are eliminated when       
Dirac      
brackets are introduced.  Nontrivial Dirac brackets  can be made simple      
 as follows \cite{A15a,A16,A21a}.  
      
The brackets involving $ p_{A A'}^{~~~~i},~\psi^A_{~~i} $ and $ \bar       
\psi^{A'}_{~~i} $ can be simplified as in the case of pure $ N = 1 $       
supergravity \cite{4} 
 by using (\ref{eq:2.22}). The $ \Phi^K $ and $ \Phi^{K^*} $ dependence of 
$ K_{I J^*} $ is responsible for  
unwanted Dirac brackets among $ \chi^I_A,~\bar\chi^{J^*}_{A},~\pi_{\Phi_L} $      
and $ \pi_{\Phi_L^*} $. In fact,  defining $ \pi_{I A} $ and       
$ \bar \pi_{I^* A'} $       
to be the momenta conjugate       
to $ \chi^{I A} $, and $ 
\bar\chi^{I^* A'} $, respectively, one has      
$   
\pi_{I A} 
 +  {i h^{{1 \over 2}}\over \sqrt 2}~K_{I J^*} n_{A A'} 
\bar \chi^{J^* A'} = 0$,  $~
\bar  \pi_{J^* A'}  +  
{i h^{{1 \over 2}} \over \sqrt 2}~K_{I J^*} n_{A A'} \chi^{I A} = 0~.      
$
  $ n^{A A'} $ is the spinor version of the unit       
 normal  $ n^\mu $, with 
$ n_{A A'} n^{A A'} = 1~,  n_{A A'} e^{A A'}_{~~~~i} = 0~.       
$
One cures this through $
\hat \chi_{I A} =  h^{1 \over 4} 
K^{1 \over 2}_{I J^*} \delta^{K J^*} \chi_{K      
A}~,  ~
\hat {\bar \chi}_{I^* A'} =  
h^{1 \over 4} K^{1 \over 2}_{J I^*} \delta^{J      
K^*} \bar \chi_{K^* A'}$.         
Here $K^{1 \over 2}_{I J^*} $ denotes a  ``square root'' of the K\"ahler       
metric, obeying      
$ K^{1 \over 2}_{I J^*} \delta^{K J^*} K^{1 \over 2}_{K L^*} =       
K_{I L}
$.
This may be found by diagonalizing $ K_{I J^*} $ via a
 unitary transformation,      
assuming that the eigenvalues are all positive. One needs to assume that       
there is an ``identity metric'' $ \delta^{K J^*} $ 
defined over the K\"ahler       
manifold.

Finally,  the brackets among $ \hat p_{A       
A'}^{~~~~i},~\lambda^{(a)}_A,~\bar\lambda^{(a)}_{A'},~\hat \chi^I_A $ and $      
\hat {\bar  \chi}^{J^*}_{A'} $ are dealt with by defining      
(see ref. \cite{A15a,A16,A21a,halli})      
$ \hat \lambda^{(a)}_A = h^{1 \over 4} \lambda^{(a)}_A~, \hat {\bar      
\lambda}^{(a)}_{A'} = h^{1 \over 4} \bar \lambda^{(a)}_{A'} $
and       
then going to the {\it time gauge} 
(cf. ref. \cite{4a}). In this case,       
 the tetrad component $ n^a $      
of the normal vector $ n^\mu $ is  restricted by      
$ n^a = \delta^a_0 \Leftrightarrow  e^0_{~~i} = 0$. 
Thus the original Lorentz rotation freedom becomes replaced by that of       
spatial rotations. In the time gauge, the geometry is described by the triad       
$ e^\alpha_{~~i} (\alpha = 1, 2, 3) $, and the conjugate       
momentum\footnote{Notice that        
$   \pi^{ij} \equiv - \frac{1}{2} p^{(ij)} = \frac{1}{2} e^{AA'(i}p_{AA'}^{j)}      
=  - \frac{1}{2} e^{\alpha(i}p_{\alpha}^{j)}$       
where the last equality       
follows from the time gauge conditions (see ref. \cite{3,4,A15a,A16,A21a}).}      
is $   p_\alpha^{~~i} $.         
All the resulting Dirac brackets are either 
the same 
as in  subsection 2.1.1 or zero, 
except  the nonzero fermionic brackets like 
\begin{eqnarray}      
\left[ \hat \lambda^{(a)}_A (x), \hat {\bar \lambda}^{(b)}_{~~A'} (x) \right]_{D} &= &       
\sqrt{ 2} i n_{A A'} \delta^{(a) (b)} \delta ( x, x' )~, 
\label{eq:2.36}\\      
\left[ \hat \chi^I_{~~A} (x),  \hat {\bar \chi}^{J^*}_{~~A'} ( x' ) \right]_{D}       
&=&  \sqrt{2} i n_{A A'} \delta^{I J^*} \delta ( x, x' )~, 
\label{eq:2.37}\\        
\left[ \psi^A_{~~i} (x), \bar \psi^{A'}_{~~j} ( x' ) \right]_{D} &= & {1 \over      
\sqrt{2}} D^{A A'}_{~~~~i j} \delta ( x, x' ).
\label{eq:2.38}\end{eqnarray}

The full Hamiltonian        contains arbitrary Lorentz rotations. 
For our present case, these ought       
surely  include   the  components         
which depend on the supermater fields. 
By employing again the redefinition  $M_{AB} \mapsto \omega_{AB0}$ and H.c., 
the Lorentz contributions   will contribute       
with {\it new}  terms of the type $\psi\chi\bar\chi,~~ \psi\lambda      
\bar\lambda$ and their Hermitian conjugates to the supersymmetry       
constraints \cite{A21a}. We notice that this last step is {\it missing} in the procedure employed in ref       
\cite{A15a,A16}. 
In the end of this section we further discuss the implications of its absence and which problems its       
presence solves.

The supersymmetry constraint $ \bar S_{A'} $ is then found to be      
\begin{eqnarray}      
\bar S_{A'} & = &       
- \sqrt 2 i   e_{A A' i} \psi^A_{~~j}  \pi^{i j}      
+    \sqrt 2      
\epsilon^{i j k} e_{A A' i}~^{3 s} \tilde {\cal D}_j \psi^A_{~~k}       
+ {1 \over \sqrt 2}       
\left [       
\pi_{J^*}      
+       
{i \over \sqrt 2} h^{{1 \over 2}} g_{L M^*} \Gamma^{M^*}_{J^* N^*} n^{B B'} 
\bar      
\chi^{N^*}_{~~~B'} \chi^L_{~~B} \right. \nonumber \\       
&- &       
 {i h^{{1 \over 2}} \over 2 \sqrt 2} K_{J^*} g_{M M^*} n^{B B'} \bar\chi^{M^*}_{~~~B'}      
\chi^M_{~~B}       
-       
{1 \over 2 \sqrt 2} \epsilon^{i j k} K_{J^*} e^{B B'}_{~~~~j} \psi_{k B} \bar
\psi_{i B'}       
 \nonumber \\       
& - & \left.       
w_{[1]}\sqrt 2 h^{{1 \over 2}} g_{I J^*} \chi^{I B} e_{B B'}^{~~~~m} n^{C B'} \psi_{m C}     
\right]       
\bar \chi^{J^*}_{~~~A'}       
\nonumber \\      
&- &      
\sqrt 2 h^{{1 \over 2}} g_{I J^*}  \tilde {\cal D}_i \Phi^I  \bar \chi^{J^*}_{B'} 
n^{B B'}      
e_{B A'}^{~~~~i} +      
w_{[2]}{i \over 2} g_{I J^*} \epsilon^{i j k} e_{A A' j} \psi^A_{~~i} \bar
\chi^{J^*      
B'} e_{B B' k} \chi^{I B} \nonumber \\        
& + &w_{[3]}{1 \over 4} h^{{1 \over 2}}       
\psi_{A i} \left( e_{B A'}^{~~~~i} n^{A C'} - e^{A C' i} n_{B A'}      
\right) g_{I J^*} \bar\chi^{J^*}_{C'} \chi^{I B}  \nonumber \\     
& -  &      
h^{{1 \over 2}} \exp ( K / 2 ) \left[ 2 P n^A_{~~A'} e_{A B'}^{~~~~i} \tilde      
\psi^{B'}_{~~i} + i ( D_I P ) n_{A A'} \chi^{I A}       
\right] \nonumber \\        
&       
- & {i \over \sqrt 2} \pi^{n (a)} e_{B A' n} \lambda^{(a) B}      
+ {1 \over 2 \sqrt 2} \epsilon^{i j  k} e_{B A' k} \lambda^{(a) B} F^{(a)}_{i j}      
+ {1 \over \sqrt 2} h^{{ 1 \over 2}} g D^{(a)} n^A_{~~A'} \lambda^{(a)}_{~~~A}       
 \nonumber \\ &      
+       
&      
w_{[4]}{1 \over 4} h^{{1 \over 2}} \psi_{A i} \left( e_{B A'}^{~~~~i} n^{A C'} - e^{A C' i} n_{B       
A'} \right) \bar  \lambda^{(a)}_{C'} \lambda^{(a) B} -
{i  \over 4} h^{{1 \over 2}} n^{B B'} \lambda^{(a)}_{~~~B} \bar\lambda^{(a)}_{~~~B'}      
K_{J^*} \bar \chi^{J^*}_{~~~A'} ,    
\label{eq:2.39}  
\end{eqnarray}      
where $ \lambda^{(a)}_A,~\bar\lambda^{(a)}_{A'} $ and $ \chi_{I A},~\bar
\chi_{I^* A'} $ should be redefined as indicated above. 
The other supersymmetry constraint $S_A$ is just the       
Hermitian conjugate of (\ref{eq:2.39}).       
The  $w_{[i]}, i=1,2,3,4$ denote numerical coefficients       
which correspond to the inclusion of the terms       
 $\psi\chi\bar\chi,~~ \psi\lambda      
\bar\lambda$ and their Hermitian conjugates to the supersymmetry       
constraints via  $\omega_{AB0}J^{AB}$ and Hermitian       
conjugate.       
In addition,      
$ 
^{3 s} \tilde {\cal D}_j \psi^A_{~~k}  =     
\partial_j \psi^A_{~~k} + ~^{3 s} \omega^A_{~~B j}      
\psi^B_{~~k}
 + {1 \over 4} ( K_K \tilde {\cal D}_j \Phi^K - K_{K^*} \tilde {\cal D} _j \Phi^{K^*} 
)      
\psi^A_{~~k}
 + {1 \over 2} g A^{(a)}_j ( I m F^{(a)} ) \psi^A_{~~k}$, 
where $ \tilde {\cal D}_i A^K = \partial_i A^K - g A^{(a)}_i X^{K (a)}$,     
with $ g $ the gauge coupling constant and $ X^{K (a)} $ the $ a $th Killing       
vector field.
$\pi^{n (a)} $ is the momentum conjugate to $ A^{(a)}_n$.

The gauge generator $ Q^{(a)} $ is given classically by      
\begin{eqnarray}      
Q^{(a)} &= -&  \partial_n \pi^{n (a)} -  f^{a b c} \pi^{n (b)} A^{(c)}_n + 
 \pi_I X^{I (a)} + \pi_{I^*} X^{I^* (a)}  \nonumber \\        
&+ & \sqrt 2 i h^{{1 \over 2}} K_{M I^*} n^{A A'} X^{J^* (a)} \Gamma^{I^*}_{J^* N^*} \bar      
\chi^{N^*}_{A'} \chi^M_A \nonumber \\        
&- & \sqrt 2 i h^{{1 \over 2}}  n^{A A'} \bar\lambda^{(b)}_{A'} \left[ f^{a b c} \lambda^{(c)}_A +      
{1 \over 2} i ( I m F^{(a)} ) \lambda^{(b)}_A \right] \nonumber \\        
&+ & \sqrt 2 i h^{{1 \over 2}}  n^{A A'} K_{I J^*} \bar\chi^{J^*}_{A'}       
\left[ {\partial X^{I (a)}      
\over \partial \Phi^J} \chi^J_A + {1 \over 2} i  I m F^{(a)}  \chi^I_A \right] 
\nonumber \\        
&-&  {i \over \sqrt 2}   I m F^{(a)}  \epsilon^{i j k} \bar \psi_{i A'} e^{A      
A'}_{~~~~j} \psi_{A k}~,
\label{eq:2.40}
\end{eqnarray}      
where $ f^{a b c} $ are the structure constants of the 
K\"ahler isometry group.         

It is worthwhile to notice that we expect now to        
obtain the correct transformation properties (cf. ref. \cite{8})       
of the       
physical fields under         
supersymmetry transformations,       
using the  brackets $\delta_{\xi}\psi^A_i       
\equiv [\bar\xi_{A'} \bar S^{A'}, \psi^A_i]_D$, etc. Here       
$\bar\xi^{A'}$ is a constant spinor parametrizing the       
supersymmetric transformation.       
In fact, that       
 was {\it not} possible for some fields, when using  the explicit form       
of  the supersymmetry constraints present in ref. \cite{A15a,A16}. 
The reasons       
are as   follows. On the one hand, the matter terms in the       
Lorentz constraints $J_{AB}, \bar       
J_{A'B'}$ were {\it not}  included in the supersymmetry       
constraints.  On the other hand, expressions       
 {\it only valid}  in pure N=1 supergravity       
were employed to simplify the supersymmetry constraints with        
supermatter. Namely, the expressions for $S_A = 0,       
\bar  S_{A'} = 0$ in pure N=1 supergravity       
were used to re-write the spatial covariant       
derivative $^3 {\cal D}_i$  in terms of its torsion free part $^{3s} {\cal D}_i$  and       
remaining terms which include the contorsion. When supermatter is present, we expect the different       
matter fields to play a role in the Lorentz constraints. Some of these 
 terms  are then  included in the supersymmetry constraints once $\omega^0_{AB}J^{AB}$
 and its Hermitian conjugate are       
employed in the canonical action.  

\subsection{Why there are {\it no} bosonic physical states }

\indent 

In this subsection we will point out how physical quantum states in pure 
N=1 supergravity can only have an infinite number of fermions \cite{C1,C2}. 
Early attempts pretended that bosonic and finite fermion number states were  
possible \cite{C6}, but subsequent analysis has shown otherwise \cite{C1,C2,C5}.
This analysis has pointed to
several incorrections, regarding   different 
approaches trying to revive the same inconsistent claims, e.g.,  \cite{C4}.

The supersymmetry constraints are  the
central issue. Since these constraints are homogeneous in the gravitino
field $\psi_{Ai}(x)$, it is consistent to look for solutions involving
homogeneous functionals of order $\psi^n$. Such states may be called 
states with 
fermion (Grassmann) number $n$.

Although the form of the
supersymmetry constraints (\ref{eq:2.25}), 
(\ref{eq:2.26})  suggests that there may be  solutions of 
definite
order $n$ in the fermion fields,  there are {\it no} such states for
any finite $n$. For $n=0$, a simple scaling argument assuredly excludes
the purely bosonic states discussed in \cite{C6,C4}. For $n>0$ (cf. \cite{C1,C2} for 
more 
details), the argument is
based on a mode expansion of the gravitino field.
 It was then  suggested that physical states in supergravity have
infinite fermion  number. This was confirmed for the free spin-3/2 field and a 
physical (but yet formal) 
state was indeed found in ref. \cite{C3} (see subsection 2.3).


Let us  then consider Lorentz invariant states.
An arbitrary state can be expanded 
in a power series as 
$
\Psi [e_{AA'i},\psi_{Ak}]=\sum_n \Psi^{(n)}[e_{AA'i},\psi_{Ak}].
$
  Note that odd $n$ states need not be considered because they are
not local Lorentz invariant. We shall refer to $\Psi^{(n)}$ as a state of
fermion number $n$.
The constraint equations  must be satisfied independently
by each term $\Psi^{(n)}[e_{AA'i},\psi_{Ak}]$.

Let us  
 consider the case where $\Psi[e_{AA'i},\psi_{Ai}]\equiv \Psi^{(0)}[e_{AA'i}]$ is a
functional only of the tetrad 
${e^a}_i$, that is $\delta \Psi^{(0)}/\delta\psi_{Ai}=0$. 
We shall refer to states of this type as {\it bosonic states}. In this
case, any Lorentz invariant state satisfying $\bar{S}^{A'}\Psi^{(0)}=0$
automatically satisfies all other constraints, since $S^A \Psi^{(0)}$ vanishes
identically.
We now show that {\it no}
such solution can exist.

The supersymmetry constraint can be written as 
\begin{equation}
\left[\Psi^{(0)}\right]^{-1}\bar{S}\Psi^{(0)}\equiv
-\varepsilon^{i j k}{e^i}_{AA'} (D_j\psi_k^{AA'})+{\hbar\kappa^2\over
2}\psi_{iAA'} {\delta(\ln \Psi^{(0)} )\over\delta e^{AA'}_i}=0 ~.
\label{eq:2.41} \end{equation}
A contradiction occurs by using an integrated
form of  (\ref{eq:2.41}) with an arbitrary continuous,   spinor test function
$\bar{\epsilon}(x)$:
\begin{equation}
\int d^3x \bar{\epsilon}(x)\left[
-\varepsilon^{i j k}{e^i}_{AA'} (D_j\psi_k^{AA'})+{\hbar\kappa^2\over
2}\psi_{iAA'} {\delta(\ln \Psi^{(0)} )\over\delta e^{AA'}_i}
\right]=0 , \label{eq:2.42}
\end{equation}
for all $\bar{\epsilon}(x)$, ${e^{AA'}}_i(x)$, and $\psi^A_{k}(x)$.

Let the integral in (\ref{eq:2.42}) be $I$, and let $I'=I+\Delta I$ be the
integral when $\bar{\epsilon}(x)$ is replaced by
$\bar{\epsilon}(x)e^{-\phi(x)}$ and $\psi^A_{i}(x)$ is replaced by
$\psi^A_{i}(x)e^{\phi(x)}$, where $\phi(x)$ is a scalar function.
Since $\bar{\epsilon}(x)\psi^A_{i}(x)$
is unchanged, the second term (with the
functional derivative) cancels in the difference between $I'$ and $I$, so
that
\begin{equation}
\Delta I=-\int d^3x
\varepsilon^{ijk}e^{AA'}_i(x) \bar{\epsilon}_A(x) \psi_{kA}(x)
\partial_j\phi(x)=-\int \omega(x)\wedge d\phi(x)
\label{eq:2.43}
\end{equation}
with the two-form
$
\omega(x)=\psi^A_{i}(x)dx^i \wedge
{e^{AA'}}_j(x)\bar{\epsilon}(x)  dx^j.
$
Notice that $\Delta I$ is independent of the state $\Psi^{(0)}$.
Clearly, it is possible to choose the arbitrary fields
$\bar\epsilon(x)$, $\phi(x)$, $e^{AA'}_i(x)$ and $\psi_{kA}(x)$ such that
(\ref{eq:2.43}) is nonvanishing. 
E.g., consider the case of a three-torus $x^i \in [0, 2\pi]$, and with 
${e^a}_i=\delta^a_i$, $\psi_{Ak}=\delta_{A1}\delta_{2k}$,
$\bar{\epsilon}_{A'}=-\delta_{A'1}f(x)$, $f=\cos x^1$,
$\phi=\sin x^1$. Then
$
\Delta I=\int d^3x f(x) \partial_1\phi(x)=\int d^3x\cos^2(x^1)
=4\pi^3\neq 0
$. 
Clearly, if $\Delta I \neq 0$, we cannot have both $I=0$ and $I'=0$,
so (\ref{eq:2.41}) {\it cannot} be satisfied {\it for all} 
 $\bar{\epsilon}(x)$, ${e^{AA'}}_{i}(x)$ and $ \psi^A_{k}(x)$.

\vspace{0.2cm}

Hence,  {\it bosonic} 
wave functionals are {\it inconsistent} with the supersymmetry
constraints of pure N=1 supergravity.

\vspace{0.2cm}

This result also suggests that there may also be {\it no}
states involving a finite nonzero number of fermion fields. That is,
all quantum wave functionals of $N=1$ supergravity
with a finite (zero or nonzero) fermion number would be
inconsistent with the supersymmetry constraints. From  the free spin-3/2 field 
case, 
it was  shown in ref. \cite{C1} 
that  all wave functionals  
  necessarily contain an infinite number of
fermion fields. An extension (under some assumptions) of this 
reasoning was then made for the case of general N=1 supergravity theory.

Overall, the conclusions in \cite{C1,C2} subdued the claims in ref. \cite{C6} 
that N=1 supergravity is finite to all orders. Moreover, recent results in 
ref. \cite{espo} seem to further establish that N=1 supergravity 
with boundary terms is fully divergent even at one-loop order.

\subsection{Solutions with {\em infinite} fermion number}

\indent

In this section we will review and summarize the method  
employed in ref. \cite{C3} which allowed for an exact 
quantum solution  of the full theory to be there determined. 
The interested reader should consult  ref. \cite{C3} for more 
details. 
Such 
  quantum state satisfies  all constraints in      
the metric representation for the general 
theory of N=1 supergravity.

The form of this solution was actually 
 conjectured  in ref. 
\cite{A21,A20}. Namely, 
that  
infinitely many physical states      
may exist having  the form  $\Psi=S^A S_A      
G(h_{pq})$. 
This  was  quite important, in order 
   to generalize the physical states      
present in  Bianchi models to full      
supergravity (see section 3).

The new elements that were employed in ref. \cite{C3}  to obtain 
physical states are the commutators 
\begin{eqnarray}      
 & &   \left[H_{AA'}(\vec{x}),S_B(\vec{y})\right]  =     
  -  i\hbar\delta(\vec{x}-\vec{y})\varepsilon_{AB}
     {\bar{D}_{A'}}^{B'C'}(\vec{x})      
     \bar{J}_{B'C'}(\vec{x})~,~
   \left[H_{AA'}(\vec{x}),\bar{S}_{B'}      
    (\vec{y})\right] =      
    i\hbar\delta(\vec{x}-\vec{y}) \nonumber \\  & &
\varepsilon_{A'B'}      
     \left[D_A^{BC}(\vec{x})      
      J_{BC}(\vec{x})  +   i\hbar\delta(0)      
      \left({\bar{E}_A}^{C'D'}(\vec{x})      
       \bar{J}_{C'D'}(\vec{x})      
        -n^{AC'}(\vec{x})h^{-1/2}(\vec{x})      
        \bar{S}^{C'}(\vec{x})\right)\right]. \label{eq:2.44}  
\end{eqnarray}   
  The coefficients $D, E$ are Grassmann-odd structure      
functions.
The divergent $\delta(0)$-factor\footnote{Fortunately, 
the last commutator in (\ref{eq:2.44})      
is not      
needed in the  solution of the constraints, and the $\delta(0)$-term      
therefore does not appear there. The only remaining  
 commutators $[H_{AA'}(x),      
H_{BB'}(y)]$ 
are obtained   from      
eqs.(\ref{eq:2.44}) via Jacobi-identities.}      
 may hide an anomaly and      
its presence reduces the result   presented in  ref. \cite{C3} 
to a formal      
one. To go beyond this level    regularization      
procedures would have to be included. 

The physical state found in ref. \cite{C3} and   conjectured   in \cite{A21,A20}    has  the form 
\begin{equation}      
\label{eq:2.45}      
 \Psi=\Pi_{(\vec{x})}S^A (\vec{x})S_A (\vec{x})      
        G (e_{iBB'}).      
\end{equation}     
It contains a formal product over all   space-points,       
with   a   bosonic functional $G $   satisfying the Lorentz constraints. The 
$S_B$ constraint and the $J_{AB}$,      
$\bar{J}_{A'B'}$ constraints are      
automatically satisfied. The $\bar{S}_{B'}$ constraint,      
after using the generator algebra and the properties of $G$, is satisfied      
if      
\begin{equation}      
\label{eq:2.46}      
\bar{S}_{A'}S_A~ G \left(e^{BB'}_i\right) =0\,.      
\end{equation}      
It is important to note that the operators $e_i^{AA'} 
\bar{S}_{A'}S_A$      
and      
$n^{AA'}\bar{S}_{A'}      
S_A$ are Lorentz-invariant.

A      
solution of eq.~(\ref{eq:2.46}) for the Bianchi models  is 
described  in subsection 3.1.1 and corresponds to the 
 restriction of the functional      
$G_0 (e^{iAA'})=\exp[-\frac{1}{2\hbar}\int d^3x      
\varepsilon^{ijk}{e_i}^{AA'}\partial_je_{kAA'}]$ to the appropriate spatially      
homogeneous tetrad.  
Remarkably,   (\ref{eq:2.46}) {\it does}      
have solutions also in the  inhomogeneous case. 
One of      
which is, surprisingly, again given by the functional $G_0$. However,      
while $\bar{J}_{A'B'} G_0=0$ is satisfied, we also have  
 that $J_{AB} G_0\neq 0$. A fully Lorentz-invariant      
amplitude $G$ is obtained from $G_0$  after
\footnote{$D\mu[\omega]$ is chosen as the
 formal direct product of the Haar      
measure of the $SU(2)$-rotation matrices.  ${\Omega_A }^{C }=      
\left[\exp i\omega\right]_{A}^{  C}$ (with      
${\omega_2}^{\ 2}=({\omega_1}^1)^*$, ${\omega_2}^1=-({\omega_1}^2)^*$) 
represent precisely the $SU(2)-$rotation matrices. 
See ref. \cite{C3} for more details.   }      
\begin{equation} 
G=\int D\mu[\omega]\exp      
\left(i\omega^{AB}J_{AB}\right) G_0 ~.
\end{equation}

The infinite product in (\ref{eq:2.45}) can then be written as  a      
Grassmannian path-integral over a Grassmann field $\epsilon^A      
(\vec{x})$.       
Applying the factors $S^A(\vec{x})$ explicitely on the functional      
$G$ 
 and using the identity      
$   \exp(i\omega^{AB}J_{AB}) $  $     
G_0 =[\exp(i\omega^{AB}J_{AB})      
G_0^2      
\exp(-i\omega^{AB}J_{AB})]      
[G_0 ]^{-1} $ 
satisfied by $G_0$,       
an  exact    physical state $\Psi$ is found  as: 
\begin{eqnarray}      
\label{eq:2.48}      
&&\Psi(\{h_{ij},{\bar{\psi}_i}^{ A'}\})=      
    \int D[\epsilon^1]D[\epsilon^2]      
     \left\{\exp\left[-\int d^3x\varepsilon^{ijk}\left(      
      \epsilon^A(\vec{x})\partial_j e_{iAA'}(\vec{x})            
       {\bar{\psi}_k}^{A'}(\vec{x})
         +\frac{1}{2\hbar}{e_i}^{AA'}(\vec{x})\partial_je_{kAA'}(\vec{x})      
          \right)\right]\right.   \nonumber\\      
&&  \int D\mu[\omega]      
     \exp      
    \left.\left[\int d^3x\varepsilon^{ijk}{\Omega_C}^A(\vec{x})      
      (\partial_j{\Omega^C}_B(\vec{x}))          
e_{iAA'}(\vec{x})      
         \left(\epsilon^B(\vec{x}){\bar{\psi}_k}^{A'}      
          (\vec{x})+\frac{1}{2\hbar}      
             e^{BA'}_k (\vec{x})\right)\right]\right\}\,.      
\end{eqnarray}      
This is precisely the solution found in ref. \cite{C3} and describes an 
exact quantum state of the full supergravity field theory.

The  amplitude      
$G$ reduces to $\sim\exp[-\frac{V}{2\hbar}m^{pq}h_{pq}]$  in the spatially homogeneous case indicating  that      
(\ref{eq:2.48}) should be interpreted as a wormhole state.  
Being   gravity and      
supergravity non-linear, it is curious that
 this result has similarities to the Gaussian form      
expected      
for the ground state of a free field (e.g.,  electromagnetic).


\section{Supersymmetric minisuperspaces}

\indent

Throughout this section we will study in some detail 
the canonical quantization of supersymmetric minisuperspaces. 
For edifying purposes  we will employ  Bianchi type-IX and 
$k = +1$ FRW models  \cite{A5}-\cite{A18a} 
and 
generalise our results to class A models \cite{A12,A21,A20,A23b}. 
There are two approaches to get supersymmetric 
Bianchi models. On the one hand,  we substitute
 a specific Bianchi ansatz directly in the 
classical action, thereby obtaining a reduced model and then quantizing it. 
On the other hand, 
  we may take the quantum constraints directly from the general 
theory and use them subject to  a Bianchi ansatz. We will make use of both 
techniques (see, e.g., \cite{schlei} for a related 
discussion on their possible equivalence).

\subsection{Models from pure N=1 supergravity}

\indent

Models obtained from pure N=1 supergravity will be 
considered in this section. A Bianchi type IX model is 
discussed in subsection 3.1.1. We  
 then proceed to the particular case of closed FRW models
in subsection 3.1.2.

\subsubsection{Bianchi class A models}

\indent

Let us consider a Bianchi type-IX model 
whose 
4-metric 
is given by      
$ g_{\mu \nu} = \eta_{a b} e^a_{~\mu} e^b_{~\nu}, $     
where $ \eta_{a b} $ is the Minkowski metric 
and the non-zero components of the tetrad $ e^a_{~\mu} $       
are given by      
\begin{equation}       
e^0_{~0}  =   N,~ e^1_{~0} = a_1 N^i E^1_{~i},~ e^2_{~0} = a_2 N^i E^2_{~i},~      
e^3_{~0} = a_3 N^i E^3_{~i}, ~
e^1_{~i}  =   a_1 E^1_{~i},~e^2_{~i} = a_2 E^2_{~i},      
~e^3_{~i} = a_3
E^3_{~i}.  \label{eq:3.1}  \end{equation}      
Here $ E^1_{~i}, E^2_{~i}, E^3_{~3} ~(i = 1, 2, 3) $ are       
a basis of    
left-invariant one-forms on the unit three-sphere \cite{A,N=2a}      
 and $ N, N^i, a_1, a_2, a_3 $ are spatially constant.       
We then write       
\begin{equation} h_{i j} = a_1^2 E^1_{~i} E^1_{~j} + a_2^2 E^2_{~i} E^2_{~j} +       
a_3^2 E^3_{~i}       
E^3_{~j}~. \label{eq:3.2}   \end{equation}      
In the calculations, we shall repeatedly      
use  the following expression for the connections:      
\begin{eqnarray}      
& & \omega_{A B i} n^A_{~~B'} e^{B B' j} =  {i \over 4} \left( {a_3 \over a_1a_2}       
+ {a_2 \over a_3      
a_1} - {a_1 \over a_2 a_3} \right) E^1_{~i} E^{1 j}  + \nonumber \\        
& & {i \over 4} \left( {a_1 \over a_2 a_3} +       
{a_3 \over a_2 a_1} - {a_2 \over a_3 a_1} \right) E^2_{~i}      
E^{2 j}  
~+ {i \over 4} \left(       
{a_2 \over a_3 a_1} + {a_1 \over a_2 a_3} - {a_3 \over a_1 a_2} \right) E^3_{~i}       
E^{3 j}.
\label{eq:3.3}
\end{eqnarray}

The 
differential equations obeyed by the   wave function are 
found by studying the quantum constraints of the full theory of supergravity 
\cite{4},  evaluated subject to a simple Bianchi ansatz \footnote{ It should 
be noted that this simple ansatz  is {\it not}  invariant under 
 homogeneous supersymmetry transformations. 
To obtain an ansatz  invariant under supersymmetry, one
must use a {\it non-diagonal} 
 triad $ e^a_{~i} = b^a_{~b} E^b_{~i} $, where $ b_{a
b} $ is symmetric $ (a, b, = 1, 2, 3 $ here), 
combined  with supersymmetry, homogeneous spatial coordinate  and 
local 
Lorentz transformations \cite{A6,A8,A9}.} on the spatial variables $ 
e^{AA'}_{~~~~i}, \psi^A_{~~i} $ and $ \bar 
\psi^{A'}_{~~i} $. 
We require  
$  \psi^A_{~~0}, \bar \psi^{A'}_{~~0}  $ be functions of time only and 
 $ \psi^A_{~~i} $ and $ \bar \psi^{A'}_{~~i} $  to be
spatially homogeneous in the basis $ e^a_{~i} $; equivalently, $
\psi^A_{~~i} e^{BB'i} $ and $ \bar \psi^{A'}_{~~i} e^{BB'i} $ are functions
of time only.

A quantum state may be described by a wave function $ \Psi 
( e^{AA'}_{~~~~i}, \psi^A_{~~i} ) $ or equivalently by a wave
function $ \bar \Psi ( e^{AA'}_{~~~~i}, \bar\psi^{A'}_{~~i} ) $. The two
descriptions are related by a fermionic Fourier transform \cite{4,A5,A6}. 
The  wave function $ \Psi ( 
e^{AA'}_{~~~~i}, \psi^A_{~i} ) $ may be expanded in even powers of 
$ \psi^A_{~i} $,  symbolically in  the form $
\psi^0, \psi^2, \psi^4 $ and $ \psi^6 $. In the representation $ \bar  \Psi ( 
e^{AA'}_{~~~~i}, \bar \psi^{A'}_{~~i} ) $, these become respectively 
of the form $ \bar \psi^6, \bar \psi^4, \bar \psi^2 $ and $ \bar \psi^0 $.
This can be better understood by noticing the 
decomposition $\psi^{~i}_A e_{BB'i} = \psi_{ABB'}$ 
with 
\begin{equation} \psi_{A B B'} = - 2 n^C_{~~B'} \gamma_{A B  C} + {2 \over 3} ( \beta_A
n_{B B'} + \beta_B n_{A B'} ) - 2 \varepsilon_{A B} n^C_{~~B'} \beta_C~,  
\label{eq:3.5}
 \end{equation}
where $ \gamma_{A B C} = \gamma_{(A B C)} $ is totally symmetric.
The $\beta^A$ and $\gamma^{BCD}$ spinors constitute 
the spin $\frac{1}{2}$ and $\frac{3}{2}$ modes of the 
gravitino fields, when these are split in irreducible representations 
of the Lorentz group \cite{6} (see also ref. \cite{4,C1} and eqs. 
(\ref{eq:3.29}), (\ref{eq:3.30}) and (\ref{eq:3.31})).  
The Lorentz constraints (\ref{eq:2.10}) 
imply that $ \Psi $ is invariant under Lorentz transformations.
 A {\it possible} ansatz for $\Psi$ would be 
\begin{equation}
\Psi = A + B \beta_A\beta^A + C\gamma^{BCD}\gamma_{BCD}
+ D \beta_A\beta^A \gamma^{BCD}\gamma_{BCD} + E \left(\gamma^{BCD}\gamma_{BCD}\right)^2 + F \beta_A\beta^A
\left(\gamma^{BCD}\gamma_{BCD}\right)^2,
\label{eq:3.4}
\end{equation}
where $A,B$, etc are functions of $a_1,a_2,a_3$.
The first term in (\ref{eq:3.4}) corresponds to the bosonic $\psi^0$ 
part, while the second and third terms in $\Psi$ represent 
the quadratic sectors. Similarly, the fourth and fifth 
correspond to the quartic sector and the last term in (\ref{eq:3.4}) 
is just the fermionic filled sector. 
A term $ ( \beta^A \gamma_{A B C} )^2 = \beta^A      
\gamma_{A B C} \beta^D \gamma_D^{~~B C} $ can be rewritten, using the       
anti-commutation of the $ \beta $'s and $ \gamma $'s, as      
$\beta^E \beta_E \epsilon^{A D} \gamma_{A B C} \gamma_D^{~~B C}       
\sim       
~ ( \beta_E \beta^E )~(\gamma_{A B C} \gamma^{A B C} ) $.     
Similarly, any quartic in $ \gamma_{A B C} $ can be rewritten as a multiple of       
$ ( \gamma_{A B C} \gamma^{A B C} )^2 $. Since there are only four independent      
components of $ \gamma_{A B C} = \gamma_{(A B C)} $, only one independent quartic       
can be made from $ \gamma_{A B C} $, and it is sufficient to check that $ (       
\gamma_{A B C} \gamma^{A B C} )^2 $ is non-zero. Now $ \gamma_{A B C} \gamma^{A B C}       
= 2      
\gamma_{000} \gamma_{111} - 6 \gamma_{100} \gamma_{011} $. Hence $ ( \gamma_{A B C}      
\gamma^{A B C} )^2 $ includes a non-zero quartic term $\gamma_{000}      
\gamma_{100} \gamma_{110} \gamma_{111} $.

Consider first the (bosonic) $ \psi^0-$part $A(a_1,a_2,a_3)$ 
of the wave function. It  automatically obeys the constraint $ S_A \Psi = 0
$, since this involves differentiation with respect to $ \psi^A_{~i} $ 
(see eq. (\ref{eq:2.26})). The 
only remaining constraint is $ \bar S_{A'} \Psi = 0 $.

Evaluated at a Bianchi IX geometry the constraint $ \bar  S_{A'} \Psi = 0 $ reads
\begin{equation}
 \epsilon^{i j k} e_{A A' i}~^{3s} \omega^A_{~B j} \psi^B_{~k} \Psi - \frac{1}{2} 
\kappa^2 \psi^A_{~i} {\delta \Psi \over \delta e^{AA'}_{~~~~i}} = 0~. \label{eq:3.6}   \end{equation}
Since the homogeneous fields $ \psi^A_{~i} $ are otherwise arbitrary, they
may be cancelled in 
eq. (\ref{eq:3.6}).
Notice that the torsion-free connections have the form
\begin{eqnarray}
^{3s} \omega^{00}_{~~i} &= & X_i + i Y_i~,  ~^{3s} \omega^{11}_{~~i} = X_i - i
Y_i~, \label{eq:3.7}  \\
^{3s} \omega^{01}_{~~i} &= &~^{3s} \omega^{01}_{~~i}   = \frac{1}{4} 
 i \left( - {a_1 \over a_2}
- {a_2 \over a_1} + {a_3^2 \over a_1 a_2} \right) E^3_{~i}~, 
\label{eq:3.8}  \end{eqnarray}
where
\begin{equation}
X_i = \frac{1}{4}  \left( {a_3 \over a_1} + {a_1 \over a_3} - {a_2^2 \over a_1a_3} \right) E^2_{~i}~,~
Y_i = \frac{1}{4} \left( {a_3 \over a_2} + {a_2 \over a_3} - {a_1^2 
\over a_2 a_3} \right) E^1_{~i}~. \label{eq:3.9}
\end{equation}

One can then contract $ e^{BA'm} $ into eq. (\ref{eq:3.6}), to obtain an equation
which  then is contracted  with 
allowed variations $ \delta h_{k m} $ of diagonal
Bianchi-IX 3-metrics. This gives for the wave function $ A (a_1, a_2, a_3)$, for
example,
\begin{equation}
{\partial (\ln A) \over \partial a_1} = \int d^3 x {\delta ( \ln A ) \over \delta h_
{k m} (x)}~{\partial h_{k m} (x) \over \partial a_1} 
= - \kappa^{- 2} a_1 \int d^3 x h^\frac{1}{2} ~,
\label{eq:3.10}  \end{equation}
where $ \int d^3 x h^\frac{1}{2}  = 16 \pi^2 $ is the volume of the compact 3-space
with $ a_1 = a_2 = a_3 = 1 $. Similar expressions for $ \partial ( \ln A ) / \partial a_2 $ 
and $ \partial ( \ln A) / \partial a_3 $ lead to
\begin{equation}  A \sim \exp ( - I )~, 
 I = \pi \left( a_1^2 + a_2^2 + a_3^2\right). 
\label{eq:3.11}  \end{equation}

Using 
equation $ S_A \bar \Psi = 0 $, one gets $ E \sim  \exp (I) $. 
 Hence the bosonic and fermionic filled 
states of the theory both have very simple {\it semi-classical} forms.

With respect to  the 
quadratic and quartic fermionic components of $\Psi$, 
ansatz (\ref{eq:3.4}), 
constitutes a rather restrictive choice.
 An unimaginable consequence of the supersymmetry constraints was   
then that {\it no} states are possible in the intermediate sectors of 
 $\psi^2$ and $\psi^4$ order 
(cf. ref. \cite{A5,A6} for more details).

The simple semi-classical form   $A \sim 
e^{- I} $ represents  a 
(Hawking-Page) wormhole \cite{13} 
quantum state.  It is certainly regular at small 3-geometries, and dies away 
rapidly at large 3-geometries. Moreover,  $ I $ is 
the Euclidean action of an asymptotically Euclidean 4-dimensional classical 
solution, outside a 3-geometry with metric  (\ref{eq:3.2}), as confirmed by studying 
the Hamilton--Jacobi equations. These give the classical
flow corresponding to the action $ I $. 
However, 
$ E \sim e^I $ is {\it not} the no-boundary (Hartle--Hawking)
\cite{12}  state. 
That conclusion can be reached by checking if $ - I $ is  
the action of a regular Riemannian solution of the classical field equations, 
with metric (\ref{eq:3.2}) prescribed on the {\it outer} boundary. 
It is quite satisfactory that the solution $ \exp ( - I ) $ gives a  wormhole
 state for Bianchi-IX but it seems very strange that the
Hartle--Hawking state is not allowed by the quantum constraints. 

It should be stressed that the above conclusions are easily 
extended to Bianchi class A models\footnote{
Only the class-$A$ models allow the spatial
sections to be compactified  by factoring (if necessary) by a
discrete subgroup of isometries \cite{A,N=2a}. Spatial compactness is needed in the 
argument used here in finding the partial derivatives 
of the wave function with respect to $ a_1, a_2 $ and $ a_3 $. However, in the
types $ II, VI_{- 1}, VII_0 $ and $ VIII $, 
the compactification might be expected to eliminate globally defined spatial
Killing vector fields, and so make it impossible to have spatially 
homogeneous triad or gravitino fields.
} \cite{A12}. 
More precisely, the physical states  
are, respectively, given by\footnote{
The solution  for $\psi^6$  is 
{\em different}  from the corresponding expression
in ref. \cite{A12}. The extra $h$ factor present in \cite{A12} {\it cannot} 
 be there though. 
The explicit form   can be 
obtained through a 
fermionic Fourier transformation \cite{4,A12}
\begin{equation}
\bar\Psi (e^{AA'}_{~~p}, \bar\psi^{A'}_{~q}) 
= D^{-1} (e^{AA'}_{~~p}) \int \Psi( e^{AA'}_{~~p}, \psi^A_{~q})
e^{-\frac{i}{\hbar}C^{~~pq}_{AA'}\psi^A_{~p}\bar\psi^{A'}_{~q}}
\Pi_{E,r}d\psi^E_{~r}~,
\label{eq:3.jap1}
\end{equation}
where 
$
C_{AA'}^{~~pq} = - \sigma \varepsilon^{pqr}e_{AA'r}~, ~
D (e_{AA'p}) = \det \left(-\frac{i}{\hbar} C_{AA'}^{~~pq}\right)$. 
This allows to obtain the wave function in the representation 
$\bar\Psi (e^{AA'}_{~~p}, \bar\psi^{A'}_{~q})$ and 
leads to a factor of $h^{-1}$ (via $
D^{-1} (e^{AA'}_{~~p})$). 
In particular, it   intertwines different representations, where  equations 
are   substantially easier to derive. 
But the inverse transformation does {\it not} involve a factor of $h$; 
see ref. \cite{4} for the reasons of this asymmetry. 
This remark should also be employed when confronting  the 
powers of $h$ in (\ref{eq:3.12}) 
coincides with the ones in ref. 
\cite{A21} and  \cite{A5,A6,A20}.} 
\begin{equation} \psi^0 \rightarrow h^{s/2} e^{ \frac{1}{2} \zeta 
m^{pq}h_{pq}}~, ~
\psi^6 \rightarrow h^{-s/2} e^{- \frac{1}{2} 
 \zeta m^{pq}h_{pq}} \Pi_{i} (\psi^{A}_i)^2. 
\label{eq:3.12}     
 \end{equation}
Here $h$ is $\det (h_{pq})$ and 
$m^{pq}$ is defined from the  
relation 
$ d w^p = {1\over 2} m^{pq} h^{-\frac{1}{2}} \epsilon_{qrs} 
\omega^r \otimes \omega^s, $
where $\omega^r$ are basis of left-invariant 1-forms on the 
space-like hypersurface of homogeneity. The constant symmetric 
matrix $m^{pq}$ is fixed by the chosen Bianchi type \cite{A}. 
In addition, the parameter $s$ specifies the {\it general} ambiguity of the 
operator ordering arising from the non-commutativity of $\psi^A_i, 
\bar \psi^{A'}_j$ and $p^{AA'}_{~~k}$.
We define $\zeta = \frac{2 \sigma}{\hbar \kappa^2}$, 
where $\sigma$ denotes the volume of spatial sections and $\kappa^2$ 
is again $8\pi^2$ times the gravitational constant. 
These results (together with the problems described in 3.2.1 
concerning the presence of a cosmological constant) were quite disconcerting. 
In fact,  the subject of supersymmetric quantum cosmology was put up to 
a harsh test and its future seemed in doubt. However, subsequent 
research present in \cite{A21,A20} provided the required 
breakthrough.

The commutators $[\bar{S}_{B'}$, $H_{AA'}]$
and $[S_{B}, H_{AA'}]$, which are 
proportional 
to  Lorentz generators, 
 are the essential new
ingredients, on which all of the following is based 

In order to show that  physical states exist in the 2-fermion
sector,  let us consider instead the wave-function
\begin{equation}
\label{eq:3.13}
    \Psi_2 = \bar{S}_{A'}\bar{S}^{A'} Y(h_{pq}),
\end{equation}
where we require, of course, that
$\bar{S}_{A'}\bar{S}^{A'} Y \ne 0 $.
This new ansatz  for the quadratic fermionic sector (see for $\psi^4$ below) 
brings {\it new} Lorentz invariants to 
$\Psi$. This is due to the presence of additional gravitational degrees of freedom
that were otherwise absent. 
A simple example of such
an additional invariant is $m^{pq}{\psi_p}^A\psi_{qA}$. 
Writing out the new  expressions present in $\psi^2, \psi^4$ 
in an explicit way it can be seen that indeed they contain
such additional invariants.
Here $Y$ is a function of the $h_{pq}$ only, and therefore, like
$\bar{S}_{A'}\bar{S}^{A'}$, a Lorentz scalar.
Therefore $\Psi_2 $ {\it automatically} satisfies the Lorentz
constraints and the $\bar{S}$ constraints

The only remaining condition is $S_A\Psi_2 = 0$, which 
reduces to
\begin{equation}
\label{eq:3.14}
  [H_{AA'} \bar{S}^{A}] Y 
   +2\bar{S}^{A}H_{AA'}Y =0.
\end{equation}
The first term is proportional to $J_{AB}$ 
and therefore vanishes. The second term vanishes if $Y$ satisfies the Wheeler-DeWitt
equation \cite{A21,A20}
\begin{equation}
\label{eq:3.15}
 {H_{AA'}}^{(0)} Y(h_{pq})=0, 
\end{equation}
where ${H_{AA'}}^{(0)}$ consists only of the bosonic
terms of $H_{AA'}$.
 Any solution of this Wheeler-DeWitt equation, which may be specified
further  by imposing, e.g.,  no-boundary, tunneling or
wormhole boundary conditions, 
 generates a solution in the 2-fermion sector.

With respect to the 4-fermion sector, the wave-function
$S^A S_A Z(h_{pq})$ 
automatically satisfies the Lorentz constraints and the $S$ constraint.
It remains to satisfy the $\bar{S}_{A'}$ constraint, which
reduces to $ 
\left(\left[H_{AA'},S^A \right]
        +2S^A H_{AA'}\right)
         Z(h_{pq})\prod_{r=1}^3(\psi_r^A)^2 = 0\,$. 
The first term in the bracket is expanded in terms containing the Lorentz generators
or $S_A$. The terms containing the Lorentz
generators vanish as they act on Lorentz scalars.   In the end,
it is   enough \cite{A21,A20} if $Z$ satisfies the
Wheeler-DeWitt equation
\begin{equation}
\label{eq:3.16}
  \left({H_{AA'}}^{(1)}-\frac{\hbar^2}{16 \pi^2 h^{1/2}}
   n^{AA'}\right)g(h_{pq}) = 0,
\end{equation}
where ${H_{AA'}}^{(1)}$ consists of those terms of
$H_{AA'}$ which remain if the terms in $p^{AA'}_i$   are brought
to the {\it left} and then equated to zero.

A  generalization of these
solutions for the case of full supergravity was 
provided in ref. \cite{C3} and briefly 
described in subsection 2.3. The
 algebra
of the constraints  has a similar form. Hence, the physical states 
found in the $\psi^2, \psi^4$ sectors are  direct minisuperspace analogues of
states in full supergravity.
While the states in the empty and filled sectors  would span at
most a 2-dimensional Hilbert space, these   physical states identified 
in the middle fermionic sectors would 
span an infinite-dimensional Hilbert space,  just as in the Bianchi
models of pure gravity.

\subsubsection{Closed FRW  models}

\indent

When considering FRW geometries in pure N=1 supergravity, 
the tetrad and  gravitino fields ought to  be chosen accordingly. 
This {\it can only be possible} 
for a suitable combination of supersymmetry, Lorentz and local 
coordinate transformation. 

Closed FRW universes have $ S^3 $ spatial sections. The tetrad of the 
four-dimensional theory can be  taken to be:
\begin{equation} e_{a\mu} = 
\left[ \begin{array}{cc} N (\tau)  & 0  \\
0 &  a (\tau) E_{\hat a i}  \end{array} 
\right] ~,~
e^{a \mu} = 
\left[ \begin{array}{cc} 
N (\tau)^{-1} &  0 \\
0 & a (\tau)^{-1} E^{\hat a i} \end{array}\right]~ ,
\label{eq:3.17}      \end{equation}
where $ \hat a $ and $ i $ run from 1 to 3.
$ E_{\hat a i} $ is a basis of left-invariant 1-forms on the unit $ S^3 $
with volume $ \sigma^2 = 2 \pi^2 $. 

This ansatz  reduces the number of degrees of freedom provided by $ e_{AA'
\mu} $. If supersymmetry invariance is to be retained, then we need an
ansatz  for $ \psi^A_{~~\mu} $ and $ \bar\psi^{A'}_{~~\mu} $ which reduces the 
number of fermionic degrees of freedom. We 
take $
\psi^A_{~~0} $ and $ \bar\psi^{A'}_{~~0} $ to be functions of time only. We 
further take
\begin{equation} 
\psi^A_{~~i} = e^{AA'}_{~~~~i} \bar\psi_{A'}~, ~
\bar\psi^{A'}_{~~i} = e^{AA'}_{~~~~i} \psi_A~,  
\label{eq:3.18}
 \end{equation}
where we introduce the new spinors $ \psi_A $ and $ \bar\psi_{A'} $ which
are functions of time only.
This means we truncate the general decomposition
$ \psi^A_{~~B B'} = e_{B B'}^{~~~~i} \psi^A_{~~i} $ in (\ref{eq:3.5}) 
at the spin$-\frac{1}{2}$ mode level. I.e., with 
$\beta^A = {3 \over 4} n^{AA'} \bar\psi_{A'} \sim \bar\psi^A$.
This constitutes 
a direct consequence of assuming a FRW geometry and it is 
 a necessary
condition for  supersymmetry invariance to be retained. 
It is also important to stress that 
auxialiary fields are also 
required to balance the number of fermionic and bosonic 
degrees of freedom.    However,  
 these auxialiary fields  can be  neglected in the end (cf. ref. \cite{A8,A9}).
The above ans\"atze preserves 
the form of the tetrad
under a {\it suitable combination}  of 
supersymmetry, 
 Lorentz and local coordinate transformations (\ref{eq:2.1})-(\ref{eq:2.6}). 
I.e., we get \cite{A9,A11,A18a}
\begin{eqnarray}\delta e^{AA'}_i  
& = &  \left( - N^{AB} + a^{-1} \xi^{AB}
+ i \kappa \epsilon^{(A} \psi^{B)} \right ) e_B^{~~A'}{}_i  \nonumber \\
~&+&  \left( - \bar N^{A'B'} +
 a^{-1} \bar \xi^{A'B'} + i \kappa \bar \epsilon^{(A'} 
\bar 
\psi^{B')} \right) e^A_{~~B'i} 
+   {i \kappa \over 2} \left( \epsilon_C \psi^C + \bar
 \epsilon_{C'} \bar \psi^{C'} \right) 
e^{AA'}_{~~~~i}~, \label{eq:3.19}
\end{eqnarray}
where $\xi^{AB}, N^{AB}, \epsilon^A$ and Hermitian 
conjugates parametrize local coordinate, Lorentz and supersymmetry 
transformations, respectively. 
Notice  that the ansatz  for the tetrad is preserved, i.e., $\delta e^{AA'}_i  \equiv 
 P_1\left[e^{AA'}_\mu, \psi^A_i\right] 
e^{AA'}_\i$, 
 provided that the relations
\begin{equation}
N^{AB} - a^{-1} \xi^{AB} - i \kappa \epsilon^{(A} \psi^{B)} = 0~, 
~\bar N^{A'B'} - a^{-1} \bar \xi^{A'B'} - i \kappa \bar
\epsilon^{(A'} \bar \psi^{B')} = 0~, 
\label{eq:3.20}
\end{equation}
between the generators of Lorentz, coordinate and supersymmetry 
transformations are satisfied.

The ansatz  for the fields $ \psi^A_{~~i} $ and $ \bar  \psi^{A'}_{~~i} $ should
also be preserved under the same combination of transformations, 
together with (\ref{eq:3.19}), (\ref{eq:3.20}). Hence \cite{A9,A11,A18a}
\begin{eqnarray}\delta \psi^A_i  
 & = & {3i \kappa \over 4} \epsilon^A \psi^B \bar \psi^{B'} e_{BB'i} + 
a^{-1} \bar \xi^{A'B'} e^A_{~~B'i} \bar \psi_{A'} \nonumber \\
~&+ & \left[ {2 \over \kappa} \left( {\dot a \over a N} + {i \over a} \right) - {i \kappa \over
2 N} \left( \psi_F \psi^F_{~~0} + \bar
 \psi_{F'0} \bar\psi^{F'} \right) \right] n_{BA'} 
e^{AA'}_{~~~~i} \epsilon^B~. \label{eq:3.21}     \end{eqnarray}
and its Hermitian conjugate.
The Ansatz  for $ \psi^A_{~~i} $ is then preserved,
i.e., $\delta \psi^A_i \equiv 
P_2 \left[e^{AA'}_\mu, \psi^A_i\right]
e^{AA'}_i\bar \psi_{A'} 
$, 
 if we impose the additional 
constraint
$ \psi^B \bar \psi^{B'} e_{BB'i} = 0 $
and take $ \bar \xi^{A'B'} = 0 $. The former constitutes basically a reduced form of the 
Lorentz constraint   in the full theory 
and is present in the two equivalent forms \cite{A8,A9}:
\begin{equation} 
J_{AB} = \psi_{(A} \bar\psi^{B'} n_{B)B'} = 0~,  
\bar J_{A'B'} = \bar\psi_{(A'} \psi^B n_{BB')} = 0~.
\label{eq:3.22}
 \end{equation}
It should be stressed that 
the invariance of $\psi^A_i, \bar \psi^{A'}_i$  
 strongly  depends on  the last term of (\ref{eq:3.21}). 
The only option was to put the other terms as equal to zero.  
 Notice 
that   solely   ${}^3 D_i \epsilon_A$ 
is able to produce $\delta \psi^A_i \equiv 
P_2 \left[e^{AA'}_\mu, \psi^A_i\right]
e^{AA'}_i\bar \psi_{A'} 
$ (see ref. \cite{A9,A18a} for further details).
By requiring that the constraint (\ref{eq:3.22})  be preserved under the same 
combination of transformations as used above, one finds equations
 which 
are satisfied provided the supersymmetry constraints $ S_A = 0,~\bar S_{A'} =
0 $  hold. 
By further requiring that the supersymmetry 
constraints be preserved, one finds additionally that the Hamiltonian 
constraint $ {\cal H} = 0 $  should hold.

By imposing the 
above mentioned symmetry conditions, we obtain a
one-dimensional (mechanical) model depending only on $t$. 
Classically, the
constraints vanish, and this set of  
constraints forms an algebra.
The constraints are functions of the 
basic dynamical variables. For the
gravitino fields, their canonical momenta produce 
(second-class) 
constraints. These are eliminated when 
Dirac
brackets are introduced \cite{4,9,4a} instead of the original Poisson brackets.

It is useful 
to make slight redefinitions of the dynamical fields. We let $ a \mapsto 
{\kappa \over \sigma} a,~ \psi^A \mapsto \sqrt {2 \over 3} {\sigma^{1/2} \over 
(\kappa a)^{3/2}} \psi^A $ and $ \bar \psi^{A'} \mapsto \sqrt {2 \over 3} 
{\sigma^{1/2} \over ( \kappa a)^{3/2}} \bar \psi^{A'}$. 
 We include the constraint $ J_{AB} = 0 $ by adding $ M^{AB} J_{AB}
$ to the Lagrangian, where $ M^{AB} $ is a Lagrange multiplier. In order to 
achieve the simplest form of the generators and 
their Dirac brackets, we make the following redefinitions of the 
non-dynamical variables $ N,~\psi^A_{~~0},~\bar \psi^{A'}_{~~0} $ and $ M^{AB} $:
\begin{eqnarray}
\hat N &= &{\sigma N \over 1 2 \kappa}~, ~
\rho^A =  {i (\kappa \sigma)^{1/2} \over 2 \sqrt 6 a^{1/2}} \psi^A_{~~0} + {i 
\sigma N \over 1 2 \kappa a^2} n^{AA'} \bar \psi_{A'}~, 
\label{eq:3.23}         
\\
L^{AB} &= & M^{AB} - {(\kappa \sigma)^{1/2} \over 3 \sqrt 6 a^{3/2}} \psi^{(A}_{~~0} 
\psi^{B)} - {2 (\kappa \sigma)^{1/2} \over 3 \sqrt 6 a^{3/2}} \bar 
\psi^{(A'}_{~~~0} \bar \psi^{B')} n^A_{~~A'} n^B_{~~B'}~,
\label{eq:3.24}
\end{eqnarray}
and Hermitian  conjugates.
Our constraints then take the rather simple form \cite{A8,A9,A11}
\begin {equation}
S_A = \psi_A \pi_a - 6 i a \psi_a,~ H = - a^{-1} ( \pi_a^2 + 
3 6 a^2 ),~  
\bar S_{A'} = \bar \psi_{A'} \pi_a + 6 i a \bar \psi_{A'},~ J_{AB} =
\psi_{(A} \bar \psi^{B'} n_{B)B'}~. 
\label{eq:3.25}
\end{equation}
The presence of the free 
parameters $ \rho_A,~\bar \rho_{A'} $ shows that this model has $ N = 4 $ 
local supersymmetry in 1 dimension. 

In solving the supersymmetry constraints $S_A, \bar S_{A'}$ in (\ref{eq:3.25}),
 note that $ J^{AB} \Psi = 0 $ 
implies that $ \Psi $ can be written as $ \Psi = A + B \psi_A \psi^A $, where 
$ A $ and $ B $ depend only on $ a $. 
The solutions are  
\begin{equation}
\Psi = C \exp [ - 3 a^2 / \hbar ] + D \exp [ 3 a^2 / \hbar ] \psi_A 
\psi^A~, 
\label{eq:3.26}
\end{equation}
where $ C $ and $ D $ are independent of $ a $ and $ \psi^A $. 
The exponential factors have a {\it semi-classical} 
 interpretation as $ 
\exp ( - I / \hbar ) $, where $ I $ is the Euclidean action for a classical 
solution outside or inside a three-sphere of radius $ {\kappa \over \sigma} a $
with a prescribed boundary value of $ \psi^A $. I.e., we get a 
Hartle-Hawking solution for $C=0$.

\subsection{Models with a cosmological constant}

\indent 

It is of interest to study more general locally
supersymmetric actions, initially in Bianchi models. Possibly the simplest
of such generalization is the addition of a cosmological constant,
$\Lambda$,  in $ N = 1 $
supergravity (see ref. \cite{A13,A14,A15,pkt} and references therein).

\subsubsection{Bianchi class A models}

\indent 

For the ansatz  (\ref{eq:3.4})
we
shall see that 
we {\it cannot} find any non-trivial 
  physical quantum states  
for Bianchi class A models \cite{A13,A14,A15,D5}.

The  action for N=1 supergravity (\ref{eq:2.7})
   includes now the additional terms 
\begin{equation}
 S = - \int d^4 x (\det e) \left[ 
( 2 \kappa^2 )^{- 1}   3 g^2 
- 
{1 \over 2} g ~( \psi^A_{~~\mu} e_{A B'}^{~~~~\mu} e_B^{~~B' \nu}
\psi^B_{~~\nu} ) 
\right] + {\rm H.c.} 
\label{eq:3.27}         \end{equation}
  Here $ g $ represents  the cosmological constant through 
the relation  $ \Lambda = 
{3 \over 2} g^2 $.  

The corresponding quantum constraints read  
\begin{equation}
\bar S_{A'} \Psi = - i \hbar g h^{1 \over 2} e^{A~~~i}_{~~A'} n_{A B'} D^{B
B'}_{~~~~j i} \left( h^{1 \over 2} {\partial \Psi \over \partial  \psi^B_{~~j}} \right)  \\
~+  \epsilon^{i j  k} e_{A A' i} \omega^A_{~~B j} \psi^B_{~~k} \Psi - {1 \over 2} 
\hbar
\kappa^2 \psi^A_{~~i} {\delta \Psi \over \delta e^{A A'}_{~~~~i}} = 0~,
\label{eq:3.28}
\end{equation}
with  the corresponding Hermitian conjugate.
We have made
the replacement $ \delta \Psi / \delta \psi^B_{~~j} \longrightarrow h^{1 \over 2} 
\partial \Psi / \partial \psi^B_{~~j} $. 
The $ h^{1 \over 2} $ factor is necessary as to ensure that each       
term has the correct weight in the equations. Namely when we take a       
variation of a  Bianchi geometry whose spatial sections are compact,       
multiplying by      
 $ \delta  / \delta h_{i j} $ and integrating over the       
three-geometry . The cause can be       
identified in the term $h^{-{1\over 2}}$ in expression (\ref{eq:2.15}).      
It  can be  checked  that the inclusion of $h^{{1 \over 2}}$       
gives the correct supersymmetry constraints in the $ k = + 1 $       
Friedmann model (see next subsection).

We will use       
the gravitino field written in terms of the $\beta$ spin$-\frac{1}{2}$ and       
$\gamma$  spin-$3 \over 2$ modes 
(see eq. (\ref{eq:3.5})),  
 and  the following expressions \cite{A21a}:        
\begin{equation} {\partial (\beta_A\beta^A)\over \partial \psi^B_{~~i}  }=  -         
n_A^{~~B'} e_{B B'}^{~~~~i} \beta^A~,~ 
{\partial ( \gamma_{A D C} \gamma^{A D C}      
) \over \partial \psi^B_{~~i}} = - 2 \gamma_{B D C}~       
n^{C C'} e^{D~~~i}_{~~C'}, 
\label{eq:3.29}        
\end{equation}      
\begin{equation} {\partial \beta_A \over \partial \psi^B_{~~j} } =       
- \frac{1}{2} n_A^{~~B'} e_{BB'}^{~~~j}~,~ 
{\partial \gamma^{ADC} \over \partial \psi^B_{~~j}}  =       
{1\over 3} \left(n^{CC'} e^{D~~~j}_{~~C'} \epsilon_B^{~~A}      
+       
 n^{AC'} e^{C~~~j}_{~~C'} \epsilon_B^{~~D}        
+      
 n^{DC'} e^{A~~~j}_{~~C'} \epsilon_B^{~~C} \right). \label{eq:3.30}  
\end{equation}      
We can also write out       
$ \beta^A $ and $ \gamma_{B D C} $ in terms of $ e^{E E'}_{~~~~j} $       
and $ \psi^E_{~~j} $ as        
\begin{equation} \beta_A = -\frac{1}{2} n_A^{A'i} e_{BA'}^i \psi^B_i ~,~ \gamma_{ABC} =       
{1 \over 3} \left( n_C^{~C'} e_{BC'}^{~~~i} \psi_{Ai} +       
 n_A^{~C'} e_{CC'}^{~~~i} \psi_{Bi}       
+  n_B^{~C'} e_{AC'}^{~~~i} \psi_{Ci}\right). 
\label{eq:3.31}
\end{equation}      

First consider the $ \bar S_{A'} \Psi = 0 $ constraint at the level $ \psi^1 $
in powers of fermions.
Since it  holds for all $ \psi^B_{~~i} $, we can take the gravitino terms 
out. From multiplying  this equation by $ e^{B A' m} $, we obtain 
 by taking a variation among the Bianchi-IX 
metrics, that 
\begin{equation} 
 \hbar \kappa^2 {\partial A \over \partial a_1} + 
16 \pi^2 a_1 A + 6 \pi^2 \hbar
g a_2 a_3  B  = 0~,
\label{eq:3.32}
\end{equation}
and two others given by cyclic permutation of $a_1, a_2, a_3$.

Next we consider the $ S_A \Psi = 0 $ constraint at order $ \psi^1 $. 
Using expressions (\ref{eq:3.29})--(\ref{eq:3.31}), 
we divide again  by $ \psi^B_{~~j} $. 
Multiplying by $ n^A_{~~D'} e^{B D' n} $, then
multiplying by different choices $ \delta h_{i m} = \partial h_{i m} / \partial a_1 $ etc. 
and integrating over the manifold, we find the constraints
\begin{eqnarray}
& & {1 \over 16} \hbar^2 \kappa^2 a_1^{- 1} \left( a_i {\partial B\over \partial a_i} \right) 
  -   {1 \over 3} \hbar \kappa^2 \left[ 3 {\partial C \over \partial a_1} - 
A^{- 1} \left( a_i 
{\partial C \over \partial a_i} \right) \right]   
-  16 \pi^2 g a_2 a_3 A  \nonumber \\ 
& - & \pi^2 \hbar a_2 a_3  \left( {a_1 \over a_2 a_3} + {a_2 \over a_1 a_3 } + 
{a_3 \over a_1 a_2} \right)
B 
 + {1 \over 3} ( 16 \pi^2 ) \hbar a_2 a_3  \left( {2 a_1 \over a_2 a_3 } - 
{a_2  \over a_3 a_1} - 
{a_3  \over a_1 a_2} \right) C = 0~,
\label{eq:3.33}        \end{eqnarray}
and two more equations given by cyclic permutation of $ a_1, a_2, a_3$.

Now consider the $ \bar S_{A'} \Psi = 0 $ constraint at order $ \psi^3 $. It 
will turn out that we need go no further than this. From this constraint we 
can set separately to zero the coefficient of $ \beta^C ( \gamma_{D E F}  
\gamma^{D E F}
) $, the symmetrized coefficient of $ \gamma_{D E F} ( \beta_C \beta^C ) $ 
and the symmetrized coefficient of $ \gamma_{F G H} ( \gamma_{C D E} \gamma^{C D E} ) $. 
From here we derive three equations and from 
following similar steps as above (cf. ref. \cite{A14,A15})  we get

\begin{equation}
{3 \over 4} 16 \pi^2  \hbar g a_1 a_2 a_3 D  + {2 \over 3} g 16 
\pi^2 ~ \left( a_1^2 + a_2^2 + a_3^2 \right) C 
+ {2 \over 3} \hbar \kappa^2 \left( a_i{\partial C \over \partial a_i} \right) = 0~,
\label{eq:3.34}. \end{equation}
and 
\begin{equation}
 3 \hbar \kappa^2 {\partial B \over \partial a_1} - \hbar \kappa^2 A^{- 1} 
\left( a_i 
{ \partial B \over \partial a_i} \right) - 16 \pi^2 a_2 a_3 
 \left( {a_3  \over a_1 a_2 } + {a_2  \over a_1 a_3 } - 2 {a_1 \over a_2 a_3} \right) 
B 
= 0~\label{eq:3.35}        \end{equation}
and two more equations given by permuting $a_1, a_2,  a_3 $ cyclically. The equation 
(\ref{eq:3.35}) also holds with $ B$ replaced by $ C $.

Consider first the equation (\ref{eq:3.35}). It 
 can be checked that these are equivalent to
\begin{equation}
 \hbar \kappa^2 \left( a_1 {\partial B \over \partial a_1} - 
a_2 {\partial B  \over \partial 
a_2} \right) = 16 \pi^2 \left( a_2^2 - a_1^2 \right) B 
\label{eq:3.36}
\end{equation}
and cyclic permutations. We can then integrate eq. (\ref{eq:3.36}) 
and ciclically, along a
characteristic $ a_1 a_2  = $ const., $ a_3  = $ const., say,  using the parametric 
description $ a_1 = w_1 e^\tau $, $ a_2 = w_2 e^{- \tau} $, to obtain
in the end 
\begin{equation}
 B = f (a_1a_2 a_3) \exp \left[ - {8 \pi^2 \over \hbar
\kappa^2} ~\left( a_1^2 + a_2^2 + a_3^2 \right) \right]~,  
\label{eq:3.37}
\end{equation}
and similarly for 
C. Substituting these back in (\ref{eq:3.33}), 
we   get  a set of equations whose only solution is $ 
C= 0 $.
The equation (\ref{eq:3.33}) and its cyclic permutations, with $ C = 0 $, must
be solved consistently with eq. (\ref{eq:3.32}) and its cyclic permutations. 
Eliminating $ A $, one finds
$ B = 0$ and subsequently $A=0$.
Then we can argue using the duality mentioned in subsection 
3.1.1   that
$D = E = F  = 0.$
Hence there are {\it no}  physical quantum states obeying the constraint equations 
in the diagonal Bianchi-IX model {\it if} $\Psi$ has the form 
given by anasatz (\ref{eq:3.4}). 
 The same conclusion can be   generalized  for the
 case of Bianchi class A models \cite{A15,D5}.

\vspace{0.3cm}

These vexatious results  motivated the research described in ref. \cite{A21,A20} 
and were only properly dealt with in ref. \cite{A23b}. 
In fact, the 
doubts thereby raised are entire legitemate: even though 
canonical quantum supergravity has more constraints 
than ordinary quantum gravity, it has surely much more 
degrees of freedom than gravity. 
According to \cite{A23b}, non trivial solutions can be  found by employing 
Ashtekar variables \cite{14}. In particular, we take 
the complexified spin-connection
${\cal A}_{pAB} $ and the tensor density $
\sigma^{pAB}= -\left(\sqrt{2}\right)^{-1}
\varepsilon^{pqr}{e_q}^{AA'}{{e_r}^B}_{A'}
$
as a canonically conjugate pair of  coordinates.

  The transformation 
from the ${\cal A}_p^{AB}$-representation to the $e_p^{AA'}$-
representation  requires the generalized Fourier-transform
\begin{equation}
\Psi \left({e_p}^{AA'}\right) = \int\left[\prod_p\prod_{(A\le B)}d{{\cal 
A}_p}^{AB}\right]e
^{-{{\cal A}_p}^{AB}{e_{qA}}^{A'}e_{rBA'}\varepsilon^{pqr}}\Psi
 \left({{\cal A}_p}^{AB}\right),
\label{eq:3.38}
\end{equation}
along a suitable 9-dimensional contour 
chosen in order to achieve convergence, and permitting
partial integration without boundary terms. Apart from this condition
the contour may still be chosen arbitrarily. There are
different possible choices corresponding to different linearly
independent solutions. In addition, a similarity tranformation has to be
performed  which takes the form
\begin{equation}
\Psi({e_p}^{AA'}) \mapsto  e^{- 8\pi^2 m^{pq}{e_p}^{AA'}e_{qAA'}}
 \Psi ({e_p}^{AA'}).
\label{eq:3.39}
\end{equation}

A special class of solutions (undoubtedly, more general solutions
 exist) is 
\begin{equation}
\Psi \sim \exp\left[i\left(F({{\cal A}_p}^{AB},{\psi_p}^A)
+G({{\cal A}_p}^{AB})\right)\right]~,
\label{eq:3.40}
\end{equation}
with
\begin{equation}
F \sim -\frac{1}{\Lambda}\left( 16 \pi^2m^{pq}{\psi_p}^A\psi_{qA}+
 \varepsilon^{pqr}{\psi_p}^A{{\cal A}_{qA}}^B\psi_{rB}\right)~,
\label{eq:3.41}
\end{equation}
and 
where $G$ is independent of the $\psi_p^A$ 
\begin{equation}
G \sim \frac{i}{\Lambda }\left(
  16 \pi^2m^{pq}{{\cal A}_p}^{AB}{\cal A}_{qAB}+\frac{2}{3}
   \varepsilon^{pqr}{{\cal A}_p}^{AB}{{\cal A}_{qB}}^C 
{\cal A}_{rCA}\right)\, ~,
\label{eq:3.42}
\end{equation}
The function
$G$ can be expressed by the Chern-Simons functional
 integrated over the spatially homogeneous 3-manifolds, 
consistently with \cite{D8,D9,15a}.

The ${\cal A}_p^{AB}$ integrals required 
for the transformation  from the Ashketar-representation to the metric
representation
need 
further specification. Not all of these integrals need to be done, because only three 
of the nine
degrees of freedom of ${{\cal A}_p}^{AB}$ are physical\footnote{ Six correspond
to gauge freedoms (three from basis changes of the 1-forms $\omega^p$, three
from Lorentz frame
rotations) which can be fixed by a choice of gauge and are not integrated
over in that gauge. }.
However, even the remaining three integrals cannot
all be performed analytically. In the
limit of vanishing cosmological constant $\Lambda\to 0$ a stationary-phase
approximation becomes possible. We then  expand $\Psi$ as
\begin{equation}
\Psi\sim \sum_{n=1}^3
 \frac{\left[iF({{\cal A}_p}^{AB},{\psi_p}^A)\right]^n}{n!}
 e^{i G({{\cal A}_p}^{AB})}\,.
\label{eq:3.43}
\end{equation}
One stationary phase point is at ${\cal A}_p^{AB} = 0$.
 The first solution is therefore defined by chosing
a suitable contour passing through this point. In the  limit of 
$ \Lambda  \rightarrow 0$ the dominant
fermion term has 6 $\psi_p^A$ -factors.
 Then we obtain from the stationary phase at
${{\cal A}_p}^{AB} = 0$
\begin{equation}
\Psi({e_p}^{AA'}) \sim \,\mbox{\rm const}\,
 \left(\prod_{p=1}^3\prod_{A=0,1}{\psi_p}^A\right)
  e^{\left(- 8\pi^2 m^{pq}{e_p}^{AA'}e_{qAA'}  \right)}.
\label{eq:3.44}
\end{equation}
The exponent corresponds to that of a   wormhole
state in the 6-fermion sector \cite{A21,A20}. 
A divergent factor
$\Lambda^{-3/2}$ has been absorbed in the prefactor.

Other stationary phase points are at ${{\cal A}_p}^{AB}\neq 0$ and further
solutions are obtained by chosing integration contours through any of
them. For the case of a Bianchi IX (cf. ref. \cite{A23b} for more details):
\begin{equation}
\Psi(b_1,b_2,b_3) \sim \sum_n\frac{i^n}{n!}F^n
  \int \,d{\cal A}_1d{\cal A}_2 d{\cal A}_3 
({\cal A}_1 {\cal A}_2 {\cal A}_3)^2
    \exp\left[\sum_q {\cal A}_{qAB}\sigma^{ABq}+iG({\cal A}_{ABq})\right]\,.
\label{eq:3.45}
\end{equation}
Points of stationary phase now
satisfy the equations
$
 2{\cal A}_1 {\cal A}_2 = 16 \pi^2 {\cal A}_3\mbox{\rm~ and cyclic}
$. 
From (\ref{eq:3.45}) we then get  in the 4-fermion sector 
the folowing amplitudes
\begin{equation}
\frac{e}{\Lambda}^{-8\pi^2[(b_1^2+b_2^2+b_3^2)+2(b_1b_2+b_2b_3+b_3b_1)]}   
 ~ ,~ 
\frac{e}{\Lambda}^{-8\pi^2 [(b_1^2,+b_2^2+b_3^2)+2(b_1b_2-b_2b_3-b_3b_1)]}~,
 \label{eq:3.46}
\end{equation} where the exponent in the former corresponds to 
 the Hartle-Hawking state
\cite{A21,A20,12} and  the other has not been discussed before.

\subsubsection{Closed FRW models}

\indent 

In a supersymmetric 
FRW  model with cosmological constant    terms, the coupling between the different 
fermionic levels `mixes up' the pattern present in 3.1.2.
From 
 $ \bar S_{A'} \Psi = 0 $ and $ S_A \Psi = 0 $   we  obtain \cite{A14,A15} 
\begin{equation}
\hbar \kappa^2 {d A \over d a} + 48 \pi^2 a A + 1 8 \pi^2 \hbar g 
a^2 B= 0~,~ 
 \hbar^2 \kappa^2 {d B \over d a} - 48 \pi^2 \hbar a B- 2 5 6
\pi^2 g a^2 A = 0~. 
\label{eq:3.47}
\end{equation}
These give second-order equations  for $ A(a) $
which have  two independent solutions, of the form
\begin{equation}
A=   c_0 + c_2 a^2 + c_4 a^4 + \ldots~, ~
A =  a^3 ( d_0 + d_2 a^2 + d_4 a^4 + \ldots ) ~, 
\label{eq:3.48}
\end{equation}
convergent for all $ a $. They obey complicated recurrence relations, where 
 $ c_6 $ is related to $ c_4,~c_2 $ and $ c_0 $.

We can look for asymptotic solutions of the type $ A \sim ( B_0 + 
\hbar B_1 + \hbar^2 B_2 + \ldots ) \exp  ( - I / \hbar ) $, and finds
$ I = \pm {\pi^2 \over g^2} ( 1 - 2 g^2 a^2 )^{3 \over 2}~, $
for $ 2 g^2 a^2 < 1 $.
The minus sign in $ I $ corresponds to taking the action of the classical 
Riemannian solution filling in smoothly inside the three-sphere, namely a
portion of the four-sphere $ S^4 $ of constant positive curvature. This gives
the Hartle--Hawking state \cite{12}. For $ a^2 > ( 1 / 2 g^2 ) $, the Riemannian
solution joins onto the Lorentzian solution
\begin{equation}
 \Psi \sim \cos \left\{ \hbar^{- 1} \left[ {\pi^2 
\left( 2 g^2 a^2 - 1 \right)^{3 \over
2} \over g^2} - {\pi \over 4} \right] \right\}~, 
\label{eq:3.49}
\end{equation}
which describes de Sitter space-time.

\subsection{Models with supermatter}

\indent 

In addressing the presence of (super)matter fields in 
minisuperspace models  we ought to choose 
the type of  action 
we will employ. 
 Following ref. \cite{8}, we 
will consider N=1 supergravity coupled to the 
more general gauged supermatter. The corresponding general theory, related 
properties and features (some of which are relevant in   minisuperspaces) were
described in subsection 2.1.2. We may 
take  the  group  $\hat{G} = SU(2)$ as the
gauge group in a $k=+1$ FRW model. In 
this case we have  the Killing potentials 
\begin{equation} D^{(1)}= \frac{1}{2} \left({{\phi + \bar \phi}\over {1 + \phi\bar\phi}}\right),
~D^{(2)}= - {i \over 2} \left({{\phi - \bar \phi}\over {1 + \phi\bar\phi}}\right),
~D^{(3)}=~- \frac{1}{2} \left({{1 - \phi\bar\phi}\over {1 + \phi\bar\phi}}\right)~,
\label{eq:3.50}
\end{equation}
and the K\"ahler potential is $ K = \ln(1 + \phi \bar \phi)$. 
The corresponding K\"ahler metric is 
$ g_{\phi \bar\phi}= { 1 \over (1 + \phi \bar \phi)^{2} }$ $
~,g^{\phi \bar\phi}=$ $ ~(1 + \phi \bar \phi)^{2} $. 
The Levi-Civita connections of the   K\"ahler manifold are just 
$ \Gamma^{\phi}_{\phi \phi} = g^{\phi \bar \phi} { {\partial g_{\phi \bar \phi} }
\over {\partial\phi}} = -2 { \bar \phi \over (1 + \phi \bar \phi)} $
and its complex conjugate. The rest of the components are zero.

The simplest  choice for the matter fields in a FRW geometry 
 is to take 
the scalar super-multiplet, consisting of a complex scalar 
field $ \phi, \bar\phi $ and  spin-$\frac{1}{2}$ field $ \chi_A, \bar\chi_{A'}$ 
as  
spatially homogeneous, depending only on time. 
Similarly, a simple choice for the 
 spin-$\frac{1}{2} $ $  
\lambda^{(a)}_A, \bar \lambda^{(a)}_{A'}  $, $(a)=1,2,3$, 
would be to take each component as an arbitrary-time dependent function.

In ordinary quantum cosmology with 
gauge fields,  it 
 is not sufficient 
for  $A_\mu^{(a)}$
to have simple 
homogenous components. 
Special ans\"atze are required for $A^{(a)}_\mu$, depending  on 
 the 
gauge group  considered, which then may also  affect the choices for  $\phi$ 
(cf. ref. \cite{mb,4m}, \cite{mem}-\cite{meob}).
A suitable ansatz   for
 $ A^{(a)}_\mu $
  requires it 
to be 
{\it invariant up to gauge 
transformations}. 
Assuminig a gauge group 
$\hat{G} = SO(3) \sim SU(2)$, 
 the 
spin-1 field is taken to be 
\begin{equation}
{\bf A}_{\mu}(t)~\omega^{\mu}  = 
\left(
{{f(t)}\over {4}}
\varepsilon_{(a)i(b)}{\cal T}^{(a)(b)}\right)\omega^i
~,\label{eq:3.51}        \end{equation}
where $\{\omega^{\mu}\}$ represents the moving coframe 
$\{\omega^{\mu}\} = \{ dt,\omega^i\}$, $\omega^i = \hat E^i_{~\hat c} dx^{\hat c}$, 
$~(i,\hat c = 1,2,3)~$,  of 
one-forms, invariant under the left action of $SU(2)$.  ${\cal T}_{(a)(b)}$
are the generators of the $SU(2)$ gauge group.
The idea of  this ansatz  for a non-Abelian 
spin-1 field is to define a homorphism of the 
isotropy group $SO(3)$ to the gauge group. 
This homomorphism defines the gauge transformation
which 
 {\em compensates}  the action of a given $SO(3)$ rotation. 
Hence, the above form 
for the gauge field,  where the $A_0$ component is taken 
to be identically zero. 
None of the 
 gauge symmetries will survive: all the available
  gauge transformations are required to 
cancel out the action of a given $SO(3)$ rotation. 
Thus, we will not have  a gauge  
constraint\footnote{{\rm 
However, in the case of larger gauge group 
some of the 
 gauge symmetries will survive. These will give rise, in the 
one-dimensional model, to local internal symmetries with 
a reduced gauge group. 
Therefore,   a gauge  
constraint can be expected to play an important role 
in that case. The  study of such a model 
would be particularly  interesting.}}
 $Q^{(a)}=0$.

\subsubsection{Canonical quantization of a FRW model}

\indent

We will then use   the ans\"atze 
(\ref{eq:3.17}), (\ref{eq:3.18}) 
 described in subsection 3.1.2, 
together with expression 
(\ref{eq:3.51}) and taking 
$\lambda^{(a)}_A = \lambda^{(a)}_A (t)$. The action of the full theory 
(Eq. (25.12) in ref. \cite{8}) is reduced to one with a 
 finite number of degrees of freedom.
Starting from the action so  obtained,
 we  study the Hamiltonian formulation of this model \cite{A17,A18}.

The contributions from the complex scalar  fields $\phi, \bar\phi$ 
to the $\bar S_{A'}$ constraint are seen to be
 \begin{eqnarray}
{1\over \sqrt{2}} \bar \chi_{A'} \left[ \pi_{\bar \phi}\right. & +
 & \left. {i \sigma^{2} a^{3} \over 2 \sqrt{2}} 
{\phi \over (1 + \phi \bar \phi)} n_{BB'} \bar \lambda^{(a)B'} 
\lambda^{(a)B} - {5i \over 2 \sqrt{2}}
{\sigma^{2}a^{3} \phi \over (1 + \phi \bar \phi)^{3}} n_{BB'} 
\bar \chi^{B'} \chi^{B} \right.
\nonumber \\ 
  & - & \left. {3i \over 2 \sqrt{2}} { \sigma^{2} a^{3} \phi \over (1 + \phi \bar \phi)} n_{BB'} \psi^{B} \bar \psi^{B'} - {3 \over \sqrt{2}} 
{\sigma^{2}a^{3} \over (1 + \phi \bar \phi)^{2}} n_{BB'} \chi^{B} \bar \psi^{B'} \right] \nonumber \\
& + & {\sigma^{2}a^{2}gf \over \sqrt{2} (1 + \phi \bar \phi)^{2}} \sigma ^{a}_{~AA'}n^{AB'} \bar 
\chi_{B'} X^{a}.
\label{eq:3.52}
\end{eqnarray}
The contributions to the $\bar S_{A'}$ constraint  from the spin-1 field are
\begin{eqnarray}
 & - & i {\sqrt{2} \over 3} \pi_{f} \sigma^{a}_{~BA'} \lambda^{(a)B} 
 + { \sigma^{2}a^{3} \over 6} \sigma^{a}_{~BA'}\lambda^{(a)B} n_{CB'} (\sigma^{bCC'} \bar \psi_{C'} \bar \lambda^{(b)B'} + 
\sigma^{bAB'} \psi_{A} \lambda^{(b)C})\nonumber \\
&+ & {1 \over 8 \sqrt{2}}\sigma^2 a^4 \sigma^{(a)C}_{~~A'} [1 - (f - 1)^2] \lambda^{(a)}_{~~C}
\nonumber \\
& + &  {1 \over 2} \sigma^{2}a^{3} \lambda^{(a)A}
\left(-n_{AB'} \bar \psi_{A'} \bar 
\lambda^{(a)B'} + n_{BA'} \psi_{A} \lambda^{(a)B} - {1 \over 2} n_{AA'} \psi_{B} \lambda^{(a)B}
 + {1 \over 2} n_{AA'} \bar \psi_{B'} 
\bar \lambda^{(a)B'}\right)~. \label{eq:3.53}        \end{eqnarray} 
The contributions from the spin-2 field and spin-3/2 field to $\bar S_{A'}$ constraint are 
\begin{equation}
 {i \over 2 \sqrt{2}}a \pi_{a} \bar \psi_{A'} -{3 \over \sqrt{2}} \sigma^{2}a^{2} \bar \psi_{A'}
+ {3 \over 8} \sigma^{2} a^{3} n^{B}_{~A'} \bar \psi^{B'} \psi_{B} \bar \psi_{B'} ~. 
\label{eq:3.54}
\end{equation}
The following terms are  also present in the  $\bar S_{A'}$   supersymmetry constraint:
\begin{eqnarray} 
 & - &{1 \over \sqrt{2}} \sigma^{2}a^{3} g D^{a} n_{AA'} \lambda^{(a)A} 
+  {\sigma^{2}a^{3} \over
 \left(1 + \phi \bar \phi\right)^{2}}\left(- n_{BA'} \bar \psi_{B'} 
+ {1 \over 2} n_{BB'} \bar \psi_{A'}\right) \chi^{B} \bar \chi^{B'}
\nonumber \\ 
& - &  {1 \over 4} \sigma^{2}a^{3}\left(n_{AB'} \lambda^{(a)A} \bar \lambda^{(a)}_{~~A'} \bar \psi^{B'} + n_{AA'} \lambda^{(a)A} \bar \lambda^{(a)}_{~~B'} \bar \psi^{B'}\right) \nonumber \\ 
&  -&  {1 \over 4 (1 + \phi \bar \phi)^{2}} \sigma^{2}a^{3} \left(n_{AB'} \chi^{A} \bar \chi_{A'} \psi^{B'} + n_{AA'}
\chi^{A} \bar \chi_{B'} \psi^{B'}\right)~. 
\label{eq:3.55}
\end{eqnarray}
The supersymmetry constraint $\bar S_{A'}$ is then 
 the sum of the above expressions. The supersymmetry constraint $S_{A}$ is just the
 complex conjugate of $\bar S_{A'}$.
Notice that  
the above expressions correspond to a 
gauge group $SU(2)$ and hence a compact K\"ahler manifold, which 
implies that the analytical potential $P(\Phi^{I})$ is zero \cite{30}.

Let us obtain expressions for the  quantum 
supersymmetry constraints.
 First we need to redefine the  fermionic fields, $ \chi_{A} $,  $ \psi_{A} $ and $\lambda_{A}^{(a)}$ 
in order to simplify the Dirac 
brackets, following  the steps described in  
\cite{3,A9,A15a,A23,4a}. We take
\begin{equation}
 \hat \chi_{A} = {\sigma a^{3 \over 2} \over 2^{1 \over 4} (1 + \phi \bar \phi)} \chi_{A}~, ~  
\hat{ \bar \chi}_{A'} = {\sigma a^{3 \over 2} \over 2^{1 \over 4} (1 + \phi \bar \phi)} \bar \chi_{A'}~. \label{eq:3.56}
\end{equation}
The conjugate momenta become
\begin{equation} \pi_{\hat \chi_{A}} = -i n_{AA'} \hat{ \bar \chi}^{A'}~, ~
 \pi_{\hat{ \bar \chi}_{A'}} = -i n_{AA'} \hat \chi^{A} ~. 
\label{eq:3.57}
\end{equation}
This pair forms a set of   (second class)
 constraints. The Dirac bracket   is

\begin{equation} \left[ 
\hat \chi_{A} , \hat{\bar \chi}_{A'} \right]_{D} = -i n_{AA'}~. 
\label{eq:3.58}        \end{equation}
Similarly for the $ \psi_{A}$  field, 

\begin{equation} \hat \psi_{A} = 
{\sqrt{3} \over 2^{1 \over 4}} \sigma a^{3 \over 2} \psi_{A}~, ~
\hat{\bar \psi}_{A'} = {\sqrt{3} \over 2^{1 \over 4}} \sigma a^{3 \over 2} \bar \psi_{A'} ~, 
\label{eq:3.59}
\end{equation}
where the conjugate momenta are 

\begin{equation} \pi_{\hat{ \psi}_{A}} = in_{AA'} \hat{\bar \psi}^{A'} ~,~
\pi_{\hat{\bar  \psi}_{A'}} = in_{AA'} \hat \psi^{A}~. 
\label{eq:3.60}
 \end{equation}
The  Dirac bracket becomes 

\begin{equation} \left[
\hat \psi_{A} , \hat{\bar \psi}_{A'}\right]_{D} = in_{AA'}~.   
\label{eq:3.61}        \end{equation}
Also for the $\lambda_A^{(a)}$ field:

\begin{equation} \hat\lambda^{(a)}_{~A} = 
{\sigma a^{ 3 \over 2} \over 2^{ 1 \over 4} }\lambda^{(a)}_{~A}, ~~
\hat{\bar \lambda}^{(a)}_{~A'} = 
{\sigma a^{ 3 \over 2} \over 2^{ 1 \over 4} }\bar \lambda^{(a)}_{~A'}  ~,
\label{eq:3.62}
\end{equation}
giving
\begin{equation}  \pi_{\hat{ \lambda}^{(a)}_{~A}} = -in_{AA'} \hat{\bar \lambda }^{(a)A'} ~,~
\pi_{\hat{\bar \lambda}^{(a)}_{~A'}} = -in_{AA'} \hat \lambda^{(a)A}~,
\label{eq:3.63}          \end{equation}
with
\begin{equation}  \left[
\hat \lambda^{(a)}_{~A}, \hat{\bar \lambda}^{(b)}_{~A'}\right]_{D} 
= - i \delta^{ab} n_{AA'}~.
\label{eq:3.64}
 \end{equation}
Furthermore,
\begin{equation} [a , \pi_{a}]_{D} = 1~, ~ [\phi, \pi_{\phi}]_{D} = 1~,
 ~[\bar \phi, \pi_{\bar \phi}]_{D} = 1~, ~[f, \pi_{f}] = 1~,
\label{eq:3.65}         \end{equation}
and the rest of the brackets are zero.

It is simpler to describe the theory using only 
 unprimed spinors, and, to this end, we define
\begin{equation} \bar \psi_{A} = 2 n_{A}^{~B'} \bar \psi_{B'}~, ~
 \bar \chi_{A} = 2 n_{A}^{~B'} \bar \chi_{B'} ~, ~\bar \lambda^{(a)}_{~A} = 2 n_{A}^{~B'}
 \bar \lambda^{(a)}_{~B'}~, 
\label{eq:3.66}         \end{equation}
with which the new Dirac brackets are 
\begin{equation} [\chi_{A}, \bar \chi_{B}]_{D} = -i \epsilon_{AB}~, ~
 [\psi_{A}, \bar \psi_{B}]_{D} = i \epsilon_{AB} ~,  ~\left[
\lambda^{(a)}_{~A}, \bar \lambda^{(a)}_{~A'}\right]_{D} =
 - i \delta^{ab} \epsilon _{AB}~. 
\label{eq:3.67}
 \end{equation}
The rest of the brackets remain unchanged. 
Quantum mechanically, one replaces the Dirac brackets by  anti-commutators if both arguments are odd 
(O) or commutators ifb
otherwise (E): 

\begin{equation} [E_{1} , E_{2}] = i [E_{1} , E_{2}]_{D} ~,~ [O , E] = i [O , E]_{D} ~,~ \{O_{1} , O_{2}\} = i [O_{1} , O_{2}]_{D} ~.
\label{eq:3,68}        \end{equation}
Here, we take units with $\hbar = 1 $. The only non-zero (anti-)commutator relations are:
\begin{equation} \{\lambda^{(a)}_{~A}, \lambda^{(b)}_{~B} \} = \delta^{ab} \epsilon_{AB} ~,~ \{\chi_{A} , \bar \chi_{B} \} = \epsilon_{AB}~, ~
\{\psi_{A} , \bar \psi_{B}\}  = - \epsilon_{AB}~,
\label{eq:3.69}
  \end{equation}
\begin{equation} [a , \pi_{a}] = [\phi , \pi_{\phi}] = [\bar \phi, \pi_{\bar \phi}] = [f, \pi_{f}] = i ~. \label{eq:3.70}         \end{equation}

We choose $ (\chi_{A} , \psi_{A} , a , \phi , \bar \phi, f) $ to be the coordinates of the configuration space, and 
$ (\bar \chi_{A}, \bar \psi_{A} , \pi_{a}$, 
$\pi_{\phi}, \pi_{\bar \phi}, \pi_f) $ to be the momentum operators in this 
representation.
Hence

\begin{eqnarray}  \lambda^{(a)}_{~A} & \rightarrow &  -{  \partial   \over   \partial   \bar \lambda^{(a)A}} ~,~
 \bar \chi^{A} \rightarrow -{  \partial   \over   \partial   \chi^{A}} ,~ \bar \psi_{A} \rightarrow {  \partial   \over   \partial   \psi^{A}} ~, \nonumber \\ 
  \pi_{a} & \rightarrow & {  \partial   \over   \partial   a} , ~\pi_{\phi} \rightarrow -i {  \partial   \over   \partial   \phi},~
 \pi_{\bar \phi} \rightarrow -i {  \partial   \over   \partial   \phi} ~,~ \pi_{f} \rightarrow -i {  \partial   \over   \partial   f} ~. \label{eq:3.71}
\end{eqnarray}

Some criteria have been presented to determine a suitable factor ordering for 
the quantum constraints  obtained from (\ref{eq:3.52})--(\ref{eq:3.55}). 
The nature of this  problem  is related to the presence of fermionic 
cubic terms.
Basically, 
$S_A, \bar S_A,$ and the Hamiltonian 
constraint $ {\cal H}$ {\it could}  be chosen by requiring that 
\cite{A9,A10,A19a,A23}: 

\vspace{0.15cm}

{\bf 1.}~ $S_A \Psi =0 $ describes the transformation properties of $\Psi$ under 
right-handed supersymmetry 
transformations (in the  $ (\chi_{A} , \psi_{A} , a , \phi , \bar \phi, f) $  
representation);

{\bf 2.}~ $\bar S_A \Psi =0 $ describes the transformation properties of $\Psi$ under 
left-handed supersymmetry 
transformations (in the  $ (\chi_{A} , \psi_{A} , a , \phi , \bar \phi, f) $  
representation);

{\bf 3.}~ $S_A, \bar S_A$ are Hermitian adjoints with respect to an adequate inner 
product [5];

{\bf 4.}~ A Hermitian Hamiltonian ${\cal H}$ is defined by consistency of the quantum algebra.

\vspace{0.15cm}

However,  not all of these criteria can be 
satisfied simultaneously (cf. \cite{A9,A11}). 
An arbitrary choice is to satisfy {\bf 1, 2, 4} 
\cite{A9,A10a,A16}-\cite{A18}, \cite{A19a}-\cite{A18a}. 
Another possibility (as in 
ref. \cite{A10,A22,A19a,A23}) is to 
go beyond this factor ordering and insist that $S_A, \bar S_A$  could still 
be related by a Hermitian adjoint operation (requirement {\bf 3}). If we adopt this,  
then there are some quantum corrections to $S_A, \bar S_A$. 
Namely, adding 
terms linear in 
$\psi_A, \chi_A$ to $S_A$ and linear in 
$\bar \psi_A, \bar\chi_A$ to $\bar S_A$ which nevertheless {\it modify}
 the transformation rules for the 
wave function under supersymmetry requirements {\bf 1, 2.}

Following the ordering used in ref. \cite{4},  we put all the fermionic derivatives in  $S_{A}$ on the right.
 In $ \bar S_{A} $  all the fermonic
derivatives are on the left. Implementing all these redefinitions, the 
 supersymmetry constraints  have the differential operator form

\begin{eqnarray} 
S_{A} & = & -{i \over \sqrt{2}} (1 +
 \phi \bar \phi) \chi_{A} {  \partial   \over    \partial   \phi}
 - {1 \over 2 \sqrt{6}} a \psi_{A} {  \partial   
\over   \partial   a} -   \sqrt{3 \over 2} \sigma^{2}a^{2} \psi_{A} 
- {5i \over 4 \sqrt{2}} \bar \phi \chi_{A} \chi^{B} 
{  \partial   \over   \partial   \chi^{B}} \nonumber \\
& - & {1 \over 8 \sqrt{6}} \psi_{B} \psi^{B} {  \partial   \over   \partial   \psi^{A}} 
- {i \over 4 \sqrt{2}} \bar \phi \chi_{A} \psi^{B} {  \partial   \over   \partial   \psi^{B}} -  {5 \over 4 \sqrt{6}} \chi_{A} \psi^{B}{  \partial   \over   \partial   \chi^{B}}
 + {\sqrt{3} \over 4 \sqrt{2}} \chi^{B} \psi_{B} {  \partial   \over   \partial   \chi^{A}}  
\nonumber \\
&+ & {1 \over 2 \sqrt{6}} \psi_{A} \chi^{B} {  \partial   \over   \partial   \chi^{B}}  +
 {1 \over 3 \sqrt{6}} \sigma^{a}_{~AB'} \sigma^{bCC'} n_{D}^{~B'} n^{B}_{~C'} \bar \lambda^{(a)D} \psi_{C} {  \partial   \over   \partial   
 \bar \lambda^{(b)B}}  \nonumber \\
&+ & {1 \over 6 \sqrt{6}} \sigma^{a}_{~AB'} \sigma^{bBA'}
 n_{D}^{~B'} n^{E}_{~A'} \bar \lambda^{(a)D} \bar \lambda^{(b)}_{~B}
{  \partial   \over   \partial   \psi^{E}}  -
 { 1 \over 2 \sqrt{6}} \psi_{A} \bar \lambda^{(a)C} {  \partial   \over \bar \lambda^{(a)C}} + {3 \over 8 \sqrt{6}} \bar 
\lambda^{a}_{~A} \lambda^{(a)C} {  \partial  
 \over   \partial   \psi^{C}}  
\nonumber \\ & + &  {1 \over 2 \sqrt{2}} \sigma^{2} a^{3} g \bar D^{(a)} \bar \lambda^{a}_{~A} - {1 \over 4 \sqrt{6}} \psi^{C} 
\bar \lambda^{(a)}_{~C} {  \partial   \over   \partial   \bar \lambda^{(a)A}}  
+  {\sigma^{2} a^{2} g f \over \sqrt{2} (1 + \phi \bar \phi)} \sigma^{a}_{~AA'} n^{BA'} \bar X^{(a)} \chi_{B}  
\nonumber \\ & + & \sigma^{a}_{~AA'} n^{BA'}  \bar \lambda^{(a)}_{~B} \left( -{ \sqrt{2} \over 3}  {   \partial   \over   \partial   f} + 
{1 \over 8 \sqrt{2}} (1 -(f-1)^{2}) \sigma^{2}  \right) 
\label{eq:3.72}
 \end{eqnarray} 
and $\bar S_A$ is just the Hermitian conjugate of (\ref{eq:3.72}) 
using (\ref{eq:3.66}).

When matter fields are taken into account the
generalisation of the $ J_{AB} $
constraint  is 
\begin{equation} J_{AB} = \psi_{(A} \bar\psi^{B'}n_{B)B'} - 
\chi_{(A} \bar\chi^{B'}n_{B)B'}
- \lambda^{(a)}_{(A} \bar\lambda^{(a)B'}n_{B)B'} =  0~. 
\label{eq:3.73}
  \end{equation}
One can justify this by observing either that it arises from the 
corresponding constraints of the full theory, or that its quantum version 
describes the invariance of the wavefunction under Lorentz transformations.
Alternatively, we could consider eq. (\ref{eq:2.22a}). The second and third 
terms in (\ref{eq:2.22a}), 
allow to  reproduce basically the last two terms in eq. (\ref{eq:3.73})
(cf. ref. \cite{A18a} for more details).

The Lorentz constraint $ J_{AB} $ implies that a physical 
wave function should be a  Lorentz scalar. 
 We can easily see that the most general form of the wave function 
is
\begin{eqnarray} 
  \Psi &= &  A  +  iB \psi^{C} \psi_{C} + C \psi^{C} \chi_{C} + iD \chi^{C} \chi_{C} + E \psi^{C} \psi_{C} \chi^{C} \chi_{C}  \nonumber \\
&  + &  c_{a} \bar \lambda^{(a)C} \chi_{C} + d_{a}  \lambda^{(a)C} \chi_{C} + c_{ab} \bar \lambda^{(a)C}  \bar \lambda^{(b)}_{~C}
+ e_{a}  \bar \lambda^{(a)C} \chi_{C} \psi^{D} \psi_{D}   \nonumber \\
& + &  f_{a} \bar \lambda^{(a)C} \psi_{C}  \chi^{D} \chi_{D} + d_{ab}  \bar \lambda^{(a)C} \chi_{C} \bar \lambda^{(a)D} \chi_{D}
+ e_{ab}  \bar \lambda^{(a)C}  \bar \lambda^{(b)}_{~C} \psi^{D} \psi_{D}  \nonumber \\
&  + &  f_{ab} \bar \lambda^{(a)C}  \bar \lambda^{(b)}_{~C} \chi^{D} \chi_{D} +g_{ab} \bar \lambda^{(a)C}  \bar \lambda^{(b)}_{~C} \chi^{D} \psi_{D}
 + c_{abc}  \bar \lambda^{(a)C}  \bar \lambda^{(b)}_{~C} \bar \lambda^{(c)D} \psi_{D}   
\nonumber \\
&  + & d_{abc} \bar \lambda^{(a)C}  \bar \lambda^{(b)}_{~C} \bar \lambda^{(c)D} \chi_{D} + c_{abcd} \bar \lambda^{(a)C}  \bar \lambda^{(b)}_{~C}
\bar \lambda^{(c)D}  \bar \lambda^{(d)}_{~D} + h_{ab}  \bar \lambda^{(a)C}  \bar \lambda^{(b)}_{~C} \psi^{D} \psi_{D} \chi^{E} \chi_{E}  \nonumber \\
& + & e_{abc} \bar \lambda^{(a)C}  \bar \lambda^{(b)}_{~C}  \bar \lambda^{(c)D} \chi_{D} \psi^{E} \psi_{E} + f_{abc} 
\bar \lambda^{(a)C}  \bar \lambda^{(b)}_{~C} \bar \lambda^{(c)D} \psi_{D} \chi^{E} \chi_{E} 
\nonumber \\ &+ & d_{abcd}
\bar \lambda^{(a)C}  \bar \lambda^{(b)}_{~C} \bar \lambda^{(c)D}  \bar \lambda^{(d)}_{~D}  \psi^{E} \psi_{E} 
 +   \nu_{3} \bar \lambda^{(1)C}  \bar \lambda^{(1)}_{~C} \bar \lambda^{(2)D} \bar \lambda^{(2)}_{~D} \bar \lambda^{(3)E} \chi_{E}
\nonumber \\ 
&  + &  e_{abcd} \bar \lambda^{(a)C}  \bar \lambda^{(b)}_{~C} \bar \lambda^{(c)D}  \bar \lambda^{(d)}_{~D} \chi^{E} \chi_{E}
+ f_{abcd} \bar \lambda^{(a)C}  \bar \lambda^{(b)}_{~C} \bar \lambda^{(c)D}  \bar \lambda^{(d)}_{~D} \psi^{E} \chi_{E}   \nonumber \\
&  + & g_{abcd} \bar \lambda^{(a)C}  \bar \lambda^{(b)}_{~C} \bar \lambda^{(c)D} \psi_{D} \bar \lambda^{(d)E} \chi_{E}  + 
  \mu_{1}  \bar \lambda^{(2)C}  \bar \lambda^{(2)}_{~C} \bar \lambda^{(3)D} \bar \lambda^{(3)}_{~D} \bar \lambda^{(1)E} \psi_{E} \nonumber \\
&+ & \mu_{2}  \bar \lambda^{(1)C}  \bar \lambda^{(1)}_{~C} \bar \lambda^{(3)D} \bar \lambda^{(3)}_{~D} \bar \lambda^{(2)E} \psi_{E}
+ \mu_{3}  \bar \lambda^{(1)C}  \bar \lambda^{(1)}_{~C} \bar \lambda^{(2)D} \bar \lambda^{(2)}_{~D} \bar \lambda^{(3)E} \psi_{E}   \nonumber \\
&  +&  \nu_{1} \bar \lambda^{(2)C}  \bar \lambda^{(2)}_{~C} \bar \lambda^{(3)D} \bar \lambda^{(3)}_{~D} \bar \lambda^{(1)E} \chi_{E}
+ \nu_{2}  \bar \lambda^{(1)C}  \bar \lambda^{(1)}_{~C} \bar \lambda^{(3)D} \bar \lambda^{(3)}_{~D} \bar \lambda^{(2)E} \chi_{E} \nonumber \\
& + & F \bar \lambda^{(1)C}  \bar \lambda^{(1)}_{~C} \bar \lambda^{(2)D} \bar \lambda^{(2)}_{~D} \bar \lambda^{(3)E} \bar \lambda^{(3)}_{~E}
+ h_{abcd} \bar \lambda^{(a)C}  \bar \lambda^{(b)}_{~C} \bar \lambda^{(c)D}  \bar \lambda^{(d)}_{~D} \psi^{E} \psi_{E} \chi^{F} \chi_{F}  \nonumber \\
& + & \delta_{1} \bar \lambda^{(2)C}  \bar \lambda^{(2)}_{~C} \bar \lambda^{(3)D} \bar \lambda^{(3)}_{~D} \bar \lambda^{(1)E} \psi_{E} 
\chi^{F} \chi_{F} +
  \delta_{2} \bar \lambda^{(1)C}  \bar \lambda^{(1)}_{~C} \bar \lambda^{(3)D} \bar \lambda^{(3)}_{~D} \bar \lambda^{(2)E} \psi_{E} 
\chi^{F} \chi_{F}  \nonumber \\
&  + & \delta_{3} \bar \lambda^{(1)C}  \bar \lambda^{(1)}_{~C} \bar \lambda^{(2)D} \bar \lambda^{(2)}_{~D} \bar \lambda^{(3)E} \psi_{E}  
\chi^{F} \chi_{F} +   \gamma_{1} \bar \lambda^{(2)C}  \bar \lambda^{(2)}_{~C} \bar \lambda^{(3)D} \bar \lambda^{(3)}_{~D} \bar \lambda^{(1)E} \chi_{E}
\psi^{F} \psi_{F}  \nonumber \\
&  +&  \gamma_{2}  \bar \lambda^{(1)C}  \bar \lambda^{(1)}_{~C} \bar \lambda^{(3)D} \bar \lambda^{(3)}_{~D} \bar \lambda^{(2)E} \chi_{E}
\psi^{F} \psi_{F} +   \gamma_{3} \bar \lambda^{(1)C}  \bar \lambda^{(1)}_{~C} \bar \lambda^{(2)D} \bar \lambda^{(2)}_{~D} \bar \lambda^{(3)E} \chi_{E}
\psi^{F} \psi_{F}  \nonumber \\
&   + &  G  \bar \lambda^{(1)C}  \bar \lambda^{(1)}_{~C} \bar \lambda^{(2)D} \bar \lambda^{(2)}_{~D} \bar \lambda^{(3)E} \bar \lambda^{(3)}_{~E}
\psi^{F} \psi_{F} +  H  \bar \lambda^{(1)C}  \bar \lambda^{(1)}_{~C} \bar \lambda^{(2)D} \bar \lambda^{(2)}_{~D} \bar \lambda^{(3)E} \bar \lambda^{(3)}_{~E}
\chi^{F} \chi_{F}  \nonumber \\
&  +&  I  \bar \lambda^{(1)C}  \bar \lambda^{(1)}_{~C} \bar \lambda^{(2)D} \bar \lambda^{(2)}_{~D} \bar \lambda^{(3)E} \bar \lambda^{(3)}_{~E}
\chi^{F} \psi_{F} +  K  \bar \lambda^{(1)C}  \bar \lambda^{(1)}_{~C} \bar \lambda^{(2)D} \bar \lambda^{(2)}_{~D} \bar \lambda^{(3)E} \bar \lambda^{(3)}_{~E}
\psi^{F} \psi_{F} \chi^{G} \chi_{G}~,
\label{eq:3.74}         \end{eqnarray}
where $A$, $B$, $C$, $D$, $E$, etc, 
  are functions of $a$, $\phi$ ,$\bar \phi$, $f$  only.  
This ansatz  contains all allowed combinations of the fermionic fields and 
 is the most general Lorentz invariant function we can write down. 

\subsubsection{FRW model with scalar supermultiplets }

\indent 

In this subsection we will restrict ourselves to a 
model where
 both the spin-1 field and its fermionic partner are set 
equal to zero. 
Such a situation will be less difficult to study as compared with 
the  more demanding case where all matter fields 
are included.
For the cases of a two-dimensional 
 spherically symmetric and flat K\"ahler geometries 
we will   find that 
the quantum states have a  simple form, but 
 {\it different}  
 from the ones presented in ref. \cite{A10}.

The most general $\Psi$ which satisfies the Lorentz constraint
$ J_{AB} = \psi_{(A} \bar \psi_{B)} - \chi_{(A} \bar \chi_{B)} $
 is
\begin{equation} \Psi = A + iB \psi^{C} \psi_{C} + C \psi^{C} \chi_{C} + iD \chi^{C} \chi_{C} + E \psi^{C} \psi_{C} \chi^{C} \chi_{C}~, 
\label{eq:3.75}         \end{equation}
where $A$, $B$, $C$, $D$, and $E$ are functions of $a$, $\phi$ and $\bar \phi$ only. The factors of $i$ are chosen for  simplicity.

The next step is to solve the supersymmetry constraints $ S_{A} \Psi = 0 $ and $ \bar S_{A'} \Psi = 0 $. We will get four equations from  $ S_{A} \Psi = 0 $ and  another four equations from $ \bar S_{A'} \Psi = 0 $:

\begin{equation} -{ i \over \sqrt{2}} (1 + \phi \bar \phi) {\partial A \over \partial \phi} = 0~,  ~   -{ a \over 2 \sqrt{6}} {\partial A \over \partial a} - \sqrt{3 \over 2} \sigma^{2} a^{2} A = 0~, \label{eq:3.76}         \end{equation}
\begin{equation} i \sqrt{2} (1 + \phi \bar \phi) {\partial E \over \partial \bar \phi} = 0~, 
 ~    {a \over \sqrt{6}} { \partial E \over \partial a} - \sqrt{6} \sigma^{2} a^{2} E = 
0~. \label{eq:3.77}
\end{equation}
We can see that (\ref{eq:3.76}) and (\ref{eq:3.77})  
constitute  decoupled equations for $A$ and $E$, respectively.
 They have the general solution.
\begin{equation} A = f(\bar \phi) \exp({-3 \sigma^{2} a^{2}})~,~
 E = g(\phi) \exp ({3 \sigma^{2} a^{2}}) ~,
\label{eq:3.78}
\end{equation}
where $ f , g $ are arbitrary anti-holomorphic and holomorphic functions of $\phi$, respectively.
The other remaining equations  are coupled equations between $B$ and $C$ and 
between $C$ and $D$, respectively.
 The first step to decouple these equations is as follows.
 Let $ B = \tilde B (1 + \phi \bar \phi)^{- {1 \over 2}} $ , 
$ C = {\tilde C  \over \sqrt{3}}(1 + \phi \bar \phi)^{- {3 \over 2}} $, $ D = \tilde D (1 + \phi \bar \phi)^{- {1 \over 2}}$.
We then get

\begin{equation} (1 + \phi \bar \phi)^{2} {\partial \tilde B \over \partial \phi} + {a \over 12} {\partial \tilde C \over \partial a} 
- {7 \over 12} \tilde C + {1 \over 2} \sigma^{2} a^{2}  \tilde C = 0~, ~
{\partial \tilde C \over  \partial \bar \phi} + a {\partial \tilde B  \over \partial a} - 6 \sigma^{2} a^{2} \tilde B - 3 \tilde B = 0, \label{eq:3.79}
\end{equation}
\begin{equation} (1 + \phi \bar \phi)^{2} {\partial \tilde D \over \partial \bar \phi} - {a \over 12} {\partial \tilde C  \over \partial a} 
+ {7 \over 12} \tilde C + {1 \over 2} \sigma^{2} a^{2}  \tilde C = 0~, ~  {\partial \tilde C \over \partial \phi} - a {\partial \tilde D \over \partial a} - 6 \sigma^{2} a^{2} \tilde D + 3 \tilde D = 0. \label{eq:3.80}        \end{equation}
From (\ref{eq:3.79}) 
we can eliminate $\tilde B$ to get a partial differential equation for 
$\tilde C$:
\begin{equation}  (1 + \phi \bar \phi)^{2} {\partial \tilde C \over \partial \bar \phi 
\partial \phi}
 - {a \over 12} {\partial \over \partial a} \left(a {\partial \tilde C \over \partial a}\right) 
+ {5 \over 6} a {\partial \tilde C \over \partial a} + \left[ 3 \sigma^{4} a^{4} + 3 \sigma^{2} a^{2} - {7 \over 4} \right] \tilde C = 0~, 
\label{eq:3.81}
\end{equation}
and from (\ref{eq:3.80}), we will get another 
partial differential equation for $\tilde C$:
\begin{equation}  (1 + \phi \bar \phi)^{2} {\partial \tilde C \over \partial \bar \phi \partial \phi}
 - {a \over 12} {\partial \over \partial a} \left(a {\partial \tilde C \over \partial a}\right) + {5 \over 6} a {\partial \tilde C \over \partial a}
 + \left[ 3 \sigma^{4} a^{4} - 3 \sigma^{2} a^{2} - {7 \over 4} \right] \tilde C = 0~.
\label{eq:3.82}
  \end{equation}
We can see immediately that  $\tilde C = 0$ because the coefficients of $\sigma^{2} a^{2} \tilde C$ are different for these two equations.
 Using this result, we find
\begin{equation} B = h(\bar \phi) (1 + \phi \bar \phi)^{- {1 \over 2}} a^{3} 
\exp({3 \sigma^{2} a^{2}})~,~
 C = 0 ~,~ D = k(\phi) (1 + \phi \bar \phi)^{- {1 \over 2}} a^{3} \exp ({-3 \sigma^{2} a^{2}} ) ~. 
\label{eq:3.83}
\end{equation}
where 
$h, k$ are holomorphic, anti-holomorphic functions of $\phi,\bar\phi$ 
respectively. 

Notice that result (\ref{eq:3.83}) 
is a direct consequence that we  could not find a consistent 
(Wheeler-DeWitt type) second-order differential equation for $C$ and hence to 
$B, D$.
Expressions  (\ref{eq:3.78}), (\ref{eq:3.83}) are obtained 
directly from (\ref{eq:3.76}), (\ref{eq:3.77}) and (\ref{eq:3.79}) 
(\ref{eq:3.80}).  
Moreover, while Lorentz invariance allows the pair $\psi_A\chi^A$ in 
(\ref{eq:3.75}), 
supersymmetry  rejects it. 
A possible conclusion could be that supersymmetry transformations 
 forbid any fermionic bound state
 $\psi_A\chi^A$ by treating the  spin-$\frac{1}{2}$ fields  
 $\psi^A, \chi^B$ differently.

These  results can be strengthened by  showing  that 
$ C = 0$ is not a consequence 
 of the particular ordering used. 
 In fact, we can try the ordering presented in ref. \cite{A10}
   such that $S_{A}$ and $\bar S_{A'}$ 
are Hermitian adjoints in the standard inner product (cf. requirements {\bf 1} -
{\bf 4} above). If one allows for the factor ordering ambiguity in $S_{A}$ 
due to the terms cubic in fermions, and insists that $\bar S_{A'}$ be the Hermitian adjoint of $S_{A}$, the {\it new} operators will have the form

\begin{equation} S_{A_{new}} = S_{A} + \lambda \psi_{A} + i \mu \bar \phi \chi_{A}~, ~\bar S_{A_{new}} = \bar S_{A} + i \left( {7 \over 4 \sqrt{2}} - \mu \right) \bar \phi \chi_{A} + 
\left({5 \over 4 \sqrt{6}} - \lambda \right) \bar \psi_{A}~.
\label{eq:3.84}
 \end{equation}
 Notice that   $S_{A}$ , $\bar S_{A}$ above 
represent the supersymmetry 
constraints but with the ordering of \cite{A19}.  
We will find another eight equations. 
We must set $\mu = {7 \over 8 \sqrt{2}}$ 
to have a consistent decoupling.
The only freedom left to get consistent equations for $\tilde C$ is from  $\lambda$. 
By setting $ B = \tilde B (1 + \phi \bar \phi)^{3 \over 8} $, 
$ C = {\tilde C  \over \sqrt{3}}(1 + \phi \bar \phi)^{- {5 \over 8}} $, 
$ D = \tilde D (1 + \phi \bar \phi)^{3 \over 8}$
 we can again get two decoupled equations for $\tilde C$. 
 Again the coeffecient of $\sigma^{2}a^{2} \tilde C$ for one equation 
is $ -{7 \over 4} $ and the coefficient of  $\sigma^{2}a^{2}  C$ for the other 
equation is ${17 \over 4}$.
 Hence,  we are led   to  $ C = 0$, showing that the two most
 interesting orderings agree.

Furthermore, this result 
 does {\it not} depend on the chosen gauge group.  
For     a two-dimensional flat K\"ahler 
manifold,  the  K\"ahler potential  would be just  $\phi \bar \phi$,  
the K\"ahler metric is $g_{\phi \bar \phi} = 1$ and the Levi-Cita connections are zero.
We will find out that 
 the structure of the supersymmetry constraints 
are again the same. 
The reason is that the K\"ahler metric and the connection only enter the Lagrangian
 through 
the spin-${1 \over 2}$ field $\chi_{A}$ and no other terms. 
So, there is only a change in 
the
 coefficient of $\bar \phi \chi_{A} \chi^{B} 
{\partial  \over \partial \chi^{B}}$ in $S_{A}$ and
 the corresponding term in $\bar S_{A}$, the rest being equivalent to 
 put $\phi \bar \phi = 0$ 
in the necessary coefficients. The supersymmetry constraints are then

\begin{eqnarray} S_{A} & = & -2 \sqrt{3} i \chi_{A} { \partial \over \partial \phi} - a \psi_{A} {\partial \over \partial a}
 - 6 \sigma^{2}a^{2} \psi_{A} - {i \sqrt{3} \over 2} \bar \phi \chi_{A} \chi^{B} {\partial \over \partial \chi^{B}}\nonumber \\ 
& - & {1\over 4} \psi_{B} \psi^{B} {\partial \over \partial \psi^{A}} - {i \sqrt{3} \over 2} \bar \phi \chi_{A} \psi^{B} {\partial \over \partial \psi^{B}}
 -{5 \over 2} \chi_{A} \psi^{B}{\partial \over \partial \chi^{B}} + 
{3 \over 2} \chi^{B} \psi_{B} {\partial \over \partial \chi^{A}}
 + \psi_{A} \chi^{B} { \partial \over \partial \chi^{B}} 
\label{eq:3.85}
\end{eqnarray}
and $ \bar S_{A}$ is the Hermitian conjugate using (\ref{eq:3.66}).

Solving  the $S_{A} \Psi = 0$ and $\bar S_{A} \Psi = 0$, 
we obtain eight equations 
between $B$ , $C$ and $D$. 
Using $B = \tilde B \exp ({-{1 \over 2} \phi \bar \phi})$ , 
 $C ={\tilde C \over \sqrt{3}} \exp ({-{1 \over 2} \phi \bar \phi})$
 and $D = \tilde D \exp ({-{1 \over 2} \phi \bar \phi})$  \cite{A10}, we get 
\begin{equation} {\partial \tilde B \over \partial \phi} + {a \over 12} {\partial \tilde C \over \partial a} 
- {7 \over 12} \tilde C + {1 \over 2} \sigma^{2} a^{2}  \tilde C = 0 ~, ~
{\partial \tilde D \over \partial \bar \phi} - {a \over 12} {\partial \tilde C \over \partial a} 
+ {7 \over 12} \tilde C + {1 \over 2} \sigma^{2} a^{2}  \tilde C = 0 ~, 
\label{eq:3.86}
\end{equation}
\begin{equation} {\partial \tilde C \over \partial \phi} - a {\partial \tilde D \over \partial a} 
- 6 \sigma^{2} a^{2} \tilde D + 3 \tilde D = 0 ~ ,~ 
{\partial \tilde C \over \partial \bar \phi} + a {\partial \tilde B \over \partial a} - 6 \sigma^{2} a^{2} \tilde B - 3 \tilde B = 0 ~. 
\label{eq:3.87}
\end{equation}
This set of equations are exactly the same as (\ref{eq:3.79}), 
(\ref{eq:3.80})  
 if we put $\phi \bar \phi = 0$ in there.
We conclude, therefore, that for the two-dimensional flat 
  K\"ahler manifold, $\tilde C =0$. 
These results seem to suggest that whatever  K\"ahler manifold one uses,
 we reach the same conclusion.

\vspace{0.15cm}

Let us now address the interpretation of the  solutions we found above. 

\vspace{0.15cm}

A Hartle-Hawking (or no-boundary proposal) \cite{12} solution can be
expressed in terms 
of a Euclidian path integral. A three-surface constitutes  the {\it only} 
boundary of a compact four-manifold, on which the  four-metric is 
$g_{\mu\nu}$ and induces $h_{ij}^0$ on the boundary, and the matter field 
is $\phi$ and matches $\phi_0$ on the boundary as well.
A path integral is performed over all such $g_{\mu\nu}$ and $\phi$ within all such manifolds.
For manifolds of the form of ${\bf R} \times \Sigma$, the no-boundary proposal 
indicates us to choose initial conditions at the initial point as to ensure the closure of 
four geometry. It basically consists in setting the initial three-surface volume $h^{1/2}$ to 
zero.

Wormholes classical solutions connect 
different asymptotic regions of a Riemannian geometry.
Such solutions can  only be found when 
certain types of matter fields are present 
However, it  seems more natural to study quantum
wormhole states, i.e., solutions of the Wheeler-DeWitt 
equation. \cite{13}.
The wormhole ground state may be defined by a path integral over all possible asymptotic 
Euclidian 4-geometries and matter fields whose 
energy-momentum tensor vanishes at  infinity. 

It is tempting to identify a 
 Hartle-Hawking wave function
   in the fermionic filled sector, say, 
with $  g(\phi) \exp ({3 \sigma^{2} a^{2}})$. However, 
 notice  that the equations obtained from 
$S_A \Psi = 0, \bar S_A \Psi = 0$ 
 are {\it not enough}  to specify 
$ g(\phi)$. 
This is particularly disappointing. 
A similar situation is also present in 
ref. \cite{A10}, although an extra multiplicative factor of  $a^5$ multiplying 
$ g(\phi)$ induces an {\it even less} 
 clear situation. In fact, no attempt was made in ref. 
\cite{A10a,A10} 
to obtain a Hartle-Hawking  solution. Being $N=1$ supergravity  
 a square root of general relativity \cite{2}, we would expect to be able 
to find solutions of the type $e^{ik\phi} e^{a^{2}}$. These would correspond 
to the Hartle-hawking state of 
a FRW model with a massless minimally coupled scalar field in 
ordinary quantum cosmology \cite{5,12}.

In principle, there are no physical arguments for wormhole 
states to be absent in N=1 supergravity with supermatter. 
In ordinary FRW quantum cosmology with scalar  fields, 
the wormhole (ground) state solution  
would have a form like $  e^{-a^2\cosh(\rho)}$ 
\cite{A10a,A10,wh1,wh2}, with $\phi = \rho \times exp (i\theta)$. However,  
such behaviour is not 
provided in eq. (\ref{eq:3.78}), (\ref{eq:3.83}). Actually, it seems quite different. 
It is very puzzling that 
the wormhole state could be absent. 
However, ref. \cite{A10} clearly represents an opposite point of view, as it 
explicitly depicts wormhole 
(Hawking-Page) \cite{13} states in a locally supersymmetric scenario.

We may then 
ask in which conditions can solutions (\ref{eq:3.78}), 
(\ref{eq:3.83}) be accomodated in order for 
Hartle-Hawking or wormhole type solutions to be obtained. 
The arbitrary functions $f, g, h, k$  do not allow to conclude unequivovally  
that they 
would be damped either at  small or  large 3-geometries,  for allowed values of 
$\phi, \bar\phi$ 
on the boundary or at  infinity. Claims were made in ref. \cite{A16,A22} 
that {\it no}  wormhole states were found. 
Moreover, the claim \cite{A16,A22}  that    a 
Hartle-Hawking solution  was identified 
is definitively {\it not}  satisfactory \cite{A19a,A23}.

\vspace{0.15cm}

Hence, the current situation is as follows. 
Hartle-Hawking and 
 quantum wormhole type solutions were found in minisuperspaces for 
pure N=1 supergravity 
\cite{A5}-\cite{A9}, \cite{A13}-\cite{A15}. However, wormhole solutions are  
absent in the literature 
\footnote{{\rm 
Notice that for pure gravity neither classical or quantum wormhole 
solutions have been produced in the literature. A matter field seems to be 
required: the ``throat'' size is proportional 
to $\sqrt{\cal J}$ where $\cal J$ represents   a (conserved) flux of 
matter fields.}}, 
concerning  pure gravity cases 
\cite{13,wh1,wh2}. Hartle-Hawking  wave functions and wormhole ground states 
are present in 
ordinary minisuperspace with matter 
\cite{5,12,13,wh1,wh2}. 
When supersymmetry is introduced 
\cite{A8}-\cite{A11}, \cite{A22}-\cite{A18a} 
we confront   problems 
  as far as Hartle-Hawking or wormhole type solutions are concerned.

\vspace{0.15cm}

Let us address the apparent absence of wormhole solutions  in 
(\ref{eq:3.78}), (\ref{eq:3.83}). As mentioned 
above, wormhole solutions were found\footnote{Notice that an attempt \cite{A10a} 
 using the constraints present in \cite{A8,A9} but the 
ordering employed in {\bf 1. - 3.} also  seemed to have failed in getting wormhole 
states.} 
 in ref. \cite{A10} for supersymmetric 
FRW models with scalar supermultiplets. But it should be 
emphasized that in \cite{A10} the re-definition 
(\ref{eq:3.23}), (\ref{eq:3.24})
for  fermionic 
non-dynamical variables
 was employed,  together with a fermionic 
ordering satisfying criteria {\bf 1, 2, 4.} 
The relevant  equations in ref. \cite{A10} are 
\begin{eqnarray}
\left( a {\partial\over \partial a} + 6 a^2 - 5 \right) A &=&  0~, 
\left( {\partial\over \partial\phi} - 9 \bar \phi \right) A = 0~, 
\label{eq:3.88}     
\\
\left( a { \partial\over \partial a} + 6 a^2 \right)  C - 2 {\partial B \over \partial\phi} &=&
0~,~~
2 \left( a {\partial\over \partial a} + 6 a^2 - 2 \right) D - {\partial C \over \partial\phi} 
= 
0~, \label{eq:3.89}     \\
2 \left( a {\partial\over \partial a} - 6 a^2 - 2 \right) i B - {\partial C \over \partial\bar
\phi} & = &  0 ~,~
\left( a {\partial\over \partial a} - 6 a^2 \right)  C - 2 {\partial D \over \partial\bar \phi}
= 0  \label{eq:3.90}       \\
\left( {\partial\over \partial\bar \phi} - 9 \phi \right) E &= & 0~,
\left( a {\partial\over \partial a} - 6 a^2 - 5 \right) E = 0 ~. 
\label{eq:3.91}     \end{eqnarray}
The four Dirac-like equations (\ref{eq:3.89}), (\ref{eq:3.90})  for the 
components $B, C, D$  lead  consistently to a set of
Wheeler--DeWitt equations for each bosonic amplitude. E.g., 
\begin{equation}
 a {\partial\over \partial a} \left( a {\partial C \over \partial a} \right) - 2 a {\partial C 
\over \partial a} - 3 6 a^4 C - {\partial^2 C \over \partial\phi \partial\bar \phi} =
0~. 
\label{eq:3.92}
\end{equation}
A wormhole state may then be obtained (cf. ref. \cite{A10})
\begin{equation}
C (a, \rho) = a \int_0^\infty d k~k~g 
(k) H^{(1)}_{\nu (k)} \left( 3 i a^2 \right)
 J_0 (k \rho)~, g (k) = \exp\left[\frac{1}{2}i\pi\nu(k)\right] ~, 
\label{eq:3.93}
\end{equation}
where $ H^{(1)}_{\nu (k)} $ and $ J_0 (k \rho)$ are Hankel and Bessel functions, respectively.

The explanation for the apparent opposite conclusions in 
\cite{A10,A22}  is that  the choice  of Lagrange multipliers 
and  fermionic derivative ordering  can 
 make a 
difference. Our arguments are as follows.

Let us first consider the choice of Lagrange multipliers and their 
possible influence. 
The quantum formulation of wormholes in ordinary 
quantum cosmology has been shown to depend on the lapse function 
\cite{wh1,wh2}. 
A similar  ambiguity has already been pointed out 
in \cite{wh7} (see also \cite{wh8}) but for generic quantum cosmology and related to 
bosonic factor ordering questions in the Wheeler-DeWitt operator. 
The self-adjointness in the  Wheeler-DeWitt operator 
involves a non-linearity in $N$. For each choice of $ N$ there is a different metric 
in minisuperspace, all these 
metrics being related by a conformal transformation \cite{wh9}. 
Therefore, for each of these 
choices, the quantization process will be different\footnote{In fact, 
consider a minisuperspace consisting 
of a FRW geometry and homogeneous scalar field.
A  conformal coupling allows a more 
general class of solutions of the Wheeler-DeWitt equation than does the minimally 
coupled case, even if a one-to-one correspondence exists between bounded states 
\cite{wh9}.}.

For some choices of $ N$ the quantization are 
 even {\it inadmissible}, 
e.g, 
when ${N} \rightarrow 0$ too fast for vanishing 
3-geometries in the wormhole case (cf. ref. \cite{wh1,wh2} for more details). 
Basically, requiring regularity for $\Psi$ at $a\rightarrow 0$ is equivalent 
to self-adjointness for the Wheeler-DeWitt operator at that point. Choices of $N$ 
that vanish too fast when $a \rightarrow 0$ will lead to problems as the 
minisuperspace measure will be infinite at (regular) configurations associated 
with vanishing three-geometries volume.

A similar effect can be expected  when local supersymmetry transformations 
are present\footnote{It 
should  be recalled that   a combination of 
two supersymmetry
transformations, generated by $S_A$ and $\bar S_{A'}$ and whose amount is represented 
by the Lagrange multipliers $\psi^A_0, \bar \psi^{A'}_0$, can 
 be   equivalent to a 
transformation generated by 
the Hamiltonian constraint and where the lapse function is the corresponding 
Lagrange multiplier.}. Besides the lapse function, we have now the 
time components of the gravitino field, $\psi^A_0$, and 
of the torsion-free connection $\omega_{AB}^0$ as Lagrange multipliers. 
At the pure N=1 supergravity level, 
the  re-definition of fermionic non-dynamical variables  
(\ref{eq:3.23}), (\ref{eq:3.24}) 
 changes the supersymmetry and Hamiltonian constraints. 
In fact, {\it no}  
fermionic terms were present in  ${\cal H} \sim \{S_A, \bar S_A\}$ and {\it no} 
cubic fermionic terms in the supersymmetry constraints. Hence, 
{\it no}  ordering problems with regard to fermionic derivatives were present. The model with matter used in \cite{A10}
was then extracted {\it post-hoc} \cite{A9,A11} from a few basic assumptions 
about their general form and supersymmetric algebra. Cubic fermionic terms
like $\psi\bar\psi\psi$ or $\psi\bar\chi\chi$ are now present but the former is  absent 
in the pure case.

Let us be more clear. 
If we just use  $\psi^A_0$, then 
the supersymmetry and Hamiltonian constraints read  (in the pure case): 

\begin{equation} S_A = \psi_A\pi_a - 6ia \psi_A + {{i}\over{2a}}n_A^{~E'}\psi^E\psi_E\overline{\psi}_{E'}, 
\label{eq:3.94}     
\end{equation}

\begin{equation} \overline{S}_{A'} = \overline{\psi}_{A'} \pi_a + 
6ia \overline{\psi}_{A'} - {{i}\over{2a}}n_E^{~A'}\overline{\psi}^{E'}
\psi_E\overline{\psi}_{E'}, 
\label{eq:3.95}
 \end{equation} 

\begin{equation} {\cal H} = -a^{-1} (\pi_a^2 + 36 a^2)       + 12a^{-1} n^{AA'}\psi_A
\overline{\psi}_{A'}. \label{eq:3.96}     \end{equation}  
Comparing with (\ref{eq:3.25}), we see that the 
redefinition (\ref{eq:3.23}), (\ref{eq:3.24}) 
imply  the the last term in eq. (\ref{eq:3.94}), (\ref{eq:3.95}), (\ref{eq:3.96}) 
to be  absent.
For 
$\Psi \sim  A_1 +  A_2 \psi_A\psi^A$, 
$A_1 = e^{-3a^2}$ and $A_2 = e^{3a^2}$ constitute 
 solutions of the equations induced by 
$S_A \Psi =0$ and $\bar S_A \Psi =0$, 
respectively. This holds 
for  the pure case if we use 
either  $\Psi^A_0$ or $\rho^A$.  
This particular $\Psi$ is also a solution of 
${\cal H}\Psi =0$,
{\it but only} for the $\cal H$ without the second term in 
(\ref{eq:3.96}). I.e., 
when (\ref{eq:3.23}), 
(\ref{eq:3.24}) are fully employed. In fact,  
$\Psi$ does not constitute a solution of  the full expression in 
(\ref{eq:3.96}):  the function $e^{3a^2}$ would have to be replaced.

Hence the choice between $\rho_A$ and $\psi^A_0$ directly affects any consistency between the 
quantum solutions of the supersymmetry constraints.

Criteria {\bf 1,2,4}   (ref. \cite{A9,A10a,A11}) 
for the fermionic derivative ordering 
were used in solving the equations above. In 
ref. \cite{A10}   an Hermitian adjoint relation 
between the supersymmetry transformations (criteria {\bf 3.})
was accomodated. In ref. \cite{A22}
all these criteria  were tested but 
with the supersymmetry and Hamiltonian constraints  directly obtained from 
$\psi^A_0, \bar \psi^{A'}_0, { N}$.
Fermionic factor ordering becomes absent for (\ref{eq:3.23}), 
(\ref{eq:3.24}). If we then use   the fermionic ordering 
employed in  \cite{A10a}
 (where we accomodate the Hermitain 
adjointness with requirements {\bf 1,2,4} up to minor 
changes relatively to {\bf 1,2}) a wormhole   state can be  found 
(cf. ref.  \cite{A10}). 

Thus, 
there seems to be a relation between a choice of Lagrange multipliers (which 
simplifies the constraints and the algebra in the pure case) and 
 fermionic factor ordering 
(which may become absent in the pure case). 
These in turn enable to  
obtain  second order consistent equations (i.e., 
 Wheeler-DeWitt type) or solutions from the 
supersymmetry constraints. The consistency failure 
found from (\ref{eq:3.81}), (\ref{eq:3.82}), e.g.,    is the 
reason why $C=0$. 
Different choices of $\psi^A_0$ or $\rho_A$, then of fermionic derivative ordering will lead to 
different supersymmetry constraints and to different solutions for the quantization of the problem.

Finally, let us write the first-order differential equations  derived from the 
supersymmetry constraints (i.e., eq. (\ref{eq:3.76}), (\ref{eq:3.77}), (\ref{eq:3.79}), 
(\ref{eq:3.80})) but with $\phi = r e^{i\theta}$ and $C=0$. This will assist us 
in getting the explicit dependence of $\Psi$ on $\phi, \bar\phi$ and 
adequately identify the Hartle-Hawking wave-function. We then get 
\begin{equation}
\frac{\partial A}{\partial r}   -  
 i\frac{1}{r}\frac{\partial A}{\partial \theta} = 0~,~ 
\frac{\partial E}{\partial r} +   
 i\frac{1}{r}\frac{\partial E}{\partial \theta} = 0~, \label{eq:2.2.20e}
\end{equation}
\begin{equation}
(1 + r^2) \frac{\partial \, B}{\partial \, r} 
- i\frac{1 + r^2}{r} \frac{\partial \, B}{\partial \, \theta} + rB 
 =  0~, 
(1 + r^2) \frac{\partial \, D}{\partial \, r} 
+ i\frac{1 + r^2}{r} \frac{\partial \, D}{\partial \, \theta} + rD
 =  0~, \label{eq:2.2.41}
\end{equation}
which provide (after integration) the general quantum state
\begin{eqnarray}
\Psi & =  & c_1 r^{\lambda_1} e^{-i\lambda_1 \theta} 
e^{- 3\sigma^2 a^2}  + 
c_3 a^3 r^{\lambda_3} e^{-i\lambda_3 \theta}(1 + r^2)^{\frac{1}{2}} 
e^{ 3\sigma^2 a^2}
  \psi^{C} \psi_{C} \nonumber \\ & + &  
c_4 a^3 r^{\lambda_4} e^{i\lambda_4 \theta}(1 + r^2)^{\frac{1}{2}} 
e^{- 3\sigma^2 a^2} 
\chi^{C} \chi_{C} + 
c_2 r^{\lambda_2} e^{i\lambda_2 \theta}
e^{3\sigma^2 a^2}
 \psi^{C} \psi_{C} \chi^{D} \chi_{D} ~,
\label{eq:2.New1}
\end{eqnarray}
where 
  $\lambda_1$...$\lambda_4$  
and $c_1$...$c_4$ are constants. 
Notice now the explicit form of $A, B, D, E$ in (\ref{eq:2.New1}) 
in contrast with previous expressions.  
If we had use $\phi = \phi_1 + i\phi_2$ then the corresponding 
first-order differential equations  would lead to 
$A = d_1 e^{-3\sigma^2 a^2} e^{k_1(\phi - i\phi_2)}$, 
$B = d_3 e^{3\sigma^2 a^2} (1 + \phi_1^2 + 
\phi^2_2) e^{k_3(\phi - i\phi_2)}$,
$B = d_4 e^{-3\sigma^2 a^2} (1 + \phi_1^2 + 
\phi^2_2) e^{k_4(\phi + i\phi_2)}$,
$E = d_2 e^{3\sigma^2 a^2} e^{k_2(\phi + i\phi_2)}$.

The bosonic coefficients in (\ref{eq:2.New1})   correspond 
to particular solutions obtainable within the framework of 
ref. \cite{Khala} {\it if} a specific factor ordering for 
$\pi_a, \pi_r, \pi_\theta$ is used in the Wheeler-DeWitt equation.   The point is that 
the supersymmetry  constraints   imply $\frac{\partial A}{\partial \phi} =0$ 
and $\frac{\partial E}{\partial \bar\phi} =0$ and 
the  Wheeler-DeWitt equation  involves a term 
$\pi_\phi \pi_{\bar\phi} \sim 
(\pi_r -i\pi_\theta)(\pi_r + i\pi_\theta) 
\mapsto 
 \frac{\partial^2 }{\partial r^2}   +  
 \frac{1}{r^2}\frac{\partial^2 }{\partial \theta^2}.$ 
But  $
\left(\frac{\partial }{\partial r}   -  
 i\frac{1}{r}\frac{\partial }{\partial \theta}\right)
 \left(\frac{\partial }{\partial r}   +   
 i\frac{1}{r}\frac{\partial }{\partial \theta}\right) 
\neq  
 \frac{\partial^2 }{\partial r^2}   +  
 \frac{1}{r^2}\frac{\partial^2 }{\partial \theta^2}$. 
Hence, the presence of supersymmetry  
{\em selects} a particular factor ordering for the canonical momenta 
in the Hamiltonian constraint. As a consequence, specific 
exact solutions\footnote{In the decomposition 
$\phi= \phi_1 + i\phi_2$ the bosonic part of the 
Hamiltonian constraint corresponds  to 
a FRW model with two independent massless 
scalar fields. The novel characteristic 
induced by supersymmetry is that the 
complex scalar fields imply via the expressions for 
$A$ or  
$E $  
that the (separation)  constant $k$ (present in 
the equations equivalent to eqs. (9)-(23) of ref. \cite{13}) 
is now $k = d_1^2 - d_1^2 = 0$. This means that the 
scalar flux associated with $\phi_1$ and $\phi_2$ as described in 
\cite{13} is now absent. Consequently, the  lower 
bound for $a$ is $a=0$. It is then not apparent  how these 
solutions can represent a wormhole connecting 
two asymptotic regions.} 
(say, $e^{-3\sigma^2 a^2} r^\lambda e^{i\lambda\theta}$) 
can be found from the Wheeler-DeWitt equation in the 
gravitational and matter sectors. The no-boundary wave function  
corresponds to the bosonic coefficient $A$.  

Notice that 
the  bosonic coefficients in (\ref{eq:2.New1}) satisfy  attractive relations
in a 3-dimensional minisuperspace (see ref. \cite{cc,essay} for 
a comprehensive description)
\begin{eqnarray}
\frac{\partial (A\cdot E)}{\partial a} + 
\frac{\partial (A\cdot E)}{\partial \theta} - ir  \left(
\frac{\partial E}{\partial r} A - \frac{\partial A}{\partial r} E\right) & = &  0,
\label{eq:2.New22a} \\
D_a (B\cdot D)  + 
\frac{\partial (\, B \cdot D)}{\partial \, \theta } 
-ir  \left(\frac{\partial \, B}{\partial \, r} D 
- \frac{\partial \, D}{\partial \, r} B \right)
   & = &   0~,
\label{eq:2.New22b}
\end{eqnarray}
with $D_a = \partial_a - 6/a$.
However, the presence of the last term in both eq. (\ref{eq:2.New22a}) and 
(\ref{eq:2.New22b}) clearly prevent us from to associate them with 
a conservation  equation of the type  $\nabla J = 0$.
It is explained in ref. \cite{cc,essay} how the presence of 
supersymmetry, $\theta$ no longer being a cyclical coordinate and 
the absence of satisfactory conserved currents are all related.

\subsubsection{FRW model with generic gauged matter}

\indent

Let us now  solve the 
supersymmetry constraints $ S_{A} \Psi = 0 $ and $ \bar S_{A'} \Psi = 0 $ 
for the more general case 
where 
all supermatter fields are present.  We will use 
the ans\"atze described in subsections 3.1.2 and 3.3.1.
The number of constraint equations will be very high. Their full 
analysis is quite tedious and to write all the terms would overburden the reader.
Let us instead show some examples of the calculations  involving 
the $S_{A} \Psi = 0$ constraint (see ref. \cite{A17,A18}).

Consider the terms linear in $\chi_{A}$: $\left[ -{i \over \sqrt{2}} (1 + \phi \bar \phi) {   \partial   A \over   \partial   \phi} \right] \chi_{A} + 
{ \sigma^{2} a^{2} g f \over \sqrt{2} (1 + \phi \bar \phi) } \sigma^{a}_{~AA'} n^{BA'} \bar X^{(a)} A \chi_{B} = 0$. 
Since this is true for all $\chi_{A}$, the above equation becomes
\begin{equation} \left[ -{i \over \sqrt{2}} (1 + \phi \bar \phi) {   \partial   A \over   \partial   \phi} \right] \epsilon_{A}^{~B} + 
{ \sigma^{2} a^{2} g f \over \sqrt{2} (1 + \phi \bar \phi) } \sigma^{a}_{~AA'} n^{BA'} \bar X^{(a)} = 0. \label{eq:3.97}      \end{equation}
Multiply  the whole equation by $n_{BB'}$ and use 
the relation $n_{BB'} n^{BA'} = {1 \over 2} \epsilon_{B'}^{~A'}$.
We can see that the two terms 
in (\ref{eq:3.97}) 
are independent of each other since the $\sigma$ 
matrices are orthogonal to the $n$ matrix.
Thus, we conclude that
$ A=0$. 

Now consider, e.g., the terms linear in $\chi_{B} \psi^{C} \psi_{C}$. We have
\begin{eqnarray} 
\left[ (1 + \phi \bar \phi) {  \partial   B 
\over   \partial   \phi} \right.& + & \left. {1 \over 2} \bar 
\phi B + i {a \over 4 \sqrt{3}} {  \partial   C \over   \partial   a} - 
i{ 7 \over 4 \sqrt{3}} C + i { \sqrt{3} \over 2 } 
\sigma^{2} a^{2} C \right] \chi_{A} \psi^{C} \psi_{C}  
\nonumber \\
& + & i  { \sigma^{2} a^{2} g f \over \sqrt{2} (1 + \phi \bar \phi) } \sigma^{a}_{~AA'} n^{BA'} \bar X^{(a)} B \chi_{B} 
\psi^{C} \psi_{C} = 0. 
\label{eq:3.98}
\end{eqnarray}
By the same argument as above, the first term is independent of the second one and we have the result
$ B = 0$.

As we proceed, this pattern keeps repeating itself. Some equations 
show that the coefficients have some symmetry 
properties. For example, let us take $d_{(ab)} = 2 g_{(ab)}$. But when these two terms
 are combined with each other, they become zero. This can be 
seen as follows,

\begin{equation} 
d_{ab}  \bar \lambda^{(a)C} \chi_{C} \bar \lambda^{(a)D} \chi_{D} +  
g_{ab} \bar \lambda^{(a)C}  \bar \lambda^{(b)}_{~C} \chi^{D} \psi_{D}  
  =  - g_{ab} \bar \lambda^{(a)C}  \bar \lambda^{(b)}_{~C} \chi^{D} \psi_{D} +
g_{ab} \bar \lambda^{(a)C}  \bar \lambda^{(b)}_{~C} \chi^{D} \psi_{D}, 
\label{eq:3.99}
\end{equation}
using the property that $g_{ab} = g_{ba}$ and the spinor identity $ \theta_{AB} = {1 \over 2} \theta_{C}^{~C} 
\epsilon_{AB} $ where $\theta_{AB} $ is anti-symmetric in the two indices.
 The same property applies to the terms with coefficients $ f_{abcd}$ and $g_{abcd}$.
 Other equations imply that the
coefficients $c_{abc}, d_{abc}, c_{abcd}, e_{abc}, f_{abc}, d_{abcd}$,
$e_{abcd},$ $h_{abcd}$ are totally 
symmetric in their indices. This then leads to  terms cancelling  each other,
 as can easily be shown. In the end, considering both the 
$S_{A} \Psi = 0$ and $\bar S_{A} \Psi = 0$  constraints, we are left with the surprising 
result that the wave function (\ref{eq:3.74}) must be zero in order to satisfy 
the quantum constraints!

\subsubsection{ FRW model with Yang-Mills fields }

\indent

We will now consider a FRW
model with Yang Mills fields 
 obtained from the more general theory of 
N=1 supergravity with gauged supermatter \cite{A18a}, 
 associated with a 
gauge group $\hat{G}=SU(2)$. We will
 put  all  scalar fields and corresponding 
supersymmetric partners equal to zero\footnote{{\rm 
It was shown  in ref. \cite{mem,me} for the case of 
a  gauge group SO(3) $\sim$ SU(2)
that  
invariance for homogeneity and isotropy as well as 
gauge transformations require 
all components of $\phi$ to be  zero. Only for SO(N), N$>$3 
we can have $\phi$ = (0,0,0, $\phi_1, ..., 
\phi_{N-3}$).
}}. 
It should be noticed that  Yang-Mills fields coupled to N=1 supergravity
can also be found in ref. \cite{ymdf}. 
We shall use the ansatz  (\ref{eq:3.51}) for $A^{(a)}_\mu$.
This implies $A^{(a)}_\mu$ to be paramatrized by 
a single effective scalar function $f(t)$. 
Ordinary FRW  cosmologies with this Yang-Mills field ansatz  are totally 
equivalent to a FRW minisuperspace with an effective 
 conformally coupled scalar field,  but with a quartic potential instead of a 
quadratic one \cite{mb,4m}, \cite{mem}-\cite{meob}.

The introduction of fermions in ordinary
quantum cosmological models with gauge fields led to 
additional {\it non-trivial} ans\"atze for the fermionic fields \cite{fob}.
These  involve 
 restrictions from  group theory,    rather than just imposing time dependence. 
However, we should notice that fermions in simple  minisuperspace models have also been considered in 
\cite{halli,isha,grek}. 
Some questions concerning the 
(in)consistency of these models were raised in \cite{isha} and an attempt to clarify them was made in \cite{grek}.

Hence, it seemed sensible that 
similarly to the case       
where only scalar fields are present,  we ought to take as       
fermionic partner for (\ref{eq:3.51}) 
a simple spin-${1 \over 2}$ field, like $\chi_A$.      
However, this would lead to some difficulties 
(see ref. \cite{A18a} for more details).
The more general ansatz for the spin-$\frac{1}{2}$ fields  
is just to take
\begin{equation}
\lambda_A^{(a)} = \lambda_A^{(a)}  (t)~,~
\label{eq:100a} \end{equation}      
and correspondingly to its Hermitian conjugate 

An important consequence of not having scalar fields and their fermionic partners is that 
the Killing potentials $D^{(a)}$ and related quantitites are now absent. In fact,   
if we had complex 
scalar fields, a K\"ahler manifold could be considered with metric 
$ g_{I J^*}$
on the space of $ ( \phi^I, \phi^{J^*} ) $. 
For $\hat{G}=SU(2)$ with  $\phi=\bar\phi = 0$ this implies $D^{(1)}=D^{(2)}=0$ and 
$D^{(3)}=~- \frac{1}{2}$. However, being the 
$D^{(a)}$ fixed up to  constants which are now arbitrarly, we can choose 
 $D^{(3)} = 0$ consistently.

The subsequent steps correspond to adapt  subsection 3.3.1 
according to the ans\"atze mentioned above. Namely,
truncatingeq.  (\ref{eq:3.72}), its H.c. and (\ref{eq:3.74}). 
 Ref. \cite{A18a} can be consulted for further details.


From $S_A \Psi = 0$  we obtain       
\begin{equation}      
-{a \over {2\sqrt{6}}} {\partial A \over \partial a} - \sqrt{{3\over 2}} \sigma^2 a^2 A = 0, 
\label{eq:3.105}     
\end{equation}
\begin{equation}
-{\sqrt{2} \over 3} {\partial A \over \partial f} + {1 \over {8\sqrt{2}}} [1 - (f - 1)^2] \sigma^2~A =0, \label{eq:3.106}         
\end{equation}          
They correspond, respectively, to terms linear in $\psi_A$ and 
$\bar\lambda^{(a)}_A$. Eq. (\ref{eq:3.105}), (\ref{eq:3.106}) 
 give the       
dependence of $A$ on $a$ and $f$, respectively. 

Solving these equations leads to       
$A = \hat A (a) \tilde A (f)$ as         
\begin{equation}      
A = e^{-3 \sigma^2 a^2} e^{{3\over 16}\sigma^2 \left(-{f^3 \over 3} + f^2\right)},      
\label{eq:3.108}
\end{equation}      
A similar relation exists for the
 $\bar S_A \Psi = 0$ equations, which from the       
$\psi_A \lambda_E^{(1)}\lambda^E{(1)}  
\lambda_E^{(2)}\lambda^E{(2)}   \lambda_E^{(3)}\lambda^E{(3)}   
 $ term give for $G = \hat G (a) \tilde G (f)$        
\begin{equation}      
G =   e^{3 \sigma^2 a^2}   
  e^{{3\over 16}\sigma^2 \left({f^3 \over 3} - f^2\right)}.      
\label{eq:3.109}
\end{equation}        
It should be emphasized   that we are       
indeed       
allowed to completely determine the dependence of $A$ and $G$ 
with respect to       
$a$ and $f$, 
differently to the case of ref. 
\cite{A10a,A10,A22,A20a}-\cite{A23}.

The solution (\ref{eq:3.109}) corresponds to the Hartle-Hawking (no-boundary)       
solution \cite{12,mb}. In fact, we basically recover solution (3.8a) of ref. \cite{mb}     
(where only ordinary quantum cosmology with Yang-Mills fields is considered) if we replace       
$f \rightarrow f + 1$. As it can be checked, this constitutes the rightful procedure       
according to the definitions employed in \cite{mem} for $A^{(a)}_\mu$. 
Solution (\ref{eq:3.109})       
is also associated with an anti-self-dual solution of the Euclidianized equations       
of motion (cf. ref. \cite{mb,4m}). It is curious that when fermions are present       
  not all the solutions present in \cite{mb} 
can be recovered.       
This applies to other anti-self-dual and self-dual solutions. Moreover, this implies       
that the Gaussian wave function (\ref{eq:3.109}), peaked around $f=1$  
 represents only one of the components of the       
Hartle-Hawking 
wave function in \cite{mb}. We should notice, however, that the Gaussian  wave function in       
\cite{mb} is peaked around the two minima of the potential due to the quadratic nature of this one.       
In our case, we have instead a Dirac-like structure for our equations and our potential terms       
 correspond rather to a square-root of the potential present in ref. \cite{mb}.
In their present form, the Dirac bracket of the supersymmetry constraints       
induces a Hamiltonian whose bosonic part contains the {\it decoupled}       
gravitational and vector field parts, in       
agreement with   \cite{mb,meob}.

Solution (\ref{eq:3.108}) 
could be interpreted as wormhole solution \cite{13,4m},       
which   has not yet  been found in  ordinary quantum cosmology.      
However, in spite of  (\ref{eq:3.108})
 being regular for $a \rightarrow 0$ and damped for       
$a \rightarrow \infty$, it may not be well behaved when $f \rightarrow -\infty$.       
This last property might constitute a drawback when attempting to identify       
it as a sucessfullquantum wormhole state \cite{13,4m,wh1,wh2}.

The remaining equations
   from
$\bar S_A \Psi = 0$ and Hermitian conjugate 
imply that  any possible solutions      
are neither the Hartle-Hawking or a wormhole state. In fact, 
we would get, say, 
 $F \sim a^5 \hat F (a) \tilde F (f)$ and similar       
expressions for other coefficients,
  with a prefactor $a^n$, $n \neq 0$. 
Hence, from their $a$-dependence equations      
these solutions cannot be       
either a Hartle-Hawking or wormhole state (cf. ref. \cite{12,13,mb,4m,wh1,wh2}). 

\subsubsection{ Bianchi type-IX model with scalar supermultiplets}

\indent

In this section we will describe 
 a  Bianchi type-IX model   with       
spatial metric in diagonal form,  using the supersymmetry constraints
(\ref{eq:2.39})       
derived in  subsection {\rm 2.1.2}. 
We restrict our case     
to a supermatter model constituted only by a scalar field and       
its spin$-{1 \over 2}$ partner with a two-dimensional       
flat K\"ahler geometry.       
The scalar super-multiplet is  chosen to be       
spatially homogeneous.
We require that the       
components      
$  \psi^A_{~~0}, \bar \psi^{A'}_{~~0}  $ be functions of time only. We      
further require that $ \psi^A_{~~i} $ and $ \bar\psi^{A'}_{~~i} $ be      
spatially homogeneous in the basis $ e^a_{~i} $ in (\ref{eq:2.1}).        

A quantum description can be made by studying  
 wave functions of the form      
$ \Psi \left[ e^{AA'}_{~~i},  \psi^A_{~~i},      
\bar\chi_A,       
\phi,       
 \bar\phi \right] $.      
The choice of $ \bar \chi_A \equiv  n_A^{~~A'} {\bar \chi}_{A'} $       
rather than $       
\chi_A $ is designed so that the quantum constraint $ \bar S_{A'} $ should       
be of first order in momenta.       
The momenta are represented\footnote{It is interesting  to notice that for the $\chi,\bar \chi$       
fields no powers of $h$ seemed to be needed        
to establish the       
equations for the coefficients in $\Psi$.}  by       
\begin{eqnarray}      
p_{A A'}^{~~~~i} &\rightarrow & - i \hbar {\delta \over \delta e^{A A'}_{~~~~i}}       
- {1 \over \sqrt{2}}       
\epsilon^{i j k} \psi_{A j} \bar \psi_{A' k} 
\label{eq:3.110}       
\\         
\pi_\phi &\rightarrow &  - i \hbar {\partial \over \partial \phi}~,        
\pi_{\bar\phi} \rightarrow   - i \hbar {\partial \over \partial \bar \phi}~,
\label{eq:3.111}
\\      
\bar \psi^{A'}_{~~i} &\rightarrow & {1 \over \sqrt 2} i \hbar D^{A A'}_{~~~~j i}       
 h^{1 \over 2}      
{\partial \over \partial      
\psi^A_{~~j}}~,       
\chi^{A} \rightarrow   - \sqrt 2 \hbar       
 {\partial \over \partial       
 \bar \chi^A}~. \label{eq:3.112}              
\end{eqnarray}      
The supersymmetry constraints become, with   $ \frac{\delta}{\delta e^{CC'j}} = 
-2 e_{CC'i} \frac{\delta}{\delta h_{ij}}$,      
\begin{eqnarray}   \bar S_{A'} &= &       
 -i\sqrt{2} \left[-i \hbar {1\over 4}\left[e_{AA'i} \psi^A_j\right] e^{CC'i}       
{\delta \over \delta e^{CC'j}}        
 -i \hbar {1\over 4}  \left[e_{AA'i} \psi^A_j\right]   e^{CC'j}       
{\delta \over \delta e^{CC'i}}        
 \right] \nonumber \\      
&      
+& \sqrt{2} \epsilon^{ijk} e_{AA'i} ~^{3(s)}\omega^A_{~~Bj} \psi^B_{~~j}      
 - {i \over \sqrt{2}}\hbar n_{CA'} \bar \chi^C {\partial \over \partial \bar \phi}      
 \nonumber \\      
&      
-&  i \sqrt{2} \hbar h e^{K/2} P(\phi) n^A_{~~A'} e_{~AB'}^i D^{CB'}_{~~~ji}       
{\partial \over \partial \psi^C_j}      
 - i \sqrt{2} \hbar h e^{K/2} D_\phi P ~ n^A_{~~A'}       
{\partial \over \partial \bar \chi^A}      
 \nonumber \\      
&      
- &{i \over 2\sqrt{2}} h^{{1 \over 2}} \hbar \phi n^{BB'} n_{CB'}       
\bar \chi^C n_{DA'} \bar \chi^D { \partial \over \partial \bar \chi^B}      
 - {i \over 4\sqrt{2}} \hbar h^{{1 \over 2}} \phi \epsilon^{ijk}       
e^{BB'}_{~~~j} \psi_{kB} D^C_{~~B'li} n_{DA'}\bar\chi^D       
{\partial \over \partial \psi^C_l}      
 \nonumber \\      
&      
- & \hbar \sqrt{2} h^{{1 \over 2}}       
e^{B~~~m}_{~~B'} n^{CB'} \psi_{mC} n_{DA'} \bar \chi^D       
{ \partial \over \partial \bar \chi^B}      
 - {i \over \sqrt{2}} \hbar \epsilon^{ijk} e_{AA'j} \psi^A_i       
n_{DB'} \bar\chi^D e^{BB'}_{~~~k}  { \partial \over \partial \bar \chi^B}      
 \nonumber \\      
&      
+ & {1 \over 2\sqrt{2}} \hbar h^{{1 \over 2}} \psi_{iA}       
(e^{B~~~i}_{~~A'} n^{AC'} - e^{AC'i} n^B_{~~A'})       
n_{DC'} \bar \chi^D { \partial \over \partial \bar \chi^B},   
\label{eq:3.113}           
\end{eqnarray}  
and  $
S_{A}$ is the Hermitian conjugate, 
where the terms containing no matter fields are       
consistent with ref. \cite{4}.  Notice that  a       
constant analytical potential is similar to the
cosmological constant term  in ref. \cite{A13,A14,A15}.        
We       
will employ the integrated form of the       
constraints, i.e.,      
${\cal H} \equiv \int d^3 x H $.

Our  general Lorentz-invariant wave       
function is then taken to be       
a polynomial of eight degree in Grassmann variables      
\begin{eqnarray}      
\Psi(a_1,a_2,a_3, \phi, \overline{\phi}) & = &  A +       
B_1\beta_A\beta^A + B_2\overline{\chi}_A \overline{\chi}^A      
+  C_1 \gamma_{ABC}\gamma^{ABC}\nonumber \\      
& + &  D_1 \beta_A\beta^A \gamma_{EBC}\gamma^{EBC} + D_2 \overline{\chi}_A 
\overline{\chi}^A      
\gamma_{EBC}\gamma^{EBC} \nonumber \\  & + &       
 F_1 (\beta_A\beta^A \gamma_{EBC}\gamma^{EBC})^2 +
 F_2 \overline{\chi}_A \overline{\chi}^A      
(\gamma_{EBC}\gamma^{EBC})^2      
\nonumber \\      
&+ & G_1 \beta_A\beta^A \overline{\chi}_B \overline{\chi}^B +       
H_1 \beta_A\beta^A \overline{\chi}_B 
\overline{\chi}^B \gamma_{EDC}\gamma^{EDC}      
\nonumber \\    &  + & E_1 
(\gamma_{ABC}\gamma^{ABC})^2      
+  I_1 \beta_A\beta^A \overline{\chi}_B \overline{\chi}^B 
(\gamma_{EDC}\gamma^{EDC})^2      
\nonumber \\       
&+ & Z_1 \overline{\chi}_A\beta^A + Z_2 
\overline{\chi}_A\beta^A \gamma_{EDC}\gamma^{EDC}      
+  Z_3 \overline{\chi}_A\beta^A (\gamma_{EDC}\gamma^{EDC})^2~.   
\label{eq:3.114}           
\end{eqnarray}

We are aware       
of its limitations as far as the middle sectors are       
concerned. In fact, we will be neglecting Lorentz       
invariants built with
gravitational degrees of freedom. The ``new'' method       
proposed in \cite{A21,A20} to 
construct the correct middle fermionic sectors would 
give the correct spectrum of 
 solutions. However,  the solutions pointed there 
were {\it not}  entirely new:  they        
were already present in the ``old''       
framework of \cite{A6,A7,A12}. Thus, our simpler Lorentz invariant construction       
could still be of some utility,       
namely in obtaining       
new {\it realistic} solutions.

The action of the constraints operators       
 $S_A, \bar S_{A'}$  on       
$\Psi$  leads to a system of coupled first order       
differential equations which the bosonic amplitude       
coefficients of $\Psi$       
must satisfy. These coefficients are functions of       
$a_1,a_2,a_3,\phi,\bar \phi$. The equations are obtained       
after eliminating the $e_i^{AA'}$ and $n^{AA'}$       
resulting in the $S_A \Psi = 0, \bar S_{A'} \Psi = 0$.
We contract  them with combinations of       
$e_j^{BB'}$ and $n^{CC'}$, followed by integraton over       
$S^3$. These equations correspond essentially to       
expressions in front of terms such as       
$\bar\chi,\beta,\gamma,\beta^A \bar\chi\bar\chi, \gamma^{ABC} \bar\chi\bar\chi$, etc,       
after the fermionic derivatives in  $S_A, \bar S_{A'}$       
have been performed.

As one can easily see, the number of obtained equations will       
be very large. Actually, its number will be $44 \times 3$,       
taking into account cyclic permutations on $a_1,a_2,a_3$. 
Their full analysis is quite tedious.        
We will  instead point here some  steps       
involved in the calculations, and          the 
interested reader is invited to follow ref. \cite{A21a} for more 
details. 
The supersymmetry constraint $\bar S_{A'}$
has fermionic terms of the  type       
$\beta^A,\gamma^{ABC}$, $\bar\chi^A,      
{\partial \over \partial  \psi^A_i},       
{\partial \over \partial \bar \chi^A}$,      
$\bar\chi\bar\chi  {\partial \over \partial \bar \chi}$ ,       
$\psi\bar\chi      
{\partial \over \partial \psi}$ ,       
$\psi\bar\chi{\partial \over \partial \bar \chi}$,  
while  $S_A$ is of second order in fermionic       
derivatives and includes  terms as       
${\partial \over \partial \psi^A_i},       
{\partial \over \partial \bar \chi^A},       
\beta^A, \gamma^{ABC},      
\bar\chi^A,       
\bar\chi {\partial \over \partial \bar \chi}       
{\partial \over \partial \bar \chi},       
\psi{\partial \over \partial \bar \chi} {\partial \over \partial \psi},      
\bar\chi {\partial \over \partial \bar \chi}       
{\partial \over \partial  \psi}$.       
Some of these fermionic terms  applied to $\Psi$       
increase the fermionic order by a factor of one (e.g, $\bar\chi$), 
while others as $\bar\chi {\partial \over \partial \bar \chi}       
{\partial \over \partial \psi}$ decrease       
it by the same amount.

In the following we will describe  two cases separately: when       
the analytic potential $P(\Phi)$ is arbitrarly and       
when is identically set to zero. We       
will begin by the former.

It is worthwhile to stress the following        
result,  which holds regardless we put $P(\phi)=0$ or not (cf. ref. \cite{A21a} 
and references therein). 
Using the symmetry properties of $e^{AA'}_i, n_{AA'},       
\gamma^{ABC}, \varepsilon_{AB}$ we can check that all       
equations which correspond to the terms       
$\gamma, \gamma\beta\beta, \bar\chi\bar\chi\gamma,       
\gamma\gamma\gamma$ , $\gamma\beta\beta\gamma\gamma$,       
$\gamma\bar\chi\bar\chi\gamma\gamma$,       
$\gamma\beta\beta\bar\chi\bar\chi$,       
$\gamma\beta\beta\bar\chi\bar\chi\gamma\gamma$       
in $S_A\Psi = 0, \bar S_{A'}\Psi=0$       
will give a similar expression for the       
the coefficients $A,B_1, B_2, C_1,$ $ D_1,$ $D_2$, $E_1$,        
$F_1$, $ F_2$, $G_1$, $H_1$, $I_1$.       
Namely,       
\begin{equation} P(a_1,a_2,a_3;\phi,\bar\phi) e^{\pm\left(a_1^2 + a_2^2 + 
a_3^2\right)}.
\label{eq:3.115}       \end{equation}      
The same does not apply       
to the $Z_1, Z_2, Z_3$ coefficients as the $\beta\bar\chi\gamma$ and       
$\beta\bar\chi\gamma(\gamma\gamma)$ terms       
from both  the supersymmetry constraints just mix them  with       
other coefficients in $\Psi$.  This can be seen, e.g., from 
 the equations corresponding to       
$\gamma_{DEF}(\beta \beta)$ term 
in $\bar S_{A'} \Psi = 0$:      
\begin{equation} 2 \epsilon^{i j k} e_{A A' i} \omega^A_{~~B j} n^D_{~~B'} e^{C B'}_{~~~~k}       
B_{1}      
 -  \hbar  n^D_{~~B'} e^{C B'}_{~~~~i} {\delta B_{1} \over \delta e^{B      
A'}_{~~~~i}}  
 +  ( B C D \rightarrow C D B ) + ( B C D \rightarrow D B C ) = 0~. 
\label{eq:3.116}
\end{equation}

Consider now the equations obtained from       
$S_A \Psi = 0$ 
with terms linear in $\beta$ and $\gamma$. 
After contraction with expressions in       
$e^{AA'}_i, n_{AA'}$ and integrating over  $S^3$, we get 
 $ C_1 = 0$.      
From the linear terms in       
$\beta$ and $\gamma$ from $\bar S_{A'} \Psi = 0$ we get 
a relation between $Z_1$ and $B_1 \simeq Y$ (see \cite{A21a}):      
\begin{equation}      
Y(a_1 a_2 a_3; \phi, \bar\phi)       
e^{ \left[  {8 \pi^2 \over \hbar}~\left( a_1^2      +
 a_2^2 + a_3^2 \right) \right]}  + 2\bar\phi Z_1       
 - 8i {\partial Z_1 \over \partial \phi} = 0.
\label{eq:3.117}
\end{equation}

For the particular case of $B_1 = 0$,        
i.e., $Y = 0$, it follows from the remaining equations       
that the only possible solution is $\Psi = 0$.       
For an arbitrarly $B_1$, eq. (\ref{eq:3.117}) allows to write an       
expression for $Z_1$ in terms of functions of       
$\phi, \bar\phi$ and $a_1, a_2, a_3$.       
If we use that expression in other equations, we       
get other formulas for other bosonic coefficients.       
From this procedure, we would get       
the general solution of this extremely complicated       
set of differential equations.       
Although apparently possible, we could not       
establish a definite result in the end       
due to the complexity of the equations involved.       
As in \cite{A9,A11}, no easy way is apparent of       
obtaining an analitycal  solution to this set of       
equations. Moreover, the exponential terms       
$e^{K/2}$ lead to some difficulties.

Let us now consider the case when we choose the analytical potential       
to be identically zero. From the equations directly obtained from       
$S_A \Psi = 0, \bar S_{A'} \Psi = 0$ we have self-contained       
groups of equations relating the 15  wave function coefficients. 
This applies to 3 groups involving $(A, B_1, B_2, C_1$, $Z_1)$,       
$(G_1, D_1, D_2, E_1, Z_2)$ and $(H_1, F_1, F_2, I_1, Z_3$).      
Moreover,       
 the equations corresponding to the terms linear in       
$\beta,\gamma,\bar\chi$ in $\bar S_{A'}\Psi = 0$ and       
$\beta\bar\chi\bar\chi(\gamma\gamma)^2, \beta\beta\bar\chi(\gamma\gamma)^2$       
in $S_A \Psi = 0$ completly determine the coefficients       
$A$ and $I_1$. In addition, $A$ and $I_1$  do not appear in any other       
equation. We have then        
\begin{equation}A = f(\phi) e^{-{{8\pi^2}\over \hbar} \left[
a_1^2 + a_2^2 + a_3^2\right]}, ~
I_1 = k({\bar\phi}) e^{-{{8\pi^2}\over \hbar}\left[
a_1^2 + a_2^2 + a_3^2\right]}       
e^{-2\pi^2\phi      
\overline{\phi}}. 
\label{eq:3.118}
\end{equation}        

The equations involving       
$B_1,B_2,C_1,Z_1$ can also be said to be self-contained in the       
same sense. They involve only these coefficients and       
no other. Moreover, these coefficients do not occur in any       
other equations. This can be   checked, namely from the       
equations for the terms linear in $\beta,\bar\chi$ in $S_A \Psi = 0$,       
$\beta\beta\bar\chi$ and $\beta\bar\chi\bar\chi$ in $\bar S _{A'} \Psi = 0$,       
$\beta\beta\gamma$ in $\bar S _{A'} \Psi = 0$,       
$\bar\chi\gamma\gamma, \beta\gamma\gamma, \gamma\gamma\gamma$ in       
$\bar S _{A'} \Psi = 0$, $\gamma$ in $S_A \Psi = 0$.       
The previous ones  in $\bar \chi\gamma\gamma, \beta\gamma\gamma,       
\gamma\gamma\gamma, \gamma$ just involve $C_1$.       
All the other equations  have $B_1, B_2, Z_1$.       
However, the $\beta\bar\chi\gamma$ equation in $\bar S_{A'} \Psi = 0$       
mixes $B_1,C_1,Z_1$. Actually, is the only equation which       
mixes $C_1$ with the remaining  bosonic coefficients in the       
corresponding group.        
The same structure of equations and relations between coefficients       
also occur, in particular for   the subsets involving       
$D_1,D_2, E_1,Z_2$ and  $F_1,F_2,I_1,Z_3$.

From the  analysis of the groups of equations       
which includes $B_1,B_2,C_1,Z_1$ ($\beta\beta\bar\chi$, $\beta\bar\chi\bar\chi$   
 equations      
from $\bar S_{A'} \Psi = 0$ and $\beta$, $\bar \chi$      
equations from $S_A \Psi = 0$) we get       
${Z}_1 = 0$.       
Consequently, the equations corresponding  to  
only $\beta$ and $\gamma$  involve just $B_1$ and       
$C_1$. These equations are then as the ones in the case of       
a Bianchi-IX with $\Lambda = 0$ and no supermatter \cite{A6,A12}.       
The only possible solution of these equations       
with respest to $a_1, a_2, a_3$ is  the       
trivial one, i.e.,  $B_1 = C_1 = 0$. The equations corresponding        
to $\bar\chi$ and  combinations of it with       
$\beta$ or $\gamma$ would give, with $B_1 = C_1 = Z_1 = 0$,       
the dependence of $B_2$ on $a_1,a_2,a_3,\phi$. This corresponds to       
\begin{equation}B_2 = h(\bar\phi)       
a_1a_2a_3 e^{-{{8\pi^2}\over \hbar}\left[
a_1^2 + a_2^2 + a_3^2\right]}       
e^{-2\pi^2\phi      
\overline{\phi}}. 
\label{eq:3.119}            
\end{equation}

This pattern repeats itself in  a similar way when we       
consider the two groups  involving $D_1,D_2,E_1$, $Z_2$ and       
$F_1,F_2,Z_3$. We get $E_1 = G_1 = H_1 = 0$ from       
$Z_2 = D_2 = 0$, $Z_2 = D_1 = 0$, and $Z_3 = F_1 = 0$.        
Hence, besides $A$ and $I_1$, only $B_2$ and       
$F_2$ will be different from zero.       
We can then write   for the       
solution of the constraints 
\begin{eqnarray}  \Psi &= &       
  f(\phi) e^{-{{8\pi^2}\over \hbar}
\left[a_1^2 + a_2^2 + a_3^2\right]}   
+ 
 h(\bar\phi) a_1a_2a_3 e^{\left[-a_1^2       
- a_2^2 - a_3^2\right]} e^{-2\pi^2 \phi      
\overline{\phi}}      
 \overline{\chi}_A\overline{\chi}^A  \nonumber \\      
&      
+ &      
 g(\phi) a_1a_2a_3 e^{{{8\pi^2}\over \hbar}
\left[a_1^2 + a_2^2  + a_3^2\right]}      
e^{-2\pi^2\phi      
\overline{\phi}}      
\beta_A\beta^A(\gamma_{BCD}\gamma^{BCD})^2 \nonumber \\       
&+ &      
  k(\bar\phi)  e^{{{8\pi^2}\over \hbar}
\left[a_1^2 + a_2^2 + a_3^2\right]}      
\overline{\chi}_A\overline{\chi}^A\beta_E\beta^E(\gamma_{BCD}\gamma^{BCD})^2. \label{eq:3.120}
\end{eqnarray}

\section{Bianchi class A models from N=2 supergravity}

\indent 

In this section we will address the canonical formulation of Bianchi class A models in 
$N=2$ supergravity, summarizing ref. \cite{A19c,A24}. 
$N=2$ supergravity \cite{6,16,20}       
 couples a graviton-gravitino pair with other pair  constituted by       
{\em another} gravitino and a Maxwell field.       
It contains a manifest $O(2)$ invariance which       
rotates the two gravitinos into each other.  We consider two cases: 
when the internal symmetry $O(2)$  
is either $(a)$ {\it global} or $(b)$ {\it local}.

The action for the general theory 
in case $(a)$
can be written as \cite{6}      
\begin{eqnarray}      
{\cal L} = & - & \frac{e}{2 \kappa^2} R[e^a_\mu,\omega(e^a_\mu, \psi^{(a)}_\nu)] -        
{e \over 2} \bar \psi^{(a)}_{\mu} 
\epsilon^{\mu \nu\rho \sigma} \gamma_5 \gamma_\nu      
D_{\rho}(\omega) \psi^{(a)}_{\sigma} \nonumber \\      
& - &  {e \over 4} F^2_{\mu \gamma} + {\kappa \over 4 \sqrt 2}       
\bar \psi^{(a)}_{\mu}[e (F^{\mu \nu} 
+ \hat F^{\mu \nu}) + {1 \over2} \gamma_5({\tilde F^{\mu \nu}}      
+ \tilde {\hat F}{}^{\mu \nu})] \psi^{(b)}_{\nu} \epsilon^{ab},      
\label{eq:4.121}
\end{eqnarray}      
where       
\begin{equation}      
\hat F_{\mu \nu} = \left( \partial_{\mu} A_{\nu} - {\kappa \over 2 \sqrt 2} \bar \psi^{(a)}_{\mu}      
\psi^{(b)}_{\nu} \epsilon^{ab} \right) - (\mu \leftrightarrow \nu),      
\label{eq:4.122}
\end{equation}      
and $\tilde {F}_{\mu \nu}$ equals $\epsilon_{\mu \nu \rho \sigma} F^{\rho \sigma}$.      
The  gravitinos $\psi^{(a)}_\mu$ are here depicted in 
4-component representation,  $(a)=1,2$ are $O(2)$ group indices and $A_\mu$ is a Maxwell field.  $\omega$ is the connection,       
$\gamma_\mu$ are Dirac matrices and $\gamma_5 = \gamma_0 \gamma_1 \gamma_2 \gamma_3$.       
Furthermore, $\epsilon^{12} = 1, \epsilon^{21} = -1$.          
           
{\it En route} to the  canonical quantization of Bianchi class A models,  we require       
the following two steps to be complied.  On the one hand, we ought to re-write the 
action (\ref{eq:4.121}) in       
2-component spinor notation. We do so using the conventions in \cite{8}       
(cf. also \cite{4,6}). On the other hand, we 
impose a consistent Bianchi anz\"atse for all fields.

Choosing a
 symmetric basis, the tetrad components 
$e^a_i$ are time dependent only, like
the spatial components of the gravitino fields 
$\psi^{A(a)}_i$ and the Maxwell field 
$A_i$. We also require the other components to 
be time dependent only. Notice that as consequence 
of choosing a symmetric basis we have 
$F_{\mu\nu} = A_{\nu ,\mu} - A_{\mu , \nu} + A_{\sigma} C^\sigma_{\mu\nu}$, where $C^\sigma_{\mu\nu} = 0$ if one or 
more indices are equal to 
zero\footnote{This simply 
means that according to the chosen 
Bianchi type we can have either a 
 pure electric, magnetic or both fields. 
See ref.\cite{N=2a} for an early review of  
Bianchi minisuperspace models in the 
presence of electromagnetic fields. 
Pure uniform magnetic or electric
 fields (but not both)  are only 
allowed for the Bianchi types 
I, II, III, VI ($h=-1$), VII ($h=0$) 
with possible constraints and are 
forbidden for the types IV, V, VI 
($h\neq -1$), VII ($h\neq 0$), VIII, 
IX. For types IV, V, VI ($h\neq -1$), 
VII ($h\neq 0$), VIII, IX, however, 
 both electric and magnetic fields are present 
but  they are parallel to each other (null Poyinting vector).}.

The momentum conjugate to the  vector field $A_i$ is      
\begin{eqnarray}      
\pi^i & =  & e h^{ij} \, \partial_0 A_j - {e \kappa \over \sqrt 2}  \, \partial_0 A_i \epsilon^{ab}      
(\bar \psi^{(a)}_{0A'} \bar \psi^{(b)iA'} + \psi_0^{(a)A} \psi^{(b)i}_{~~~A}) \nonumber\\      
& + & {i e \kappa \over 2 \sqrt 2} \, \partial_0 A_i \, \epsilon^{ijk} \epsilon^{ab}      
(\bar \psi^{(a)}_{jA'} \bar \psi_k^{(b)A'} - \psi_j^{(a)A} \psi^{(b)}_{kA}).      
\label{eq:4.123}
\end{eqnarray}

Notice that we do not get any       
gauge -- central charge -- constraint term of the form       
$A_0 Q$. This is due to our 
specific homogeneous ans\"atze  choices 
above mentioned and also to
 the choice of  {\it global} invariance . Had we considered a mininal coupling,       
i.e., gauging the $O(2)$ transformations \cite{20}       
then  a   $A_0 Q$ term 
could be present in the Hamiltonian.

The Dirac bracket relations are now:      
\begin{equation}      
\left[ e_i^{~AA'}, {\hat p}^j_{BB'} \right]_D  =  \delta_i^j \delta_B^A \delta_{B'}^{A'},~
\left[ \psi_{i}^{(a)A}, \bar \psi_{j}^{(b)A'} \right]_D  =  - \delta^{ab} D^{AA'}_{~~ij},   ~   
\left[ A_i, \pi^j \right]_D  =  \delta_i^j, \label{eq:4.124}
\end{equation}      
where we have defined      
$     
{\hat p}^i_{AA'} = p^i_{AA'} - {1 \over 2} \epsilon^{ijk} \psi^{(a)}_{jA} \bar \psi^{(a)}_{kA'}.      
$
The Lorentz constraints are      
$
J_{AB} = p_{i(A}^{~~~A'}e^i_{B)A'} + \psi^{(a)}_{i(A} \pi^{(a)i}_{B)},      
$
and its conjugate.       
 Multiplying  the Lorentz constraints by $\omega_{0AB}$ and $\bar \omega_{0A'B'}$ and       
 adding them in this way  to the supersymmetry      
constraints, the supersymmetry constraints take the form      
\begin{equation}      
\bar S^{(a)}_{A'} = -i p^{i}_{AA'} \psi^{(a)A}_{i} + \kappa \epsilon^{ab} \left[ \pi^i - i h^{\frac{1}{2} } {\kappa \over 4}       
\epsilon^{imn} \epsilon^{cd} \psi_m^{(c)B} \psi_{nB}^{(d)} + i h^{\frac{1}{2} } \frac{1}{8} F_{jk} \epsilon^{ijk}  \right] \bar \psi^{(b)}_{iA'}      
\label{eq:4.125}      
\end{equation}          
and its Hermitian conjugate.

The promotion of the $O(2)$ internal symmetry to a gauge transformation \cite{6} 
implies that the following terms (already written in 2-component spinor notation) 
should be added to the Lagrangian in (\ref{eq:4.121}):      
\begin{eqnarray}      
- e \Lambda & -   & e \sqrt{-\frac{2 \Lambda}{3}} \left[ \psi^{A (a)}_\mu 
e_{AB'}^{\mu} e_{B}^{B' \nu} \psi^{B (a)}_\nu + \bar \psi^{ (a)}_{ A' \mu} e^{BA' 
\mu} e_{BB'}^{ \nu} \bar \psi^{B' (a)}_\nu \right] \nonumber \\ & - &  \frac{1}{2} e 
\sqrt{-\frac{\Lambda}{6}} \epsilon ^{\mu\nu\rho\sigma} \left[ \psi_\mu^{A (a)} 
e_{\nu AA'} A_\rho \bar \psi_\sigma^{A' (b)} - \bar \psi_{\mu A'}^{(a)} e_{\nu}^{ AA'} A_\rho  \psi_{\sigma A}^{ (b)}\right] \epsilon^{ab},      
\label{eq:4.126}      
\end{eqnarray}      
where the cosmological constant $\Lambda$ is related to the gauge coupling constant, 
${\rm g}$ by $\Lambda = -6 {\rm g}^2$. 
Eq. (\ref{eq:4.126}) is 
a consequence of coupling {\em minimally}  
the Maxwell field to the fermions.         
From local invariance, we now get a gauge -- central charge -- 
constraint term $A_0 Q$ in the action, which for the case of our Bianchi models takes 
the form       
\begin{equation}      
Q =  h^{\frac{1}{2}} \sqrt{-\frac{\Lambda}{6}}\epsilon^{ab}\epsilon^{ijk} 
\psi_i^{(a)A} e_{j AA'} \bar \psi^{(b) A'}_k.      
\label{eq:4.127}      
\end{equation}        
Furthermore,  the $\bar S^{(a)}_{A'}$  supersymmetry constraint 
(\ref{eq:4.125}) gets the additional contributions,       
\begin{equation}      
h^{\frac{1}{2}} \sqrt{-\frac{ \Lambda}{6}} \epsilon^{ab} \epsilon^{ijk} e_{i AA'} 
A_j \psi_k^{(b)A} - h^{\frac{1}{2}} \sqrt{-\frac{2 \Lambda}{3}} e^{A~~i}_{~A'} 
n_{AB'} \bar \psi_i^{(a) B'}.      \label{eq:4.128}
\end{equation}

Quantum mechanically, the bracket relations become      
\begin{equation}      
\left[ e_i^{~AA'}, {\hat p}^j_{BB'} \right]  =  i \hbar \delta_i^j \delta_B^{~A} 
\delta_{B'}^{~A'},~
\left\{ \psi_{i}^{(a)A}, \bar \psi_{j}^{(b)A'} \right\}  =  -i \hbar \delta^{ab} 
D^{AA'}_{~~ij},   ~    
\left[ A_i, \pi^j \right]  =  i \hbar \delta_i^j.      
\label{eq:4.129}
\end{equation}      
Choosing $\left(A_i, e_{iAA'}, \psi^{(a)A}_i\right)$ as the  coordinates variables in 
our       
minisuperspace, we consequently employ       
\begin{equation}      
{\hat p}^j_{AA'} \rightarrow -i \hbar {\partial \over \partial e_j^{AA'}},~~~       
\bar \psi_j^{(a)A'} \rightarrow -i \hbar D^{AA'}_{~~ij} {\partial \over \partial \psi_i^{(a)A}},~~~      
\pi^i \rightarrow -i \hbar {\partial \over \partial A_i}.      
\label{eq:4.130}
\end{equation}        
After all the simplications, we finally obtain the quantum supersymmetry constraints    as  
\begin{eqnarray}      
\bar S^{(a)}_{A'} & = & - \hbar \psi^{(a)A}_i {\partial \over \partial e_i^{AA'}}      
- {1\over2} \hbar \epsilon^{ijk} \Gamma^{ab} \psi_i^{(a)A} \psi^{(b)}_{jA} 
D^B_{~A'lk}       
{\partial \over \partial \psi_l^{(b)B}} \nonumber\\      
& + & \hbar  \kappa \epsilon^{ab} D^c_{~A'mi} \left[ \hbar {\partial \over \partial 
A_i} - h^{\frac{1}{2}}{\kappa \over 4}      
\epsilon^{ijk} \epsilon^{cd} \psi_j^{(c)B} \psi_{kB}^{(d)} + \frac{i}{8} h^{\frac{1}{2}} \epsilon^{ijk} F_{jk} \right] {\partial \over \partial \psi_m^{(b)C}}~,      
\label{eq:4.131}
\end{eqnarray}      
plus the terms in (\ref{eq:4.128})
and    $  
S^{(a)}_{A}$ is the  Hermitian conjugate, 
where $\Gamma^{12} = \Gamma^{21} = 1$ and the remaining are zero.

We now address the physical states which are solutions of       
the above constraints. The quantum states may be described by the wave function $\Psi 
(e_{AA'i}, A_j, \psi^{(a)}_{Ai})$. From the Lorentz 
constraint,  $\Psi$ must  be expanded in 
even powers of $\psi^{(a)}_{Ai}$, symbolically represented by $\psi^0, \psi^2$ up to 
$\psi^{12}$. This is due to 
the   anti-commutation relations of the six spatial components of 
the {\it two} types of gravitino.

An important consequence of eq. (\ref{eq:4.131}) is that 
{\it neither} of  the supersymmetry constraints 
$\bar S_{A'}^{(a)}$ and $S_A^{(a)}$ conserves       
fermion number. In fact,  a mixing between fermionic 
modes occurs for $\Psi$. This is due to  terms involving  ${\partial \over \partial A_i}       
{\partial \over \partial \psi_m^{(b)C}}$      
and ${\partial \over \partial A_i} \psi^{(b)}_{iA}$ or the ones 
associated with $F_{jk}$. While the remaining
 fermionic terms in  $\bar S_{A'}^{(a)}$  (\ref{eq:4.131}) 
 act on $\Psi$ by 
increasing the fermionic order by a factor of one, ${\partial \over \partial A_i}       
{\partial \over \partial \psi_m^{(b)C}}$ decrease it by the same amount. Concerning the $S_A^{(a)}$ constraint,  the situation is precisely the reverse.           
The nature of this problem can also be better understood  as follows.

 Let us consider the two fermion level and follow the guidelines described in 
\cite{A20}. 
Since we have 12 degrees of freedom associated with the       
gravitinos, we may expect to have up to $66$ terms in this fermionic sector. Thus, the       
 two-fermion level of the more general ansatz  of the wave function can be written as      
\begin{equation}      
\Psi_2 = \left( C_{ijab} + E_{ijab} \right) \psi^{(a)iB} \psi^{(b)j}_{~~~B}      
+ \left( U_{ijkab} + V_{ijkab} \right) e^i_{~AA'} n_B^{~A'} \psi^{(a)jA} \psi^{(b)kB}, 
\label{eq:4.132}                 
\end{equation}      
where $C_{ijab} = C_{(ij)(ab)}$, $E_{ijab} = E_{[ij][ab]}$, $U_{ijkab} = U_{i(jk)[ab]}$ and      
$V_{ijkab} = V_{i[jk](ab)}$.      
When  $\Psi$ is truncated in the second fermionic order, 
we obtain a set of equations, 
relating 
 $\frac{\partial \Psi_0}{\partial a_i}$   with       
 $\frac{\partial C_{ijab}}{\partial A_i}$, $\frac{\partial E_{ijab}}{\partial A_i}$, $\frac{\partial U_{ijkab}}{\partial A_i}$, $\frac{\partial V_{ijkab}}{\partial A_i}$      
(from $\bar S^{(a)}_{A'}$)       
and       
 $\frac{\partial C_{ijab}}{\partial a_i}$, $\frac{\partial E_{ijab}}{\partial a_i}$, 
$\frac{\partial U_{ijkab}}{\partial a_i}$, $\frac{\partial V_{ijkab}}{\partial a_i}$      
 with       
$\frac{\partial \Psi_0}{\partial A_i}$       
(from $ S^{(a)}_{A}$). Here $a_i, i=1,2,3$,      
stand for scale factors in a Bianchi class A model, $\Psi_0$ denotes  the       
bosonic sector and $\Psi_0, C_{ijab}, E_{ijab}, U_{ijkab}, V_{ijkab}$       
are functions of $A_j, a_i$ solely.        
Moving to the equations corresponding to higher fermionic terms 
this pattern keeps repeating itself, with algebraic terms added to it.

However, the present situation is rather different from the one in FRW models in 
$N=1$ supergravity with supermatter (cf. ref. \cite{A22}). 
In the FRW case, the mixing occurs only within each fermionic 
level 
and decoupled from other Lorentz invariant fermionic  sectors with different order.
 In the present case, the mixing   is between  fermionic 
sectors of the same and {\it any}  different adjacent order. This situation is 
quite 
similar to 
what a
cosmological constant causes.

It is tempting to see the relations between       
  gradient  terms such as $\frac{\partial \Psi_0}{\partial a_i}$ with       
 $\frac{\partial \Psi_2}{\partial A_i}$ (from $\bar S^{(a)}_{A'}$) and       
 $\frac{\partial \Psi_2}{\partial a_i}$ with       
$\frac{\partial \Psi_0}{\partial A_i}$ (from $ S^{(a)}_{A}$) 
as a consequence of $N=2$ supergravity \cite{6,16,20} realizing Einstein's 
dream of       
unifying gravity with electromagnetism. 
These relations establish a duality between 
the  coefficients of $\Psi$ 
in fermionic sectors of adjacent order relatively  to 
the intertwining of the 
derivatives 
$\frac{\partial}{\partial A_i}, \frac{\partial}{\partial a_i}$. 

When considering  $O(2)$ local invariance 
a gauge constraint (\ref{eq:4.127}) appears. 
In spite of the additional difficulties caused now by  cosmological constant and 
gravitinos mass-terms, the  presence of (\ref{eq:4.127}) allows us to extract some 
information concerning the form of the wave function.         
Quantum mechanically, the gauge constraint takes the form      
\begin{equation}      
\hat Q \sim \varepsilon^{ab} \psi_m^{(a)A} \frac{\partial}{\partial \psi^{(b)A}_m}.      
\label{eq:4.133}      
\end{equation}      
Notice that the gauge constraint has no factor ordering problem due to the presence of $\varepsilon^{ab}$.       
It is then simple to verify that the following fermionic expressions satisfy the gauge constraint operator above:       
\begin{equation}      
\hat Q S^{mn} \psi^{(c)}_{mA} \psi_n^{(c)A} = 0~, ~
\hat Q T_{i[jk]} e^i_{AA'} n_B^{~A'} \psi^{(a)jA} \psi^{(a)kB} = 0.      
\label{eq:4.134}
\end{equation}      
where       
$S^{mn} = S^{(mn)}$.        
Hence, the most general solution of the quantum gauge constraint   can be written as       
\begin{equation}      
\Psi = \Psi \left[ e^i_{AA'}, A_j, S^{mn} \psi^{(c)}_{mA} \psi_n^{(c)A}, T_{i[jk]} e^i_{AA'} n_B^{~A'} \psi^{(a)jA} \psi^{(a)kB}\right].      
\label{eq:4.135}
\end{equation}              
However,       
obtaining non-trivial solutions of the supersymmetry constraints in the metric 
representation for Bianchi class A models with $\Lambda \neq 0$ proved too difficult.

Alternatively, a simplified 
adaptation of the method  outlined in ref. \cite{A23b}
was employed in \cite{A24}. We 
will use a solution previously obtained in 
\cite{16a} for the   case of the general theory 
of $N=2$ supergravity. For the case of Bianchi class A models, it becomes     
$      
\Psi = e^{i(F + G) +H},      
$
where      
$
F  \sim  -\frac{1}{\sqrt{\Lambda}} $$\left[ m^{ij} \psi^{(a)A}_i \psi^{(a)}_{jA} +       
\epsilon^{ijk} \psi^{(a)A}_i \psi^{(a)}_{kB} {\cal A}_{jA}^B \right],      
$    
$G  \sim  \frac{i}{\Lambda} $ $  \left[ m^{ij} {\cal A}_i^{AB} {\cal A}_{jAB} +       
\frac{2}{3} {\cal A}_i^{AB} {\cal A}_{jB}^C {\cal A}_{kCA} \right],       
$
$H  \sim  -\left[  \varepsilon^{ijk} 
\psi^{(a)i}_A\right.$ $ \psi^{(a) A k} $ $\left.  A_j \right. $ $ + \left. 
\varepsilon^{ijk} A_i F_{jk} \right],      
$
where ${\cal A}_{iAB}$ are the complexified spin-connections.

In the metric representation, and for the case of a Bianchi-IX model, 
the 12th  fermion level term, with half for each gravitino type involves a bosonic coefficient as 
\begin{equation}       
\Psi_{12} (e_{AA'i}, A_i) \sim (A^2 + A^4 + A^6) e^{-m^{ij}e_i^{CC'}e_{jCC'}} e^{\varepsilon^{ijk}A_i F_{jk}},    \label{eq:4.136}              
\end{equation}      
where $A^2 = A_i A^i$. Other    bosonic coefficients for the 8th fermionic order 
would include        
\begin{eqnarray}      
\Psi_8 & \sim & A^4 e^{-(a_1^2 + a_2^2 + a_3^2) + (a_1 a_2 + a_2 a_3 + a_3 a_1)} e^{\varepsilon^{ijk}A_i F_{jk}}, \label{eq:4.137}            \\      
\Psi^\prime_8 &\sim &  A^4 e^{-(a_1^2 + a_2^2 + a_3^2) + (a_1 a_2 - a_2 a_3 - a_3 a_1)} e^{\varepsilon^{ijk}A_i F_{jk}},      \label{eq:4.138}
\end{eqnarray}      
half for each gravitino type.

The Chern-Simons functional constitutes       
an exact solution to the Ashtekar-Hamilton-Jacobi equations of       
general relativity with non-zero cosmological constant \cite{15a}. Furthermore,       
the exponential of the Chern-Simons functional provides a semiclassical       
approximation to the no-boundary   wave function in some minisuperspaces \cite{N=2b}.      
However, the      
exponential of the Chern-Simons functional has also been shown {\em not} to be a       
proper quantum state, because it is non-normalizable \cite{mena}
(see ref. \cite{A24} and references therein for related discussions).

\section{ Assessment: results achieved and further research}

\indent 

The purposes of the present report were twofold.

\vspace{0.15cm}

On the one hand, we were committed to describe in some detail the fundamental  
elements and results so far achieved in the canonical 
quantization of supergravity theories. We restricted ourselves to a 
metric  and fermionic differential 
operator representation  \cite{1}-\cite{4}, \cite{A5}-\cite{R}, 
\cite{C1}-\cite{C4}.
We did so within the possible length assigned to these reviews.

A particular emphasis was put on supersymmetric Bianchi 
minisuperspaces. These are simple models obtained from a 
truncation on the full theory of supergravity. In spite of such limitations, 
these models can  provide useful guidelines concerning the 
general theory. But the canonical 
quantization of supersymmetric Bianchi models   
was soon confronted with troublesome prospects: few or no 
physical states seemed to be allowed 
  \cite{A6}-\cite{A12}, \cite{A13}-\cite{A15}. This 
apparently  
implied that such 
minisuperspaces were useless models as far as the general 
theory was concerned. But a refreshing breakthrough 
has been recently proposed \cite{A21,A20,A23b}, providing the 
{\it correct} spectrum of solutions. Hence, the subject 
gained new momentum.

The inclusion of matter in supersymmetric 
FRW and Bianchi models 
brought further difficulties: results 
were apparently incompatible or incomplete \cite{A10,A22,A21a,A23,A18a} and 
no states could be found in other cases \cite{A17,A18}. 
However, some interesting results and recent improvements are 
described in ref. \cite{A23,cc,essay}. As far as (more complicated) 
Bianchi models derived from N=2 supergravity are concerned, these were 
shown 
{\it not} to preserve fermionic number due to the 
presence of the Maxwell field \cite{A19c,A24}.

The canonical quantization of the general theory was also addressed here. 
Premature claims for the existence of states with 
finite fermionic number \cite{C6,C4} were   opposed 
by unavoidable objections \cite{C1,C2,C5}. In fact, it was  
shown in ref. \cite{C1,C2} that quantum states could only have an {\it infinite} fermionic 
number, in accordance with what was predicted in \cite{C5}. Such a solution was presented in 
ref. \cite{C3} and shown to correspond to a wormhole solution in a minisuperspace 
sector.

The following table summarizes the type of solutions found 
so far within the canonical 
quantization of $N=1$ and $N=2$  supergravities. The 
initials/symbols $HH, WH, CS$ and $``?''$ stand for 
no-boundary (Hartle-Hawking), wormhole (Hawking-Page), 
Chern-Simmons solutions and {\it not yet found}, respectively. 
For further details, the interested reader ought to 
consult sections 2, 3 and 4  and the references thereby mentioned.

\[
\begin{tabular}{|c||c|c|c|}
\hline
{\bf Models} ${\rightarrow}$  & $k=+1$ {\sf FRW} & 
{\sf Bianchi class-A} & {\sf Full theory} \\
\cline{1-1} {\bf Solutions} ${ \searrow}$ & & & \\
\cline{1-1} {\bf Supergravity theory} ${\downarrow}$ & & & \\
\hline\hline
{\sf Pure N=1} & HH, ``WH'' & HH, WH & WH \\
\hline
{\sf N=1 with} $\Lambda$ & HH & CS $\mapsto$ ``WH'', HH & ? \\
\hline
{\sf N=2} & --- & CS & CS \\ 
\hline
{\sf N=1 with scalar fields}  & {\it not quite} HH or WH & Not WH or HH & ? \\
\hline
{\sf N=1 with vector fields} & HH, WH & ? & ? \\
\hline
{\sf N=1 with general matter} & $\Psi = 0$ & ? & ? \\
\hline
\end{tabular}
\]

\vspace{0.15cm}

But on the other hand, we also wanted to motivate further research. The canonical 
quantization of supergravity theories is by no means a closed book. There are 
still many open (and serious) problems. 
These are  waiting for  adequate    explanations 
in order to safeguard the future of the subject.  
Hence the subtitle {\it shaken not stirred}\footnote{British 
agent James Bond (OO7) usual motto when asking 
for his dry martini.}: supersymmetric quantum cosmology 
may seem afflicted by current disapointments but if 
substantial energy and committment are invested, we may still 
achieve a sucessful outcome. 
We thus reach the end of this review. However, this is not the 
end for supersymmetric quantum cosmology. In fact, let us consider 
this point as an intermediate  stage and  bequeath a series of further  
tempting  challenges\footnote{Where the author and collaborators have been  
currently involved regarding some of them.} for the 
canonical quantization of supergravity:

\begin{itemize}

\item Why a comprehensible 
identification of the 
Hartle-Hawking solution  for FRW models in N=1 
supergravity with scalar supermultiplets \cite{A23} has 
been so problematic. For recent improvements see ref. \cite{cc,essay};

\item Why  there are {\it no}  physical states in a locally supersymmetric 
FRW model with gauged supermatter \cite{A17,A18} 
 but we can find them in  a   
FRW model with   Yang-Mills fields \cite{A18a}(see also ref. \cite{rosu}); 

\item Obtaining conserved currents in supersymmetric 
minisuperspaces from  
$\Psi$ \cite{A17a}-\cite{cc}. It seems that this is not possible 
unless vor very simple cases (see ref. \cite{cc,essay} for more details);

\item The validity of the 
minisuperspace approximation in locally supersymmetric models;

\item Including larger gauge groups in supersymmetric FRW models with 
supermatter;

\item In the approach of  ref. 
\cite{A21,A20} the correct 
spectrum of solutions  can only  be achieved from 
an associated Wheeler-DeWitt equation. Could we 
regain the same results but directly 
 from the supersymmetry constraint 
  equations in \cite{A20a}?

\item The approach of ref. \cite{A23b} produces {\it different} amplitudes 
in the same fermionic sector. But  in \cite{A21,A20} the amplitude in the 
same fermionic sector are re-expressed in just  a  {\it single one},  
which satisfies  a Wheeler-DeWitt--type equation. Any relation with the 
above point or something else deserves further exploration?

\item Obtain a satisfactory supersymmetric FRW model with 
just gauge fields from suitable ans\"atze for the vector and 
fermionic fields \cite{A18a};

\item Study the canonical quantization of black-holes in 
N=2 and N=4  supergravities \cite{R}  (see ref. 
\cite{T,n=2bh} for related issues);

\item Perform the 
canonical quantization of FRW models in N=3 supergravity;

\item Analyse  the Chern-Simons states 
\cite{A23b}, \cite{D8}-\cite{16a}, \cite{15a,N=2b,mena} 
(Recent contributions present in ref. \cite{CSRG,CSRGa,CSadyc} may 
prove to be useful in addressing this issue). 
Are these enough  and physically valid? Should we (and 
{\it how}) consider  other solutions?

\item Describe the results and features present in ref. \cite{6,21} concerning finite 
probabilities for photon-photon scattering in N=2 supergravity but now from 
a canonical quantization point of view;

\item Try to 
obtain a no-boundary (Hartle-Hawking) solution, as well 
as other solutions corresponding to 
gravitons (in the same sector)   or pairs of gravitinos 
(in sectors differing by an even fermion number)  as 
quantum states in the full theory \cite{C3}. It would also
be important to consider the case where  
supermatter is present;

 \item Deal properly with any divergent factor present in section 2.3 
and ref. \cite{C3};

\item Study the canonical quantization of supergravity theories in 
$d>4$ dimensions;

\item The study of FRW supersymmetric minisuperspaces within a 
a superspace formalism has been recently brought back 
\cite{SSOO} (see ref. \cite{A11} for a former description). 
It would be particularly interesting (as mentioned in \cite{SSOO}) to 
consider Bianchi models and FRW models with supermatter. 
Namely, to see if the problems afflicting the FRW models mentioned in this 
review still remains. 

\item Another issue of interest 
is that  the action of pure 
N=1 supergravity {\it with boundary terms} currently used 
\cite{4}  is {\it not}
 fully invariant under 
supersymmetry transformations. But 
 a particular fully invariant 
action  
 has been presented in ref. \cite{A19} 
 for the case of Bianchi class A models. A generalization of this action 
for the full theory would be most welcomed \cite{aussie} 
(see ref. \cite{Teit} for a related discussion in the context of 
general relativity). 
  Then proceed with the corresponding 
quantization and obtain physical states;



\item It would also  be particularly interesting to address 
the following (fundamental) issues of quantum gravity but now 
within a supersymmetric scenario: the problem of time and how 
classical properties may emerge. See ref. \cite{A19} and subsequently 
in \cite{HughRG} for an analysis of the problem of time in 
quantum supergravity. Ref. \cite{cc,essay} introduce and discuss 
the issue of retrieving classical properties from supersymmetric 
quantum cosmologies;

\end{itemize}
 
\vspace{0.15cm}

{\large\bf  Acknowledgments}

The author  gratefully 
acknowledges 
for the support of  
 a JNICT/PRAXIS-XXI fellowship (BPD/6095/95). 
In addition, he is  most thankfull to 
Dr. K K Phua
(WSPC -- IJMP-A) for his  congenial invitation to write this review 
and to Ms. Lim Feng Nee  (WSPC -- IJMP-A)  for  assistance. 
 Estimable suggestions from  a IJMP-A editor/referee were also 
most welcomed. 
Comments from D. Freedman, R. Graham, H. Luckock, O. Obregon 
 and R. Kallosh are also
acknowledged. Special thanks go to 
A.D.Y. Cheng, A. Yu. Kamenshchik, C. Kiefer and J. Turner 
for  their most valuable suggestions and comments. 
The author's knowledge and insight on the subjects of supersymmetry, 
supergravity and 
quantum cosmology are  enfolded in this review;  
discussions and conversations  
enjoyed with  A.D.Y. Cheng, P. D Eath, D. Freedman, G.W. Gibbons, 
S.W. Hawking, J.M. Izquierdo, H. Luckock, O. Obregon, 
G. Papadopoulos, 
M. Perry, R. Graham and 
P. Townsend since his stay in University of Cambridge 
were influential in his introduction and 
progress in those subjets.
Finally, and  {\em most important} , 
the author is endlessly grateful for his wife, Teresa, constant and 
unfailling support.

\end{document}